\documentclass[aps,pra,notitlepage,singlecolumn,superscriptaddress,nofootinbib,
amsmath,amssymb,
]{revtex4-1}

\usepackage{epsfig}
\usepackage{soul,xcolor} 
\usepackage{graphicx} 

\usepackage{amsthm}
\usepackage{amsmath}
\usepackage{amssymb}
\usepackage{bbm,amsfonts}
\usepackage{nomencl}
\usepackage{amstext} 
\usepackage{setspace}

\usepackage{url}            
\usepackage{booktabs}       
\usepackage{nicefrac}       
\usepackage{microtype}      
\usepackage{lipsum}
\usepackage{dsfont} 

\usepackage{comment}

\usepackage{algorithm}
\usepackage[noend]{algpseudocode}

\usepackage{hyperref}
\hypersetup{colorlinks=true,citecolor={blue}}

\usepackage{dcolumn}
\usepackage{bm}

\usepackage{multirow}

\usepackage{tikz}
\usetikzlibrary{quantikz}

\ExplSyntaxOn

\keys_define:nn { miguel/label }
 {
  label   .tl_set:N = \l_miguel_label_tl,
  unknown .code:n   = \clist_put_right:Nx \l_miguel_label_clist
                       { \l_keys_key_tl = \exp_not:n { #1 } }
 }
\clist_new:N \l_miguel_label_clist
\box_new:N \l_miguel_label_box
\box_new:N \l_miguel_label_image_box

\NewDocumentCommand{\xincludegraphics}{O{}m}
 {
  \tl_clear:N \l_miguel_label_tl
  \clist_clear:N \l_miguel_label_clist
  \keys_set:nn { miguel/label } { #1 }
  \tl_if_empty:NTF \l_miguel_label_tl
   {
    \miguel_includegraphics:Vn \l_miguel_label_clist { #2 }
   }
   {
    \hbox_set:Nn \l_miguel_label_image_box
     {
      \miguel_includegraphics:Vn \l_miguel_label_clist { #2 }
     }
    \hbox_set:Nn \l_miguel_label_box
     {
      \skip_horizontal:n { -9pt }
      \fcolorbox{white}{white}{\footnotesize \tl_use:N \l_miguel_label_tl}
     }
    \leavevmode
    \box_use:N \l_miguel_label_image_box
    \skip_horizontal:n { -\box_wd:N \l_miguel_label_image_box }
    \hbox_overlap_right:n
     {
      \box_move_up:nn
       {
        \box_ht:N \l_miguel_label_image_box - 
        \box_ht:N \l_miguel_label_box/5
       }
       { \box_use_drop:N \l_miguel_label_box }
     }
    \skip_horizontal:n { \box_wd:N \l_miguel_label_image_box }
   }
 }

\cs_new_protected:Nn \miguel_includegraphics:nn
 {
  \includegraphics[#1]{#2}
 }
\cs_generate_variant:Nn \miguel_includegraphics:nn { V }

\ExplSyntaxOff

\newcommand{\Expectation}{\mathbb{E}}
\newcommand{\Var}{\mathrm{Var}}
\newcommand{\Cov}{\mathrm{Cov}}

\newcommand{\ptest}{p_{\mathrm{test}}}

\newcommand{\readoutSignal}{\mathbf{c}}

\newcommand{\ketbra}[2]{|{#1}\rangle\langle {#2}|}
\newcommand{\braketLR}[1]{\langle {#1} \rangle}

\newcommand{\braketExp}[3]{\left\langle {#1} | {#2} | {#3} \right\rangle}

\newcommand{\query}{x}

\newcommand{\prepOperator}{U}

\newcommand{\querySpace}{\mathcal{Q}}

\newcommand{\labelSpace}{\mathcal{Y}}
\newcommand{\measSpace}{\mathcal{M}}
\newcommand{\prepSpace}{\mathcal{U}}
\newcommand{\timeSpace}{\mathcal{T}}
\newcommand{\fisherInfo}{\mathcal{I}}

\DeclareMathOperator{\Tr}{Tr}

\newcommand{\threecdots}{\cdot\cdot\cdot}

\newtheorem{problem}{Problem}[section]

\global\long\def\argmin{\operatornamewithlimits{argmin}}

\begin{document}


\title{Active Learning of Quantum System Hamiltonians yields Query Advantage}

\author{Arkopal Dutt}
\affiliation{MIT-IBM Watson AI Lab, Cambridge, Massachusetts 02142, USA}
\affiliation{Center for Ultracold Atoms, Massachusetts Institute of Technology, Cambridge, Massachusetts 02139, USA}
\affiliation{Department of Physics, Co-Design Center for Quantum Advantage, Massachusetts Institute of Technology, Cambridge, Massachusetts 02139, USA}

\author{Edwin Pednault}
\affiliation{IBM Quantum, IBM T.J. Watson Research Center, Yorktown Heights, New York 10598, USA}

\author{Chai Wah Wu}
\affiliation{IBM Research AI, IBM T.J. Watson Research Center, Yorktown Heights, New York 10598, USA}

\author{Sarah Sheldon}
\affiliation{IBM Quantum, IBM T.J. Watson Research Center, Yorktown Heights, New York 10598, USA}

\author{John Smolin}
\affiliation{IBM Quantum, IBM T.J. Watson Research Center, Yorktown Heights, New York 10598, USA}
\affiliation{Department of Physics, Co-Design Center for Quantum Advantage, Massachusetts Institute of Technology, Cambridge, Massachusetts 02139, USA}

\author{Lev Bishop}
\affiliation{IBM Quantum, IBM T.J. Watson Research Center, Yorktown Heights, New York 10598, USA}

\author{Isaac L. Chuang}
\affiliation{Center for Ultracold Atoms, Massachusetts Institute of Technology, Cambridge, Massachusetts 02139, USA}
\affiliation{Department of Physics, Co-Design Center for Quantum Advantage, Massachusetts Institute of Technology, Cambridge, Massachusetts 02139, USA}


\begin{abstract}
Hamiltonian learning is an important procedure in quantum system identification, calibration, and successful operation of quantum computers. Through queries to the quantum system, this procedure seeks to obtain the parameters of a given Hamiltonian model and description of noise sources. Standard techniques for Hamiltonian learning require careful design of queries and $O(\epsilon^{-2})$ queries in achieving learning error $\epsilon$ due to the standard quantum limit. With the goal of efficiently and accurately estimating the Hamiltonian parameters within learning error $\epsilon$ through minimal queries, we introduce an active learner that is given an initial set of training examples and the ability to interactively query the quantum system to generate new training data. We formally specify and experimentally assess the performance of this Hamiltonian active learning (HAL) algorithm for learning the six parameters of a two-qubit cross-resonance Hamiltonian on four different superconducting IBM Quantum devices. Compared with standard techniques for the same problem and a specified learning error, HAL achieves up to a $99.8\%$ reduction in queries required, and a $99.1\%$ reduction over the comparable non-adaptive learning algorithm. Moreover, with access to prior information on a subset of Hamiltonian parameters and given the ability to select queries with linearly (or exponentially) longer system interaction times during learning, HAL can exceed the standard quantum limit and achieve Heisenberg (or super-Heisenberg) limited convergence rates during learning.
\end{abstract}

\maketitle

\section{Introduction} \label{sec:intro}
Hamiltonian learning constitutes the problem of learning the Hamiltonian governing the dynamics of a quantum system given finite classical and quantum resources. This is a fundamental problem encountered in identification of quantum systems \cite{cole2005identifying,cole2006identifying,burgarth2012quantum}, operation of quantum information devices \cite{valenti2019hamiltonian,wang2017experimental}, validation of theoretical physical models, and has implications for computational bounds on quantum algorithms \cite{farhi1998analog,demkowicz2015quantum,liu2016power}. In the calibration of quantum computers alone, it is a significant step in each of the following tasks: system characterization, learning device parameters, different sources of noise, gate design \cite{innocenti2020supervised} and control strategies for implementing robust quantum gates with high fidelity. Moreover, a quantum computer typically requires frequent recalibrations to account for drift in parameters over time requiring multiple iterations of some Hamiltonian learning routine. 

The resource requirements for learning a generic many-body Hamiltonian rise exponentially with the system size \cite{mohseni2008quantum}. Even for a fixed system size, however, the achievable learning error $\epsilon$ is fundamentally limited by the number of queries $N$ made to the quantum system. In particular, through repeated system queries, $N$ in general scales as $\epsilon^{-2}$ as a consequence of the central limit theorem. This is commonly referred to as the standard quantum limit (SQL) or shot noise limited scaling. Using quantum resources, however, a number of approaches have shown a much better \textit{Heisenberg limited} scaling of $N \sim \epsilon^{-1}$. The Heisenberg limit is known to be fundamental \cite{giovannetti2004quantum,zwierz2010general,giovannetti2011advances,toth2014quantum,pezze2018quantum}, under a wide range of assumptions \cite{imai2007geometry,berry2015quantum,szczykulska2016multi,gorecki2020pi}, but it is typically only saturated with the help of quantum resources.

This has been achieved using entanglement \cite{giovannetti2006quantum} such as NOON states \cite{bollinger1996optimal,lee2002quantum}, but can also be accomplished without entanglement under certain circumstances. For example in the problem of phase estimation \cite{nielsen2002quantum}, one has to estimate the phase $\phi$ in an unitary operator of form $U=\exp(-i \phi H)$ where $H$ is a Hermitian operator and $\phi$ can also be interpreted as the strength of coupling in a Hamiltonian. It has been shown that Heisenberg limited scaling can be achieved using multi-round protocols using both adaptive measurements \cite{higgins2007entanglement,wiseman2009adaptive} and predetermined non-adaptive measurement sequences \cite{higgins2009demonstrating,kimmel2015robust}. In contrast, there has not been such a detailed study in the case of learning a general many-body or a multi-parametric Hamiltonian. This motivates understanding how might just classical resources be employed to solve the Hamiltonian learning problem with a scaling which surpasses the SQL, and ideally achieves the Heisenberg limit. 

Even if scalings higher than SQL cannot be achieved, it is still desirable to minimize resource requirements. This may be accomplished by changing the estimation procedure used for Hamiltonian learning in combination with engineered experiments. Fast Fourier Transform (FFT) and linear regression are some of the traditional estimation methods which still form the powerhouse of modern Hamiltonian learning strategies \cite{arute2020observation}. It has been shown that adopting alternate estimation methods such as Bayesian estimation \cite{evans2019scalable}, stochastic estimation \cite{krastanov2019stochastic}, and neural-network based Hamiltonian reconstruction \cite{valenti2019hamiltonian} can reduce resource requirements and improve scalability. We will call the reduction in resource requirements achieved by replacing one Hamiltonian learning strategy with another as \textit{query advantage}. This will obviously depend on the two strategies being compared. We will call the strategy being replaced as the baseline. The baseline and our proposed replacements will be discussed in detail later. This brings us to the primary question we tackle in this work: what is a common framework for Hamiltonian learning that can achieve query advantage even if scalings higher than SQL are not achieved, just using classical resources?

One effective framework for surpassing central limit theorem bounds when possible and/or achieving query advantage is \textit{active learning}, e.g.\ using optimal experiment design. In \cite{kosut2004optimal}, this idea was explored for quantum state tomography, process tomography and Hamiltonian learning given a model, but in an offline manner. This has also been used to reduce experiment budget and propose different control schemes \cite{geremia2002optimal}. Active learning of a Hamiltonian is a more challenging problem than for quantum state tomography \cite{nunn2010optimal} where one optimizes over different measurements and process tomography \cite{gazit2019quantum} where one additionally optimizes over initial states, due to the additional control parameter of system time evolution. Active learning thus provides a general framework for making adaptive queries to the quantum system, comprised of initial state, measurements and system time evolution during Hamiltonian learning. In fact, \cite{yuan2015optimal} has shown that with adaptive feedback control, Heisenberg limited scaling can be reached in principle, but a recipe for this is only given for estimation of a single Hamiltonian parameter and the procedure requires prior information of the parameter. This was later extended to multi-parameter Hamiltonians by \cite{kura2018finite}. A common ingredient of these works and the earlier mentioned multi-round protocols for phase estimation, is trading the cost of using physical quantum resources for cost in time resources \cite{giovannetti2006quantum}. 

In this work, we introduce a Hamiltonian active learning algorithm (HAL) based on the criteria of Fisher information, which is a way of measuring the information content of different queries and naturally appears in the bound on the errors achieved by a Hamiltonian learner. This resulting variant of HAL is called HAL-FI. We also introduce another variant of HAL for the task of predicting queries to the Hamiltonian, which uses the criteria of Fisher information ratio (FIR). We call the resulting active learning algorithm HAL-FIR. We demonstrate the performance of HAL-FI experimentally on IBM Quantum devices which are based on the superconducting cross-resonance (CR) gate. Compared with passive learning which scales as SQL, we show that HAL-FI with a fixed space of queries also has an asymptotic scaling of SQL but is able to achieve a constant reduction of $99.1\%$ in number of queries required for a desired learning error in learning the two-qubit CR Hamiltonian on a 20-qubit IBM Quantum device. HAL-FI can achieve up to a $99.8\%$ reduction in number of queries required when compared to current standard methods used for Hamiltonian learning which use estimation procedures based on FFT and regression. We finally show that queries involving exponentially growing system evolution time to the quantum devices suffices during learning to achieve Heisenberg limited scaling with HAL-FI when prior information is available. This is another example of trading physical quantum resources with time resources as highlighted before. 

The paper is organized as follows. In Sec \ref{sec:Hamiltonian_Learning}, we formally describe the problem of Hamiltonian learning, and the concept of an active learner. In Sec.~\ref{sec:HAL}, we present the HAL algorithms of HAL-FI and HAL-FIR. To illustrate the performance of HAL-FI, we consider the example of calibrating CR gates on IBM Quantum devices. In Sec. \ref{sec:CR_Gate_Model_and_Setup}, we describe our experimental setup, provide a theoretical description of the Hamiltonian model of the CR gate and physical models of the different noise sources affecting the quantum devices. Further, we provide details of our experiments on evaluating the performance of HAL-FI. Finally in Sec. \ref{sec:Results}, we describe the amount of query advantage that can be obtained using HAL-FI and specify the conditions under which Heisenberg limited rate of convergence or even super-Heisenberg limited rate of convergence can be achieved. Specifically for CR gates, we show that HAL-FI can be used to learn an accurate Hamiltonian using only a fraction of the queries required by currently used methodologies, resulting in reduction of queries of around two or three orders of magnitude over currently used methods for particular learning tasks.
\section{Hamiltonian Learning} \label{sec:Hamiltonian_Learning}

In this section, we present a formal description of the problem of Hamiltonian learning for a general quantum system (Section~\ref{sec:hlproblem}), and in the presence of different noise sources (Section~\ref{sec:hlproblem_noise}). Further, we introduce the concept of an active learner in this context (Section~\ref{sec:activelearning}). Notation used for this work will be defined as introduced and is also summarized in Appendix~\ref{app_sec:notation}.

\subsection{Problem Statement} \label{sec:hlproblem}
This subsection of the paper describes the unknown Hamiltonian of interest, specifies our query setting and formally describes the different Hamiltonian learning tasks.

\subsubsection{Unknown Hamiltonian} \label{subsec:unknown_hamiltonian}
Let $H$ be a partially or fully unknown system Hamiltonian over $n$ qudits (e.g., $d=2$ for a qubit, $d=3$ for a qutrit, etc.) As $H$ is a linear operator acting within a linear vector space with a well-defined inner product (Hilbert space), we can define $H$ in terms of the tensor product of the system qudits with Hilbert space $SU(d)$. $H$ can thus be represented by a $d^n \times d^n$ matrix and is constrained to be Hermitian.  

We assume that we have a model for the unknown Hamiltonian and belongs to the the family of models $\mathcal{H}=\{H(\boldsymbol{\theta})|\boldsymbol{\theta}\in \boldsymbol{\Theta}\} \subset \mathbb{R}^m$ parametrized by the real vector $\boldsymbol{\theta}$. We use $H(\cdot)$ to denote the model formulation and $\Theta$ as the space over parameters. The model may be derived from first principles based on the understanding of the physics of the quantum system or may be a complete blackbox such as a neural network. We further assume that the system Hamiltonian is time-independent and any unitary operator implemented on the system can be approximated by the evolution of $H$ for sufficiently long evolution times.

Let the unknown Hamiltonian of the quantum system be $H^\star$ and the true Hamiltonian parameters be $\boldsymbol{\theta}^\star$. We refer to the quantum system of whose Hamiltonian we wish to learn as an \textit{oracle} that we can query and which returns measurement outcomes upon querying. We denote the random variable of query to the oracle by $\mathrm{x}$ and the resulting output by the oracle as the measurement outcome random variable $\mathrm{y}$ corresponding to a single shot of the qubits being readout in the standard computational basis. The pair of a query and its corresponding output is called an example and is denoted by $(x,y)$. The alphabet of query $\mathrm{x}$ is referred to as the query space and we denote it by $\mathcal{Q}$. The distribution from which queries are sampled from $\querySpace$ is referred to as the \textit{query distribution} and denoted by $q$. Commonly, $y \in \labelSpace = \{0,1\}^{n_r}$ where $n_r$ is the number of qubits being readout. Our goal is to then learn the parameters $\bm{\theta}$ of the Hamiltonian with error $\epsilon$ from examples of the form $\{(x^{(i)},y^{(i)})\}_{i=1}^N$ while minimizing the number of queries made to the oracle or the query complexity $N$.

\subsubsection{Specifying a Query}
\begin{figure}[ht!]
\centering
\begin{quantikz}
    \lstick{$\ket{0}^{\otimes n}$} & \gate{U} & \gate{e^{-iHt}} & \gate{M} & \meter{}
\end{quantikz}
\caption{Quantum circuit picture of how the query $x=(M,U,t)$ is used. The input to the quantum circuit is the zero state $\ket{0}^{\otimes n}$. Application of the preparation operator $U$ on $\ket{0}^{\otimes n}$ is used to create an initial state before interacting with the system Hamiltonian $H$ for time $t$. Finally, a measurement operator $M$ is applied to the evolved state before measuring in the usual computational basis denoted by the meter. The output is a single shot of qubits being readout.}
\label{fig:query_specification}
\end{figure}
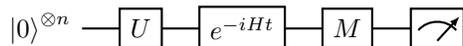

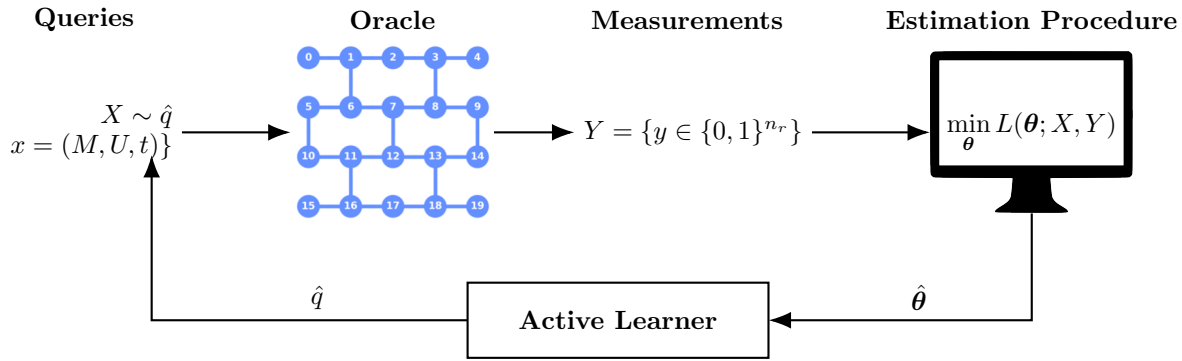
\begin{figure}[ht!]
\centering{
\begin{tikzpicture}
    \node (A) [align=right] at (-4.0,0.0) {$X \sim \hat{q}$\\$x=(M,U,t)\}$};
    \node (caption) at (-4.1,1.5) {\textbf{Queries}};
    \node[inner sep=0pt] (ibmq) at (0,0.0)
    {\includegraphics[width=.15\textwidth]{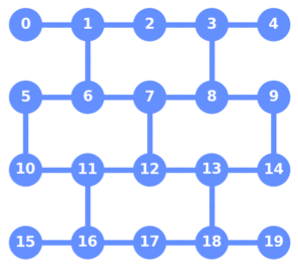}};
    \node (caption) at (0,1.5) {\textbf{Oracle}};
    \node (B) at (4.0,0) {$Y=\{y \in \{0,1\}^{n_r}\}$};
    \node (caption) at (3.92,1.5) {\textbf{Measurements}};
    \node (caption) at (8.5,1.5) {\textbf{Estimation Procedure}};
    \node[inner sep=0pt] (opt) at (8.5,0.0) {\includegraphics[width=.15\textwidth]{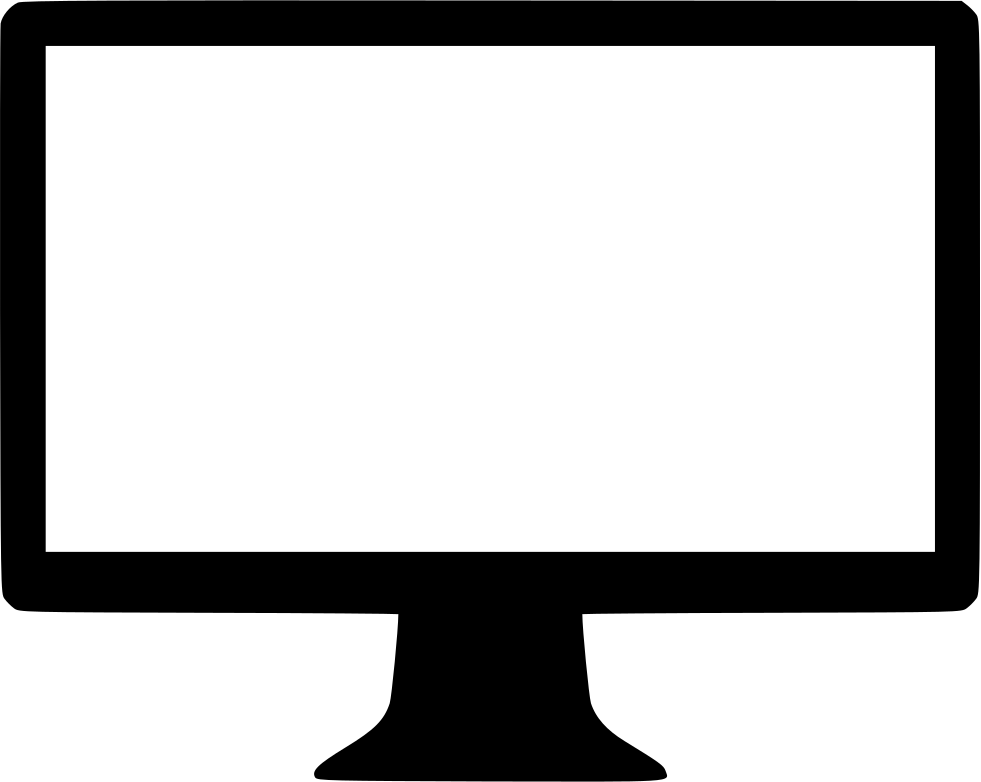}};
    \node (caption) at (8.5,0) {$\min \limits_{\bm{\theta}} L(\bm{\theta};X,Y)$};
    \node (caption) at (3,-2.5) {\textbf{Active Learner}};
    \node (caption) at (7,-2.2) {$\hat{\boldsymbol{\theta}}$};
    \node (caption) at (-1,-2.2) {$\hat{q}$};
    
    \draw [-{Latex[length=3mm]}][thick] (A) edge (ibmq);
    \draw [-{Latex[length=3mm]}][thick] (ibmq) edge (B);
    \draw [-{Latex[length=3mm]}][thick] (B) edge (opt);
    \draw [-{Latex[length=3mm]}][thick] (opt) |- (5,-2.5);
    \draw [-{Latex[length=3mm]}][thick] (1,-2.5) -| (-3.2,-0.3);
    
    \draw [thick] (1,-2) rectangle (5,-3);
\end{tikzpicture}    
}
\caption{Schematic of Hamiltonian learning with an active learner: The \textit{oracle} constitutes the unknown Hamiltonian and noise sources, with the true parameter vector of $\boldsymbol{\theta}^\star$ which is unknown and needs to be learned. Here, we show the device coupling map of the 20-qubit IBM Quantum device \textit{ibmq\_boeblingen}. Learning is carried out on training examples of the form $(x,y)$ where $x$ are queries inputted to the oracle specifying the measurement observable $M$, preparation operator $U$ and system time evolution $t$, and $y$ are the corresponding measurement outcomes outputted by the oracle. An illustrative quantum circuit picture of the oracle in the noiseless case is shown in Figure~\ref{fig:query_specification}. The set of queries inputted to the oracle is denoted as $X$ and the corresponding set of measurement outcomes outputted by the oracle as $Y$. An estimation procedure is run on the training examples of $(X,Y)$ to learn a model parameter estimate $\hat{\bm{\theta}}$. The complete top row corresponds to how a passive learner operates and is also called open-loop Hamiltonian learning. We add a feedback loop to introduce an active learner which uses the current estimate of model parameters $\hat{\boldsymbol{\theta}}$ to prescribe the distribution $\hat{q}$ from which queries to the oracle should be sampled from next. This process is then repeated during learning until an accurate estimate $\hat{\bm{\theta}}$ is obtained. The Hamiltonian Active Learning (HAL) algorithm introduced in this work comprises the estimation procedure and active learner shown in this schematic.  The active learner is described in Section~\ref{sec:activelearning} and the HAL algorithm in Section~\ref{sec:HAL}.}
\label{fig:hamiltonian_learning_schematic}
\end{figure}

We first describe what we mean by a query before specifying the different learning objectives. The query comprises three parts: measurement observables $M$, initial state preparation operators $U$, and control parameters $t$. A schematic of how the query is used in a quantum circuit to evolve a quantum system is shown in Figure~\ref{fig:query_specification}. In Figure~\ref{fig:hamiltonian_learning_schematic}, we show the oracle of a quantum device receiving the input of queries of form $x=(M,U,t)$ and returning the corresponding outputs of a single shot of the qubits on the device being readout denoted by a binary string of length $n_r$: $y \in \{0,1\}^{n_r}$.

\paragraph{Measurement Observables}\label{subsubsec:HT_meas_obs} The measurement observables $M \in \measSpace$ specify the basis set $\{\ket{\psi_m}\}$ used to generate the final measurement observation results. The permissible set of observables is typically constrained to be simple operators acting on just one qubit at a time, because that is experimentally realistic. It would be unrealistic to allow any arbitrary operator, because specific operators could be constructed to measure individual parameters of $H$, making the problem too simple, and the measurement operators too hard to realize in practice. One standard option for specifying $M$ is as a complete set of orthogonal projectors $\{\ketbra{\psi_m}{\psi_m}\}$ where $\sum_m \ketbra{\psi_m}{\psi_m} = I$. This is known as a projective measurement. For qubit systems, one example of such a measurement basis set is the computational basis.

\paragraph{Initial State Preparation Operators}\label{subsubsec:HT_init_state_opers}
Together with the measurement observable, an initial state $\ket{\psi_0}$ must also be specified, in order to determine a measurement result from $H$. This initial state can in principle be any unit vector in the Hilbert space, but much like for measurement observables, it only makes sense to allow a subset of possible states to be specified. We consider $\ket{\psi_0}= \ket{00\threecdots0}$ and allow for realistic single qubit unitary transforms. Thus, a common initial state specification is to provide a unitary operator $U~=~U_n~\otimes~U_{n-1}~\otimes~\threecdots~\otimes~U_1~\otimes~U_0$ as a tensor product of single qubit unitary transforms, acting on each of the $n$ qubits in the system. This then determines the $\prepOperator$ element of $\query$. We denote the set of all considered preparation operators as $\prepSpace$

\paragraph{Control Parameters}\label{subsubsec:HT_control_params}
The control parameters are the last element of $\query$, and are typically a set of classical numbers. For example, one canonical control parameter is the time $t$ for which $H$ should be applied. Other control parameters may modulate interactions between qubits. In this work, we consider only the Hamiltonian evolution time $t$ as a control parameter and denote the set of all possible evolution times as $\timeSpace$.

After defining the set of measurement operators $\measSpace$, the set of preparation operators $\prepSpace$, and the set of Hamiltonian evolution times $\timeSpace$, the resulting query space is given by $\querySpace = \measSpace \times \prepSpace \times \timeSpace$. We require $\querySpace$ to be \textit{complete}, i.e., there exist queries in $\querySpace$ that are informative about each of the Hamiltonian parameters $\bm{\theta}_i$ so that Hamiltonian learning can succeed. As we will see later in our discussion on active learning in Section~\ref{sec:AL_query_criteria}, completeness of $\querySpace$ is equivalent to the condition of there existing a set of queries or query distribution for which the resulting Fisher information matrix is full rank and invertible.

\subsubsection{Learning Framework} 

Having fully described a query, we are now in a position to formalize the problem of Hamiltonian learning. While doing so, we draw parallels to and introduce language from machine learning and statistical learning theory. 

We are given a query space $\querySpace=\measSpace \times \prepSpace \times \timeSpace$ constructed as discussed above and a label (or measurement outcome) space $\mathcal{Y}=\{0,1\}^{n_r}$. The labels $y \in \mathcal{Y}$ are not deterministic but are probabilistic in nature. We consider the class of Hamiltonian models $\mathcal{H}$ which map $\querySpace$ to $\mathcal{Y}$. In particular we assume the availability of a model description $\mathcal{H}=\{H(\boldsymbol{\theta})|\boldsymbol{\theta}\in \boldsymbol{\Theta}\} \subset \mathbb{R}^m$ parametrized by the vector $\boldsymbol{\theta}$ and where $\Theta$ is considered to be the space over the parameters. When no prior information is available to constrain the parameter space, we can consider $\Theta = \mathbb{R}^m$. We denote the true Hamiltonian as $H^\star$ and assume that $H^\star \in \mathcal{H}$ i.e., it is realizable. We denote the parameter vector of the true Hamiltonian as $\bm{\theta}^\star$. Our goal is to learn an estimate of $\bm{\theta}^\star$ which we denote by $\hat{\bm{\theta}}$. With regards to notation, we use $\bm{\theta}$ whenever we make statements that are true for any parameter in $\Theta$.

For the parameter vector $\boldsymbol{\theta}$, the label (or measurement outcome) $y$ given query $x=(M,U,t)$ is produced with the conditional probability
\begin{equation}
    p_{\mathrm{y}|\mathrm{x}}(y|x;\boldsymbol{\theta}) = \sum_z \left| \braketExp{yz}{M e^{-iH(\boldsymbol{\theta})t}U}{0^{\otimes n}} \right|^2
    \label{eq:borns_rule}
\end{equation}
as per Born's rule and assuming the absence of noise. The summation is over the hidden measurement outcomes of the $(n-n_r)$ qubits that are not read out which we denote by $z$ and $\left|\cdot \right|$ is used to denote the absolute value.  

Given the dataset of $N$ training examples $D = \{(x^{(i)},y^{(i)})\}_{i \in [N]}$, the goal is to learn the parameter vector $\boldsymbol{\theta}$ which characterizes the conditional probability (Eq.~\ref{eq:borns_rule}) that maps probabilistically $\querySpace$ to $\mathcal{Y}$. This is a supervised learning task for which a suitable loss function is
\begin{equation}
    L(\boldsymbol{\theta}; D) = \frac{1}{N} \sum \limits_{i=1}^N \mathcal{L} \left(x^{(i)},y^{(i)};\boldsymbol{\theta}\right) + R(\boldsymbol{\theta})
    \label{eq:general_loss_function_ML}
\end{equation}
where ${\cal L}$ is a suitable error function for assessing the chosen model on the training dataset, and $R$ is a regularization function penalizing model complexity. The former is usually a $p$-norm. The latter is chosen to incorporate prior information, enforce conditions such as sparsity through $\ell_1$-norm and generalize to unseen data. The parameter estimate is determined by minimizing the loss function $\hat{\boldsymbol{\theta}} = \argmin_{\boldsymbol{\theta} \in \boldsymbol{\Theta}} L(\boldsymbol{\theta}; D)$. After obtaining an estimate $\hat{\boldsymbol{\theta}}$, it is common to evaluate the performance against a testing dataset which includes examples not seen during training. This testing dataset which we denote as $D_{\mathrm{test}} = \{((x^{(i)},y^{(i)})\}_{i \in [N_{\mathrm{test}}}$ contains $N_{\mathrm{test}}$ queries and their corresponding outputs. The $N_{\mathrm{test}}$ queries are sampled from the query space $\querySpace$ according to a testing distribution which we denote by $\ptest$. This allows us to further distinguish between the problems of \textit{model inference} and \textit{prediction} against $\ptest$. 

In model inference, the goal is to learn a parameter estimate $\hat{\boldsymbol{\theta}}$ without making special considerations for the testing distribution $\ptest$ which may change or may be unknown. We use root mean squared error (RMSE) as $\mathcal{L}$ in Eq.~\ref{eq:general_loss_function_ML}
\begin{equation}
    \text{RMSE}(\hat{\bm{\theta}};\bm{\xi}) = \left( \sum_{i=1}^m \Expectation \left[ \left( \frac{\hat{\theta}_i}{\xi_i} - \frac{\theta^\star_i}{\xi_i} \right)^2 \right] \right)^{1/2}
\end{equation}
where $\boldsymbol{\theta}^\star$ is the parameter vector corresponding to the unknown Hamiltonian $H^\star$, and $\boldsymbol{\xi}$ is the vector to non-dimensionalize and normalize $\boldsymbol{\theta}$. This is done to account for different relative magnitudes of each parameter component ${\theta}_i^\star$ and to ensure that the RMSE does not explode due to contributions of few parameter components. These normalization factors $\bm{\xi}$ can be obtained during estimation or be available through prior information. Moreover, we restrict estimators to be unbiased; this is commonly desired and also, a universal optimal estimator that minimizes the RMSE cannot generally be defined. Under this condition,  $\mathcal{L}$ may be reduced to being the square root of the variance, $\left[ \sum_{j} \Var(\theta_j/\xi_j) \right]^{1/2}$.

A suitable consistent estimator for reducing the variance is then the maximum likelihood estimator (MLE) which determines an estimate of the parameters $\hat{\bm{\theta}}$ from the data $D$ through
\begin{equation}
    \quad \hat{\boldsymbol{\theta}} = \argmin_{\boldsymbol{\theta} \in \boldsymbol{\Theta}} \frac{1}{N} \sum \limits_{i=1}^N  - \log p_{\mathrm{y}|\mathrm{x}} \left(y^{(i)}|x^{(i)};\bm{\theta} \right)
    \label{eq:general_MLE}
\end{equation}
where $- \log p_{\mathrm{y}|\mathrm{x}} \left(y^{(i)}|x^{(i)};\bm{\theta} \right)$ is the negative log-likelihood of the measurement outcome $y^{(i)}$ given the query $x^{(i)}$ and when considering the model parameters $\bm{\theta}$. Throughout this work, we often denote the negative log-likelihood in shorthand by $L(y|x;\bm{\theta})$ for a general measurement outcome $y$ and query $x$ given model parameters $\bm{\theta}$.

In prediction against a testing distribution $\ptest$, the goal is to learn an estimate $\hat{\boldsymbol{\theta}}$ that will allow us to perform well on predicting the likelihood function of measurement outcomes given queries sampled from $\ptest$. In such a scenario, $\mathcal{L}$ is still the RMSE but the performance of the estimate is assessed through the testing error
\begin{equation}
    \text{Testing Error} =  \Expectation_{x \sim \ptest, y \sim p_{\mathrm{y}|\mathrm{x}} \left(\cdot|x;\bm{\theta}^\star \right)} [ L(y|x;\hat{\bm{\theta}}) - L(y|x;\bm{\theta}^\star)]
\end{equation}
which we have chosen as the expectation of the difference in negative log-likelihood when using the estimate $\bm{\theta}$ and truth $\bm{\theta}^\star$ with respect to the testing distribution $\ptest$. This difference in negative log-likelihood is also often called the log-likelihood ratio \cite{sourati2017asymptotic}. It is then desired to learn $\hat{\bm{\theta}}$ such that the testing error is lower than a given error $\epsilon$. The error $\epsilon$ here may need to be specified relative to how the negative log-likelihood scales which will depend on the Hamiltonian of interest and the queries themselves.

We note that Eq.~\ref{eq:general_MLE} is an optimization problem that may be solved in several ways. The different programs or algorithms to solve the optimization problem are commonly called \textit{optimization procedures} or \textit{estimation procedures}. Different procedures have different properties such as rate of convergence and optimality such as guarantee of a local or global minimum. In order to improve convergence or search direction, it is common for procedures to require first or second gradient information. The estimation procedure will need to be specified along with the formulated optimization problem at hand for a complete specification of the learning algorithm.

Given $N$ training examples, we then say that we have succeeded at model inference (prediction) if we are able to produce an estimate $\hat{\boldsymbol{\theta}}$ such that the RMSE (testing error) is bounded by some error parameter $\epsilon$. We want to accomplish these learning tasks by minimizing the resource requirements or number of queries $N$. This is formalized in the following respective problem statements.

\begin{problem}[Model Inference] \label{prob:HL_model_inference}
Suppose we are given access to a quantum system with an unknown time-independent $2^n \times 2^n$ dimensional Hamiltonian $H$ over $n$ qubits, parameterized by $\bm{\theta}$ and the vector of normalization factors $\bm{\xi}$. Suppose we fix the error parameter $\epsilon > 0$. What is then the number of queries $x \in \querySpace$ (where each query is of the form $x \in \{(M,U,t) | M \in \measSpace, U \in \prepSpace, t \in \timeSpace\}$) required to learn an estimate $\hat{\bm{\theta}}$ of the true parameters $\bm{\theta}^\star$ such that the root mean squared error $\mathrm{RMSE}(\hat{\bm{\theta}};\bm{\xi}) \leq \epsilon$?
\end{problem}

\begin{problem}[Prediction Against Testing Distribution] \label{prob:HL_prediction_testing}
Suppose we are given access to a quantum system with an unknown time-independent $2^n \times 2^n$ dimensional Hamiltonian $H$ over $n$ qubits, parameterized by $\bm{\theta}$. Suppose we fix the error parameter $\epsilon > 0$. What is then the number of queries $x \in \querySpace$ (where each query is of the form $x \in \{(M,U,t) | M \in \measSpace, U \in \prepSpace, t \in \timeSpace\}$) required to learn an estimate  $\hat{\bm{\theta}}$  of the true parameters $\bm{\theta}^\star$ such that $\Expectation_{x \sim \ptest, y \sim p_{\mathrm{y}|\mathrm{x}}(y|x;\bm{\theta}^\star)} [L(y|x,\hat{\bm{\theta}}) - [L(y|x,\bm{\theta}^\star)] \leq \epsilon$ where $L(y|x,\bm{\theta})$ is the negative log-likelihood?
\end{problem}

We distinguish between the two problems of model inference and prediction against $\ptest$ as the ideal training dataset for these two learning tasks may come from two different distributions. This will influence how the training data is chosen when we introduce an active learner into Hamiltonian learning, as we will discuss in Section~\ref{sec:HAL}. The performance of the Hamiltonian learning algorithms in tackling Problem~\ref{prob:HL_prediction_testing} will also be useful for cross-validation and testing the robustness of these methods.

\subsection{Learning in the Presence of Noise} \label{sec:hlproblem_noise}
Previously, in Section~\ref{sec:hlproblem}, we gave formal statements of the Hamiltonian learning problem in the absence of noise assuming an ideal oracle. However, the oracle is usually noisy due to the presence of different noise sources affecting the quantum system. In this section, we describe the problem of learning in the presence of these noise sources. 

Common noise sources include readout noise, decoherence, and imperfect control of the quantum system. State preparation and measurement (SPAM) errors are already accounted for by considering these noise sources. Classical SPAM errors encountered are included in the readout noise and the errors in implementing state preparation or measurement operators fall under imperfect control. 

\paragraph{Readout Noise}
The readout line of the qubit measurement outcomes $y$ is also a classical communication channel and hence suffers from bit-flip errors. The readout noise can then be modeled as a classical bit flip channel where the true measurement outcome $y_i \in \{0,1\}$ of the $i$th qubit may be flipped. We denote the observed measurement outcomes' random variable by $\tilde{\mathrm{y}}$ which can be interpreted as a noisy observation of $\mathrm{y}$. We denote the conditional probability of observing $\tilde{y}$ given measurement outcome $y$ which is hidden from us as $p_{\tilde{\mathrm{y}}|\mathrm{y}}(\tilde{y}|y)$. Note that in general we expect the readout channel to be asymmetric i.e.,  $p_{\tilde{\mathrm{y}}|\mathrm{y}}(1|0) \neq p_{\tilde{\mathrm{y}}|\mathrm{y}}(0|1)$, and this is accounted for in the readout noise model. The probability of observing a noisy measurement outcome $\tilde{y}_i$ of the $i$th qubit given query $x$ is then given by
\begin{equation}
    p_{\tilde{\mathrm{y}}_i|\mathrm{x}}(\tilde{y}_i|x;\bm{\theta}) = p_{\tilde{\mathrm{y}}_i|\mathrm{y}}(\tilde{y}_i|\tilde{y}_i) p_{\mathrm{y}_i|\mathrm{x}}(\tilde{y}_i|x;\bm{\theta}) + p_{\tilde{\mathrm{y}}_i|\mathrm{y}}(\tilde{y}_i|1-\tilde{y}_i) p_{\mathrm{y}_i|\mathrm{x}}(1-\tilde{y}_i|x;\bm{\theta})
    \label{eq:prob_readout_noise}
\end{equation}
where $p_{\mathrm{y}|\mathrm{x}}(y|x;\boldsymbol{\theta})$ is given by Eq.~\ref{eq:borns_rule}.

\paragraph{Imperfect Control Strategies}
In general, a Hamiltonian can be decomposed as 
\begin{equation}
    H(t) = H_d + \sum_{k=1}^{K} H_k(t, u_k(t))
\end{equation}
where $H_d$ is the drift/free part of the Hamiltonian that is internal to the system and cannot be controlled externally. The terms $H_k$ corresponds to the parts of the Hamiltonian that can be controlled using the control function $u_k(t)$. Depending on the quantum architecture, the control function can be realized as microwave pulses for superconducting qubits, optical pulses for ion traps, etc.

Target operators $U_f$ are then obtained starting from the identity matrix $I$ by implementing controls $u_k(t)$ over time duration $[0,T]$ as
\begin{equation}
    U_f = \mathbb{T} \left[\exp \left(-i \int_0^T \left[ H_d + \sum_{k=1}^{K} H_k(t, u_k(t)) \right] dt \right) \right]
\end{equation}
where $\mathbb{T}$ is the time-ordering operator. These target operators $U_f$ can be preparation operators, measurement operators or different gates e.g., the Clifford gates. Imperfections in these pulses will lead to errors in $U_f$. For example, the strength of the qubit-qubit interactions is sensitive to variations in the pulse amplitudes. Strong driving can lead to leakage of states outside of the computational subspace. Moreover, bandwidth effects or dispersion can cause leading or trailing edge distortions in the pulse shapes that can lead to errors in the unitary operator implemented. 

\paragraph{Decoherence}
Let us consider the unitary operator of $U(t) = e^{-iHt}$. The application of this unitary operator is accompanied by decoherence on a real quantum system due to interactions with its environment. We model this as a depolarizing channel $\mathcal{E}$
\begin{equation}
    \mathcal{E}(\rho(t)) = (1-p_d(t))\rho(t) + p_d(t) \frac{I}{2^n}
    \label{eq:depolarization_error}
\end{equation}
where $\rho(t)=U(t)\rho(0)U(t)^\dagger$ is the state obtained on application of the unitary $U(t)$ to the initial state $\rho(0)$, $p_d(t)$ is the probability of the state being depolarized and $I/2^n$ is the maximally mixed state. We have assumed the case of complete depolarization here. To obtain a functional form of $p_d(t)$, we note that up to first order, depolarization events can be assumed to arrive under a Poisson process with rate $\mu$. By a depolarization event, we refer to the occurrence of an error that completely randomizes the quantum state. As the time between Poisson events follows an exponential distribution, we can write
\begin{equation}
    p_d(t) = 1 - \exp\left( - \frac{t-t_0}{\mu} \right)
\end{equation}
where $t_0$ denotes the starting time of the experiment. Denoting the probability of measurement outcomes $y$ given a query $x$ under depolarization errors as $p_{\mathcal{E}}(y|x;\boldsymbol{\theta})$, we have
\begin{equation}
    p_{\mathcal{E}}(y|x;\boldsymbol{\theta}) = \exp\left( - \frac{t-t_0}{\mu} \right) p_{\mathrm{y}|\mathrm{x}}(y|x;\boldsymbol{\theta}) + \frac{1}{2^n} \left( 1 - \exp\left( - \frac{t-t_0}{\mu} \right) \right)
    \label{eq:prob_complete_depolarization}
\end{equation}
where $p_{\mathrm{y}|\mathrm{x}}(y|x;\boldsymbol{\theta})$ is the probability of measurement outcome $y$ assuming no depolarization given by Eq.~\ref{eq:borns_rule}. The rate $\mu$ can be related to the amplitude relaxation time $T_1$ and dephasing time $T_2$ of the qubits on the quantum system. We will describe this in Section~\ref{sec:noise_sources} for the IBM Quantum devices considered for application of Hamiltonian learning.

Let the collective set of parameters associated with the different noise models so far described be denoted by $\bm{\zeta}$. The measurement outcomes $y$ from the oracle will then be a function of the queries $x$ and the true parameters $(\bm{\theta}^\star,\bm{\zeta}^\star)$. In order to obtain estimates for $\hat{\zeta}$ in addition to $\hat{\bm{\theta}}$, we solve the following modified optimization problem over the training examples
\begin{equation}
    \left( \hat{\boldsymbol{\theta}}, \hat{\boldsymbol{\zeta}} \right) = \argmin_{\boldsymbol{\theta} \in \boldsymbol{\Theta}, \boldsymbol{\zeta} \in \mathrm{Z}} \frac{1}{N} \sum \limits_{i=1}^N \mathcal{L} \left(x^{(i)},y^{(i)};\boldsymbol{\theta},\boldsymbol{\zeta}\right) + R(\boldsymbol{\theta},\boldsymbol{\zeta})
    \label{eq:general_noise_loss_function_ML}
\end{equation}
However, this increases the computational cost of the estimation procedure due to the increase in the number of parameters and the corresponding search space. In practice, prior calibration data can be used to obtain estimates of $\hat{\boldsymbol{\zeta}}$ which are then used in Eq.~\ref{eq:general_loss_function_ML} to obtain $\hat{\boldsymbol{\theta}}$.

Hence, the Hamiltonian learning Problems~\ref{prob:HL_model_inference} and \ref{prob:HL_prediction_testing} can be restated assuming additional information of the noise model parameters $\hat{\boldsymbol{\zeta}}$ when we have access to a noisy oracle. When describing our application of Hamiltonian learning, we will describe both the Hamiltonian of interest in Section~\ref{sec:CR_hamiltonian} and the specific noise sources in Section~\ref{sec:noise_sources}.

\subsection{Active Learning of Hamiltonians} \label{sec:activelearning}
This subsection introduces the concept of active learning and how an active learner can be used in the context of Hamiltonian learning. We begin by giving a quick overview of different types of learners, and describe how active learning differs from passive learning and online learning. This is followed by an overview of the different active learning (AL) strategies that have been proposed in the literature. We describe which AL strategy seems best suited for Hamiltonian learning. AL strategies based on Fisher information (FI) for Problem~\ref{prob:HL_model_inference} and Fisher information ratio (FIR) for Problem~\ref{prob:HL_prediction_testing} seem to be notably appropriate. In doing so, we establish criteria for evaluating the performance of an AL algorithm, and observe that AL algorithms are known to solve some learning problems faster than passive learners. Moreover, under some circumstances, AL algorithms can exceed the central limit theorem bounds. 

\subsubsection{Types of Learners}
An overview of different learners often considered for learning tasks is given below. We will use the passive learner as a baseline and the active learner is the focus of our work. A description of an \textit{offline learner} and \textit{online learner} is given for completeness and to distinguish an active learner from them.

\paragraph{Offline Learner} In offline learning, all of the training data is given to the learner at once and a model is learned. The training dataset may be made available to the learner or collected by sampling queries from an arbitrary distribution and then obtaining outputs from an oracle using these as inputs. This is the learning paradigm under which most machine learning tasks operate in.

\paragraph{Online Learner} In online learning, the training data is made available in a sequence, typically one at a time by a referee. A query $x_t$ is made available to the learner at the $t$'th round in a sequence after which the learner constructs an estimate of the output $\hat{y}_t$ to this query. The learner provides $\hat{y}_t$ to the referee, and suffers a loss that depends on $\hat{y}_t$ and the actual output. The learner is then provided with feedback by the referee which the learner can then use to update the model. In this case, the queries that are provided to the learner by the referee may be adversarial or adaptive to the learner's behavior. The learner has no control over the query distribution from which these queries arrive.

\paragraph{Passive Learner}\label{sec:passive_learner}
In the learning problems described in Sec.~\ref{sec:hlproblem}, we did not specify the distribution from which the queries $x$ are sampled from. In open-loop Hamiltonian learning, the query distribution remains fixed during learning and all the queries to the oracle are sampled from this distribution. When no prior information is available, it is common to set the query distribution to the uniform distribution over the query space $\querySpace$. We will refer to this setting as \textit{passive learning} through
out this paper. Combined with a specification of an estimation procedure, the passive learner will serve as a baseline to the active learner which we introduce in the next section.

\paragraph{Active Learner}\label{sec:active_learner}
In active learning (AL), the learner has access to the query space $\querySpace$ and the ability to select queries or decide the query distribution during training using the current estimates of the model parameters $\hat{\boldsymbol{\theta}}$. This is accomplished by introducing a feedback into the open loop Hamiltonian learning approach shown earlier in Figure~\ref{fig:hamiltonian_learning_schematic}. Based on the current estimate $\hat{\boldsymbol{\theta}}$ and the queries made so far combined with their respective outcomes, the active learner proposes a query distribution from which queries should be selected from to send to the oracle. These queries may be sent to the oracle in a sequential manner one at a time or in batches. In Section~\ref{sec:AL_query_criteria}, we discuss different criteria used for query selection or proposing query distributions.

Here, we distinguish a passive learner from an offline learner only in the number of rounds of training data collection. An offline learner is given access to a complete training dataset or is allowed to collect it through queries to an oracle in one round. On the other hand, a passive learner has continued access to the oracle and is allowed to collect training data until the learning task has been accomplished. The primary difference between the active learner and the online learner is that the former has control over the query distribution from which it will select queries to input to the oracle.

\subsubsection{Query Criteria for Active Learner} \label{sec:AL_query_criteria}
There have been multiple criteria proposed for query selection but are usually subdivided into the two categories of informativeness and representativeness. Criteria based on informativeness aim to select queries that will reduce the uncertainty of the statistical model and include uncertainty sampling \cite{lewis1994sequential,settles2009active}, query-by-committee \cite{seung1992query,burbidge2007active}, and margin \cite{balcan2007margin}. On the other hand, the goal of representativeness \cite{yu2006active,flaherty2006robust} is to ensure selection of queries that exploit the structure of the underlying distribution and are diverse. There has also been exploration into combining the criteria of informativeness and representativeness \cite{du2015exploring, huang2014active}.

Multiple query criteria used in active learning in practice are based on heuristics and empirical evidence \cite{settles2012active}. Here, we choose the informativeness criteria of Fisher information (FI) and Fisher information ratio (FIR) as they have direct relationships with the different learning problems we introduced in Sec.~\ref{sec:hlproblem}. This allows us to provide guarantees on the performance of our AL strategy. Later, we discuss how we can ensure that representative queries are also selected.

We introduce some notation before discussing the query criteria for our AL strategy. Let us denote the Fisher Information matrix of a particular query $x \in \querySpace$ as $\fisherInfo_x(\boldsymbol{\theta})$ where the $(i,j)$th element of the matrix is given by
\begin{equation}
    \fisherInfo_x(\boldsymbol{\theta})[i,j] = \Expectation \left[ \frac{\partial \log p_{\mathrm{y}|\mathrm{x}}(y|x;\boldsymbol{\theta)}}{\partial \theta_i} \frac{\partial \log p_{\mathrm{y}|\mathrm{x}}(y|x;\boldsymbol{\theta)}}{\partial \theta_j} \right]
    \label{eq:fisher_information_query}
\end{equation}
and where the expectation is taken with respect to $p(y|x;\boldsymbol{\theta})$. The Fisher Information matrix is equivalently written as $\fisherInfo_x(\boldsymbol{\theta})=\Expectation[SS^T]$ where $S=\partial \log p_{\mathrm{y}|\mathrm{x}}(y|x;\boldsymbol{\theta)}/\partial \boldsymbol{\theta}$ is commonly called the score vector. Instead of selecting one query at a time, we often require the active learner to select a distribution over the query space $\querySpace$ which we will call the query distribution. The Fisher Information matrix corresponding to $q$ will then be given by
\begin{equation}
    \fisherInfo_q(\boldsymbol{\theta}) = \Expectation_q [ \fisherInfo_x(\boldsymbol{\theta}) ] = \sum_{x \in \querySpace} q(x) \fisherInfo_x(\boldsymbol{\theta})
    \label{eq:fisher_information_query_distribution}
\end{equation}
where the summation can be replaced by an integral in the case of a continuous query space. If the parameters describing the Hamiltonian model $\boldsymbol{\theta}\in \mathbb{R}^m$, then $\fisherInfo_q \in \mathbb{R}^{m\times m}$. Let us now describe the different query criteria in the context of the different learning tasks. 

\paragraph{RMSE of Parameters}
If the learning objective is to learn the parameters with small RMSE (Problem~\ref{prob:HL_model_inference}), a natural query optimization strategy is obtained by noting the Cramer-Rao bound for unbiased estimators: \cite{cover2012elements}:
\begin{align}
    \Cov(\boldsymbol{\theta}) &\geq \frac{1}{N} \fisherInfo_q^{-1}(\boldsymbol{\theta}) \\
    \sum_i \Var(\theta_i) & \geq \frac{1}{N} \Tr(\fisherInfo_q^{-1}(\boldsymbol{\theta}))
    \label{eq:cramer_rao_bound}
\end{align}
Combining the Cramer-Rao bound with the fact that the parameter estimates converges in probability $\hat{\bm{\theta}} \rightarrow \bm{\theta}^\star$ for an unbiased estimator such as MLE and the distribution converges in law as $\sqrt{N}(\hat{\boldsymbol{\theta}} - \boldsymbol{\theta}^\star) \rightarrow \mathcal{N}\left(0,\fisherInfo_q^{-1} (\boldsymbol{\theta}^\star) \right)$ \cite{lehmann2006theory,wasserman2013all}, we get that MLE is an asymptotically efficient estimator with the efficiency equal to the Fisher information over the training examples \cite{sourati2017asymptotic}. An optimal query distribution can then be obtained through the following query optimization:
\begin{equation}
    q^\star = \arg \min_{q \in \mathcal{P}} \Tr(\fisherInfo_q^{-1}(\boldsymbol{\theta}))
    \label{eq:query_optimization_variance_of_params}
\end{equation}
where $\mathcal{P}$ is the family of all valid probability distributions over the query space $\querySpace$. We also note that for $q^\star$, $\fisherInfo_{q^\star}$ must necessarily be invertible and hence full rank. This ensures that the queries inform us about all the Hamiltonian parameters of interest. Also for an unbiased and consistent estimator, the Cramer-Rao bound is likely to be saturated in the limit of large number of queries.

\paragraph{Testing accuracy against testing distribution}
When the learning objective is to minimize the expected log-likelihood error against a testing distribution $\ptest$ (Problem~\ref{prob:HL_prediction_testing}), we use an active learning strategy based on Fisher Information ratio (FIR) $\Tr(\fisherInfo_q^{-1}(\boldsymbol{\theta}) \fisherInfo_{\ptest}(\boldsymbol{\theta}))$. The name FIR comes from the scalar case where it can be viewed a ratio of the Fisher information corresponding to the query distribution and that corresponding to the testing distribution. The use of FIR for this learning task can be motivated by noting the following inequality \cite{sourati2017asymptotic}
\begin{equation}
    \mathbb{E}_{x \sim \ptest, y \sim p_{\mathrm{y}|\mathrm{x}}(\cdot|x;\boldsymbol{\theta})} [\text{Var}_{\text{D} \sim q(x)p_{\mathrm{y}|\mathrm{x}}(y|x;\boldsymbol{\theta})} [ \mathcal{L}(\hat{\boldsymbol{\theta}}_{\text{D}};x,y) - \mathcal{L}(\boldsymbol{\theta};x,y) ]] \leq \frac{1}{N} \Tr(\mathcal{I}_q^{-1}(\boldsymbol{\theta}) \mathcal{I}_{\ptest}(\boldsymbol{\theta}))
\end{equation}
where the left hand side is the expected variance of the asymptotic distribution of the log-likelihood ratio which can be viewed as a testing error and the right hand side involves the FIR. Minimizing the upper bound would then allow us to control the testing error and hence this suggests using the following query distribution:
\begin{equation}
    q^\star = \arg \min_{q \in \mathcal{P}} \Tr(\fisherInfo_q^{-1}(\boldsymbol{\theta}) \fisherInfo_{\ptest}(\boldsymbol{\theta}))
    \label{eq:query_optimization_testing_error}
\end{equation}
We note the Fisher information ratio is related to the Fisher information matrices of the query distribution $\fisherInfo_q$ and testing distribution $\fisherInfo_{\ptest}$ through the following inequality:
\begin{equation}
    \Tr(\mathcal{I}_q^{-1}(\boldsymbol{\theta}) \mathcal{I}_p(\boldsymbol{\theta})) \leq \Tr(\mathcal{I}_q^{-1}(\boldsymbol{\theta})) \cdot \Tr(\mathcal{I}_{\ptest}(\boldsymbol{\theta}))
\end{equation}
Thus, we recover the query optimization of Eq.~\ref{eq:query_optimization_variance_of_params} when the testing distribution is unknown.

In quantum tomography, such query criteria have been applied in optimal experiment design (OED) or adaptive quantum tomography. Fisher information has been used as a query criteria for offline OED in quantum state tomography \cite{nunn2010optimal}. Even earlier, an active learner based on Shannon entropy aka maximum uncertainty sampling was considered in \cite{fischer2000quantum}. An AL strategy based on Shannon information combined with Bayesian estimation was proposed in \cite{huszar2012adaptive} for the selection of measurement operators during quantum state tomography. Fisher information was again used in \cite{gazit2019quantum} where OEDs were analyzed for a family of qubit channels over different design problems. In Hamiltonian learning, Fisher information has been used to comment on heuristic strategies for OED \cite{ferrie2013best} and has been combined with Bayesian estimation to produce an active learner \cite{granade2012robust}.

\subsubsection{Active Learning Strategy}
Implicit in the above descriptions of the query criteria is the idea of proposing a query distribution rather than selecting one query at a time during active learning. This demarcates sequential active learning where one query is chosen at a time from batch mode active learning where a batch of queries sampled from a query distribution are selected to be inputted to the oracle. Moreover, combining FI/FIR query criteria with batch mode active learning \cite{hoi2006batch} ensures that representative queries are chosen as well.

In what follows, we describe the batch-mode active learning scheme that forms the basis of the AL algorithm for Hamiltonian learning that we discuss in Section~\ref{sec:HAL}. Given a budget of $N$ queries, the training is divided into multiple rounds. We index each round of the training process as $i$ and denote the batch size as $N_b$. The number of queries made till the $i$th round (inclusive) is denoted as $N_{tot}^{(i)}$. In each round, a batch of queries is sampled from the optimal query distribution based on the current parameters' estimate $\hat{\boldsymbol{\theta}}$ and then this estimate is updated using the measurement outcomes of the queries. This is then repeated until all of the budget has been expended. We denote the estimate in the $i$th round by $\hat{\bm{\theta}}^{(i)}$ and the optimal query distribution in the $i$th round based on $\hat{\bm{\theta}}^{(i)}$ by $q^{(i)}$. The query distribution at the very beginning of the training process $q^{(0)}$ is determined from any available prior information of the parameters or else set to be the uniform distribution over the query space.

What should be the size of the initial set of queries $N_{tot}^{(0)}$? Some suggestions are given in \cite{chaudhuri2015convergence} based on a finite-sample analysis for logistic regression but these do not suffice for the application considered in this paper. Qualitatively, one hopes that $N_{tot}^{(0)}$ is high enough such that the parameter estimate $\hat{\boldsymbol{\theta}}^{(0)}$ lies close to the true parameter value $\bm{\theta}^\star$ and in a convex basin of the asymptotic negative log-likelihood loss function. However, setting $N_{tot}^{(0)}$ to a very high value may not allow us take advantage of the presence of an active learner and the savings it can provide. 

Additionally, it may be advantageous to adaptively change the query space for exploration from one batch to the next. It is not necessary for the query space to remain static or unchanged \cite{settles2012active} during active learning. In fact, there is an element of adaptively changing the \textit{search} space in many prominent algorithms. In sparse fast Fourier transform \cite{hassanieh2012nearly,hassanieh2012simple}, the bins of frequencies are randomly chosen with each iteration in the algorithm. In the related machine learning tool of reinforcement learning, action spaces are changed to eliminate actions \cite{zahavy2018learn}, and to generalize over time by parameterizing them \cite{masson2016reinforcement} or embedding them in a continuous action space \cite{dulac2015deep}. 

Adaptively changing the query space is particularly compelling for the application of active learning to the Hamiltonian learning problem, because evidence suggests that it may result in so-called \textit{Heisenberg}-limited scaling of the number of queries as we noted in Section~\ref{sec:intro}. We will also make a case for why we expect the active learner which we will introduce for Hamiltonian learning in Section~\ref{sec:HAL} to achieve Heisenberg-limited scaling when possible.

The computational cost of the batch-mode AL scheme is determined by the number of rounds of batches issued and the computational cost of solving the query optimization problems of Equations~\ref{eq:query_optimization_variance_of_params} and \ref{eq:query_optimization_testing_error}. Solving these directly can be challenging but fortunately, the query optimizations can be reformulated as semidefinite programs (SDP) \cite{sourati2017asymptotic} under the assumptions of differentiability of the log-likelihood function, and invertibility of the Fisher information matrix for query distributions in a compact space around the optimal query distribution and around the uniform distribution. When using the query criteria of Fisher Information ratio (FIR) (Eq.~\ref{eq:query_optimization_testing_error}), the optimization problem \cite{chaudhuri2015convergence, sourati2017asymptotic} is
\begin{align}
    & \arg \min_{\alpha_1,...,\alpha_d} \sum \limits_{i=1}^{m} \alpha_i \quad
    \text{such that } \sum_{x \in \querySpace}q(x) = 1, \, \text{and }
    \begin{bmatrix} \fisherInfo_q(\boldsymbol{\theta}) & \mathbf{e}_j \\ \mathbf{e}_j^T & \alpha_j \end{bmatrix} \succcurlyeq 0, \, j \in [m] 
    \label{eq:SDP_query_optimization}
\end{align}
where we have introduced $m$ auxiliary variables $\alpha_1,...,\alpha_m$, and $\mathbf{e}_j$ are the eigenvectors of $\fisherInfo_p(\hat{\boldsymbol{\theta}})$. Recall that the parameter vector $\bm{\theta}$ has $m$ components. To obtain the SDP program for the query optimization when using the query criteria of Fisher information (FI) (Eq.~\ref{eq:query_optimization_variance_of_params}) in our AL strategy, we replace $\mathbf{e}_j$ by the eigenvectors of the identity matrix. In this case, $\mathbf{e}_j$ are $m$-dimensional canonical vectors with $1$ in the $j$th component and zero elsewhere. The computational cost of solving the above SDP programs with a barrier interior-point method is $\mathcal{O}(n_{\mathcal{Q}}^2 m^3 + n_{\mathcal{Q}}m^4 + m^5)$ where we have denoted $n_{\mathcal{Q}}= |\mathcal{Q}|$ as the size of the query space of interest.

\subsubsection{Query Advantage} \label{sec:def_query_advantage}
To compare resource requirements of different Hamiltonian learning (HL) methods for accomplishing a learning task, we introduce the concept of \textit{query advantage}. The query advantage (QA) of an HL method in achieving a learning error of $\epsilon$ is:
\begin{equation}
    \text{QA} = 1 - \frac{\text{Number of queries required by method}}{\text{Number of queries required by baseline}}
\end{equation}
As discussed before, QA measures the amount of query reduction obtained by selecting a HL method over a baseline strategy. In this work, we consider the passive learner equipped with an appropriately chosen estimation procedure as the baseline. We will specify the estimation procedure when discussing a QA result or it will be clear from context.

The benefits of quantifying QA for an HL method are twofold. Firstly, it allows us to comment on the performance boost obtained by using one HL method over another for accomplishing a learning task. Moreover, we can comment on learning tasks that can be achieved using one HL method but is unattainable by another. Secondly, it gives us a direct way to select a particular HL method based on minimal resources required, from a set of methods for a particular learning task by choosing the method with the highest QA.  

As we will see in the next Section~\ref{sec:HAL}, an active learning strategy is a framework for achieving query advantage and higher learning rates of convergence when possible.

\section{Hamiltonian Active Learning Algorithms} \label{sec:HAL} \label{sec:ouralgorithm}
We are now in a position to describe how to adapt probabilistic pool-based batch-mode active learning with query criteria of FI and FIR for Hamiltonian learning. The resulting algorithms are collectively called Hamiltonian Active Learning algorithms. We call HAL combined with the query criterion of Fisher Information as HAL-FI and that with Fisher Information Ratio as HAL-FIR. We discuss how the resulting algorithms are expected to achieve query advantage over the baseline and Heisenberg limited scaling in query complexity with access to prior information if possible.

\subsection{Algorithm} \label{sec:HAL-FI_algo_description}
The HAL algorithm is summarized in Algorithm~\ref{algo:AL_hamiltonians}. We assume that the unknown Hamiltonian is time-independent and a model parameterized by $\boldsymbol{\theta}^\star$ for the oracle (i.e., Hamiltonian with noise sources) is available to us. Inputs to the HAL algorithm include the initial query distribution $q^{(0)}$ to be used for sampling the initial set of $N_{tot}^{(0)}$ queries and the query optimization algorithm (QOA). When the query criterion is Fisher information (FI), the corresponding QOA is given by Algorithm~\ref{algo:QOA_FI} which solves Eq.~\ref{eq:query_optimization_variance_of_params}. Similarly when the query criterion is Fisher information ratio (FIR), the corresponding QOA is given by Algorithm~\ref{algo:QOA_FIR} which solves Eq.~\ref{eq:query_optimization_testing_error}. Recall that the choice of query criteria and hence QOA depends on the learning task at hand: we consider FI for learning Hamiltonian parameters with low RMSE (Eq.~\ref{eq:query_optimization_variance_of_params}) and FIR for minimizing the testing error (Eq.~\ref{eq:query_optimization_testing_error}). Additional inputs include the maximum number of batches $i_{max}$ that will be issued during learning which we will denote by $N_b$ and the total experimental budget available.

Using the notation for batch mode active learning as discussed in Sec.~\ref{sec:activelearning}, we denote the batch of queries that are sampled at the $i$th round with respect to the query distribution $q^{(i)}$ as $X_q^{(i)}$ and the corresponding measurement outcomes from inputting these queries to the \textit{oracle} as $Y_q^{(i)}$. The set of all queries made so far at any round is denoted by $X^{(i)}$ and their corresponding measurement outcomes as $Y^{(i)}$. We note that $|X_q^{(0)}|=|Y_q^{(0)}|=N_{tot}^{(0)}$ which is typically larger than $|X_q^{(i)}|=|Y_q^{(i)}|=N_b$, when prior information about the parameters is not available. The initial set of training examples is used to determine an initial value of the parameters which is used for determining an informative albeit suboptimal query distribution $q^{(1)}$ through the given QOA. Note that in the QOA of Algorithms~\ref{algo:QOA_FI},\ref{algo:QOA_FIR}, we modify the query distribution $q^{(i)}$ obtained through solving Eq.~\ref{eq:query_optimization_variance_of_params} or Eq.~\ref{eq:query_optimization_testing_error}, by mixing it with the uniform distribution over the query space $p_U$ i.e., $\mu q^{(i)} + (1-\mu)p_U$ where $0 \leq \mu \leq 1$ is the mixing coefficient. This is done to encourage exploration and is analogous to epsilon-greedy policies in reinforcement learning \cite{sutton2018reinforcement}. The value of $\mu$ typically depends on the number of queries made so far, and we set it to $\mu = 1 - 1/|X^{(i)}|^{1/6}$ as often used for such active learning algorithms \cite{sourati2017asymptotic,chaudhuri2015convergence}.

Deviating from a vanilla batch model AL scheme, we require an additional input of the query space $\{\querySpace^{(i)}\}_{i \in [i_{\text{max}}]}$ during training. We allow the query space to adaptively change from one batch to the next. How do we decide how the query space changes from one batch to the next? We remind ourselves that the query distribution in $\fisherInfo_q$ is a joint probability distribution over the measurement operators, preparation operators and evolution times. Conditioned on a particular evolution time, we would not expect changing $\measSpace$ or $\prepSpace$ to help in reducing the query complexity. The query space is then adaptively grown by growing $\timeSpace$ linearly or exponentially with each batch. We note that $\querySpace^{(1)} \subset \querySpace^{(2)} \subset ... \subset \querySpace^{(i_{max})}$.

The output of the algorithm is an estimate $\hat{\bm{\theta}}$ of the true Hamiltonian model parameters $\bm{\theta}^\star$ learned through active learning using $N$ queries. We use the unbiased maximum likelihood estimator in the HAL algorithm. HAL-FI produces an estimate $\hat{\bm{\theta}}$ that has the minimum RMSE possible using a budget of $N$ queries, and batches of size $N_b$. Similarly, HAL-FIR produces an estimate $\hat{\bm{\theta}}$ that performs best in prediction of queries to the Hamiltonian against the testing distribution $\ptest$.

\begin{algorithm}[H]
	\caption{Hamiltonian Active Learning (HAL)}
	\small
	\textbf{Input}: Initial number of queries $N^{(0)}_{tot}$, Batch size $N_{b}$, Initial query distribution $q^{(0)}$, Maximum number of batches $i_{\text{max}}$, Adaptively growing query space $\{\querySpace^{(i)}\}_{i \in [i_{\text{max}}]}$, oracle, query optimization algorithm (QOA) \\
	\textbf{Output}: $\hat{\boldsymbol{\theta}}$
	\begin{algorithmic}[1]
		\State Sample $N^{(0)}_{tot}$ queries $X_q^{(0)}$ from $\querySpace^{(0)}$ according to $q^{(0)}$
		\State Obtain measurement outcomes $Y_q^{(0)}$ by sending queries $X^{(0)}$ to oracle
		\State Set $X^{(0)}=X_q^{(0)}$ and $Y^{(0)}=Y_q^{(0)}$
		\State Compute MLE estimate: $\hat{\boldsymbol{\theta}}^{(i-1)} = \arg \min \limits_{\boldsymbol{\theta}} L(\boldsymbol{\theta};X^{(0)},Y^{(0)})$
		\For{$i = 1:i_{\text{max}}$}
		\State Solve $q^{(i)}$ through QOA (Algorithm~\ref{algo:QOA_FI} or \ref{algo:QOA_FIR}) \label{lst:line:hal_query_optimization}
		\State Sample $N_{b}$ queries $X_q^{(i)}$ from $\querySpace^{(i)}$ w.p. $q^{(i)}$ \label{lst:line:halfi_sampling_queries}
		\State Update number of queries: $N_{tot}^{(i)} = N_{tot}^{(i-1)} + N_{b}$
		\State Obtain measurement outcomes $Y^{(i)}$ by issuing queries $X^{(i)}$ to oracle
		\State Set $X^{(i)}=X^{(i-1)} \bigcup X_q^{(i)}$ and $Y^{(i)}=Y^{(i-1)} \bigcup Y_q^{(i)}$
		\State Compute MLE estimate: $\hat{\boldsymbol{\theta}}^{(i)} = \arg \min \limits_{\boldsymbol{\theta}} L(\boldsymbol{\theta};X^{(i)},Y^{(i)})$
		\EndFor
		\State \Return $\hat{\boldsymbol{\theta}}^{(i_{\text{max}})}$
	\end{algorithmic}
	\label{algo:AL_hamiltonians}
\end{algorithm}

\begin{algorithm}[H]
	\caption{Query Optimization based on Fisher Information (FI)}
	\small
	\textbf{Input}: Number of queries made so far $N^{(i-1)}_{tot}$, Batch size $N_{b}$, Query space $\querySpace^{(i)}$, Current parameter estimates $\hat{\boldsymbol{\theta}}^{(i)}$ \\
	\textbf{Output}: $q^{(i)}$
	\begin{algorithmic}[1]
		\State Set $N^{(i)}_{tot} = N^{(i-1)}_{tot} + N_b$ 
		\State Solve $q^{(i)} = \arg \min \limits_{q} \Tr(\fisherInfo_q^{-1}(\hat{\boldsymbol{\theta}}^{(i)}))$ subject to $\sum \limits_{x \in \querySpace^{(i)}} q(x) = 1$, and  $0 \leq q(x) \leq 1, \forall x \in \querySpace^{(i)}$ \label{lst:line:halfi_query_optimization}
		\State Obtain uniform distribution over query space: $p_U = 1/|\querySpace^{(i)}|$
		\State Set mixing coefficient: $\mu = 1 - 1/|N^{(i-1)}_{tot}|^{1/6}$
		\State Modify query distribution: $q^{(i)} = \mu q^{(i)} + (1 - \mu) p_U$
		\State \Return $q^{(i)}$
	\end{algorithmic}
	\label{algo:QOA_FI}
\end{algorithm}

\begin{algorithm}[H]
	\caption{Query Optimization based on Fisher Information Ratio (FIR)}
	\small
	\textbf{Input}: Number of queries made so far $N^{(i-1)}_{tot}$, Batch size $N_{b}$, Query space $\querySpace^{(i)}$, Current parameter estimates $\hat{\boldsymbol{\theta}}^{(i)}$, Testing Distribution $\ptest$ \\
	\textbf{Output}: $q^{(i)}$
	\begin{algorithmic}[1]
		\State Set $N^{(i)}_{tot} = N^{(i-1)}_{tot} + N_b$ 
		\State Compute model fisher information corresponding to $\ptest$: $\fisherInfo_{\ptest}(\hat{\boldsymbol{\theta}})$
		\State Solve $q^{(i)} = \arg \min \limits_{q} \Tr\left(\fisherInfo_q^{-1}(\hat{\boldsymbol{\theta}}^{(i)}) \fisherInfo_{\ptest}(\hat{\boldsymbol{\theta}})\right)$ subject to $\sum \limits_{x \in \querySpace^{(i)}} q(x) = 1$, and  $0 \leq q(x) \leq 1, \forall x \in \querySpace^{(i)}$ \label{lst:line:halfir_query_optimization}
		\State Obtain uniform distribution over query space: $p_U = 1/|\querySpace^{(i)}|$
		\State Set mixing coefficient: $\mu = 1 - 1/|N^{(i-1)}_{tot}|^{1/6}$
		\State Modify query distribution: $q^{(i)} = \mu q^{(i)} + (1 - \mu) p_U$
		\State \Return $q^{(i)}$
	\end{algorithmic}
	\label{algo:QOA_FIR}
\end{algorithm}

\subsection{Comment on Heisenberg Limited Scaling} \label{sec:comment_HS_scaling_HAL}
We claim that HAL-FI with an appropriately chosen adaptively growing query space can achieve Heisenberg limited scaling in Hamiltonian parameters where possible. If the query space is not \textit{rich} enough, it will not be possible to determine a sequence of queries to achieve Heisenberg limited scaling. Without an adaptively growing query space (i.e., fixed query space), the scaling of the number of queries $N$ with error parameter $\epsilon$ is $O(1/\epsilon^2)$ as is dictated by the Cramer-Rao bound. The query complexity by using an active learner only improves by a constant factor in such a case. If one chose to adaptively grow the query space during training without an active learning strategy and chose for example an uniform distribution over the new query space (i.e., carry out passive learning), Heisenberg limited scaling would not be expected as it would become exponentially more unlikely to sample an informative query. Moreover, Heisenberg limited scaling is expected as long as the evolution time is below the decoherence time. Beyond the decoherence time, the oracle will start losing its quantum behavior. 

If the so chosen adaptively growing query space cannot be used to achieve Heisenberg limited scaling, it should still be possible to achieve Heisenberg limited scaling for a subset of the Hamiltonian parameters provided that the goal is to now learn these parameters and we are provided information about the other parameters. This learning task often occurs in practice during recalibrations of quantum devices when it is required to learn a subset of parameters that are known to fluctuate significantly with time but information about the other parameters can be used from previous calibrations. We provide empirical evidence for this claim in Section~\ref{sec:Results} where we consider the application of the HAL-FI to cross-resonance Hamiltonians. The results of HAL-FI for this particular application are summarized in Table~\ref{tab:query_complexity_summary}.

What distinguishes HAL-FI from other methods such as Floquet calibration \cite{arute2020observation} which have been shown to achieve Heisenberg limited scaling is that it does not require prior specification of experiments and their order of implementation. This is decided by the HAL-FI during learning. Moreover, HAL-FI utilizes single-shot outimes from queries, instead of requiring expectation values. In practice, this can result in multiple orders of magnitude reduction in queries required. Finally, HAL-FI can achieve query advantage over passive learners for complete query spaces even when Heisenberg limited scaling cannot be achieved. 

\subsection{Computational Cost and Extensions}
A consequence of adaptively growing the query space over rounds during learning is that the SDP programs (Eq.~\ref{eq:SDP_query_optimization}) corresponding to the query optimizations of Eqs.~\ref{eq:query_optimization_variance_of_params} and \ref{eq:query_optimization_testing_error} (with computational cost $\mathcal{O}(n_{\mathcal{Q}}^2 m^3 + n_{\mathcal{Q}}m^4 + m^5)$) become increasingly more computationally expensive to solve over rounds. If $n_\querySpace$ grows exponentially, each iteration of the query optimization problem becomes more exponentially expensive to solve. This can be circumvented by reducing the number of queries to optimize over using uncertainty filtering \cite{wei2015submodularity}, thereby effectively reducing the size of the query space $n_\querySpace$ over which the query optimization is carried out. Uncertainty filtering for our application of Hamiltonian learning to the cross-resonance Hamiltonian is discussed in Appendix~\ref{app_sec:uncertainty_filtering}.

The HAL-FI and HAL-FIR algorithms can be generalized to different experimental setups or requirements. The HAL algorithm presented in Algorithm~\ref{algo:AL_hamiltonians} uses the stopping criterion of maximum number of batches of queries issued during learning but other stopping criteria such as the $\ell_2$ norm of the differences in consecutive parameter values $||\hat{\boldsymbol{\theta}}^{(i)}-\hat{\boldsymbol{\theta}}^{(i-1)}||_2$ could also be used. We have also restricted our discussion to the maximum-likelihood estimator but the HAL algorithm can be combined with any unbiased consistent estimator.

\begin{table}[ht!]
\centering
\begin{tabular}{|c|c|c|}
\hline
\textbf{Adaptivity in Query Space} &  \textbf{Scaling of N} &  \textbf{Scaling of N (Recalibration)} \\ \hline
None & $O(1/\epsilon^2)$ & $O(1/\epsilon^2)$ \\ \hline
Linearly growing $\timeSpace$  & $O(1/\epsilon^2)$ & $O(1/\epsilon^{2/3})$ \\ \hline
Exponentially growing $\timeSpace$ & $O(1/\epsilon^2)$ & $> O(1/\epsilon)$ \\ \hline
\end{tabular}
\caption{\label{tab:query_complexity_summary} Query complexity of HAL-FI for Hamiltonian learning (Problem~\ref{prob:HL_model_inference}) under different conditions on a real quantum device. The case of fixed query space corresponds to adaptivity in query space being \textit{none}.}
\end{table}

\section{Hamiltonian Learning for a Two-Qubit Superconducting CR Gate: Model and Setup} 
\label{sec:CR_Gate_Model_and_Setup}
To assess the performance of the HAL algorithms described in Section~\ref{sec:HAL} over a passive learner and empirically verify our claims, we consider the application of learning cross-resonance Hamiltonians on superconducting IBM Quantum devices. In this section, we describe the model of the two-qubit cross-resonance Hamiltonian, and the set of queries that we can make to it. This is followed by a description of the different noise sources that affect the quantum system and how this modifies the MLE of Hamiltonian parameters. A description of the IBM Quantum devices employed for assessing the performance of the HAL algorithms can be found in Appendix~\ref{app_sec:details_cr_hamiltonians}.

\subsection{Cross-Resonance Hamiltonian}\label{sec:CR_hamiltonian}
The cross resonance (CR) gate is a two-qubit entangling gate for superconducting qubits requiring only microwave control which allows for the use of fixed-frequency transmon qubits \cite{rigetti2010fully,chow2011simple}. Using appropriate pulse sequences such as multi-pulse echos and cancellation tones \cite{sheldon2016procedure}, the CR gate can be transformed to a locally equivalent CNOT gate \cite{zhang2003geometric}. Combined with arbitrary single qubit gates, this then forms a complete set of gates for universal quantum computation.

The Hamiltonian of the cross-resonance (CR) gate has the following structure
\begin{align}
    H_{CR} &= \frac{\sigma_Z \otimes A}{2} + \frac{\sigma_I \otimes B}{2} \\
    A &= c_{ZI} \sigma_I + c_{ZX} \sigma_X + c_{ZY} \sigma_Y + c_{ZZ} \sigma_Z \\
    B &= c_{IX} \sigma_X + c_{IY} \sigma_Y + c_{IZ} \sigma_Z
    \label{eq:general_cr_hamiltonian_def}
\end{align}
where $\{\sigma_I,\sigma_X,\sigma_Y,\sigma_Z\}$ are the single-qubit Paulis and $c_{ab} \in \mathbb{R}$ are real coefficients of the Pauli product terms $\sigma_a \otimes \sigma_b$. The above time-independent Hamiltonian description of the cross-resonance gate can be obtained through theoretical models based on effective block-diagonal Hamiltonian techniques \cite{magesan2018effective}. In our experiments which we describe in Section~\ref{sec:experimental_setup}, the CR gate is implemented without using an echo \cite{sheldon2016procedure} to refocus the $\sigma_I \sigma_X$, $\sigma_Z \sigma_Z$ and $\sigma_Z \sigma_I$ terms. However, we measure only the target qubit through our queries and thus effectively neglect the $\sigma_Z \sigma_I$ term. The target qubit is typically chosen to be qubit $1$ in a $(0,1)$ qubit pair and is specified for the different quantum devices we consider in Appendix~\ref{app_sec:quantum_device_description}. The effective removal of the $\sigma_Z \sigma_I$ term from Eq.~\ref{eq:general_cr_hamiltonian_def} results in the following simplified CR Hamiltonian which we consider throughout the rest of our work:
\begin{equation}
	H = \sum \limits_{\substack{a \in \{I,Z\} \\ b \in \{X,Y,Z\}}} J_{ab} \sigma_a \otimes \sigma_b
	\label{eq:cr_hamiltonian_def}
\end{equation}
where we have used the parameter vector $\mathbf{J} = [J_{IX},J_{IY},J_{IZ},J_{ZX},J_{ZY}, J_{ZZ}]^T$ to denote the non-zero coefficients of the corresponding Pauli product terms and have omitted the subscript CR. Learning the unknown Hamiltonian of the two-qubit CR gate is then reduced to estimating the unknown parameter vector $\mathbf{J}$.

Noting the block-diagonal structure of $H$, we can express it in the usual computational basis of $(\ket{00}, \ket{01}, \ket{10}, \ket{11})$ as
\begin{equation}
	H = \begin{bmatrix} a_0 & \beta_0^\star & 0 & 0 \\ \beta_0 & -a_0 & 0 & 0 \\ 0 & 0 & a_1 & \beta_1^\star \\ 0 & 0 & \beta_1 & -a_1 \end{bmatrix}, \text{ where } 
	\begin{matrix} 
	   a_j = J_{IZ} + (-1)^j J_{ZZ}  \\
	   \beta_j = (J_{IX} + (-1)^j J_{ZX}) + i(J_{IY} + (-1)^j J_{ZY})
    \end{matrix} 
    \label{eq:cr_hamiltonian_block_form}
\end{equation}
where the subscript $j \in \{0,1\}$ is used to refer to the two different blocks. The two blocks have similar structure with their elements differing in the sign of $J_{ZX}$, $J_{ZY}$, and $J_{ZZ}$. Using $a_j$, and $\beta_j$ for $j \in \{0,1\}$ from Eq.~\ref{eq:cr_hamiltonian_block_form}, we define the following parameters
\begin{align}
	\omega_j &= \sqrt{a_j^2 + |\beta_j|^2} \\
	\delta_j &= \sin^{-1}\frac{a_j}{\omega_j} \\
	\phi_j &= \arg(\beta_j)
	\label{eq:Lambda_Hcr_parameterization}
\end{align}
where the subscript $j=0$ is used to denote the preparation operator $U=\sigma_I \sigma_I$ and $j=1$ to denote $U= \sigma_X \sigma_I$. We then arrive at an alternate parameterization (related to the spectral decomposition of $H$) which turns out to be useful for simplifying expressions for probability, likelihood and Fisher information: $\boldsymbol{\Lambda} = (\omega_0, \delta_0, \phi_0, \omega_1, \delta_1, \phi_1)^T$. The unitary operator $U(t)=e^{-iHt}$ in the usual computational basis is then given by
\begin{equation}
    \small
    U(t) = \begin{bmatrix} \cos(\omega_0 t)-i \sin(\delta_0) \sin(\omega_0 t) & -i e^{-i \phi_0} \cos(\delta_0) \sin(\omega_0 t) & 0 & 0 \\ -i e^{i \phi_0} \cos(\delta_0) \sin(\omega_0 t) & \cos(\omega_0 t)+ i \sin(\delta_0) \sin(\omega_0 t) & 0 & 0 \\ 0 & 0 & \cos(\omega_1 t)-i \sin(\delta_1) \sin(\omega_1 t) & -i e^{-i \phi_1} \cos(\delta_1) \sin(\omega_1 t) \\ 0 & 0 & -i e^{i \phi_1} \cos(\delta_1) \sin(\omega_1 t) & \cos(\omega_1 t)+i \sin(\delta_1) \sin(\omega_1 t) \end{bmatrix}
    \label{eq:unitary_cr_hamiltonian_block_form}
\end{equation}
where we have used the parameter vector of $\bm{\Lambda}$ defined in Eq.~\ref{eq:Lambda_Hcr_parameterization}. Moreover, the different components of $\bm{\Lambda}$ can be bounded based on their definitions and these bounds will help us later while solving the MLE problems. By construction, we have $-\frac{\pi}{2} \leq \delta_{0,1} \leq \frac{\pi}{2}$ and $\phi_{0,1}$ can be bounded within any interval of size $2\pi$ e.g., $-\pi \leq \phi_{0,1} \leq \pi$. Assuming $\delta t$ is the average time increment between distinct ordered values of evolution times $t \in \timeSpace$, $\omega_{0,1}$ can be bounded based on the Nyquist sampling theorem as $0 \leq \omega_{0,1} \leq \frac{\pi}{\delta t}$.

\begin{figure}[h!]
    \centering
    \includegraphics[scale=0.23]{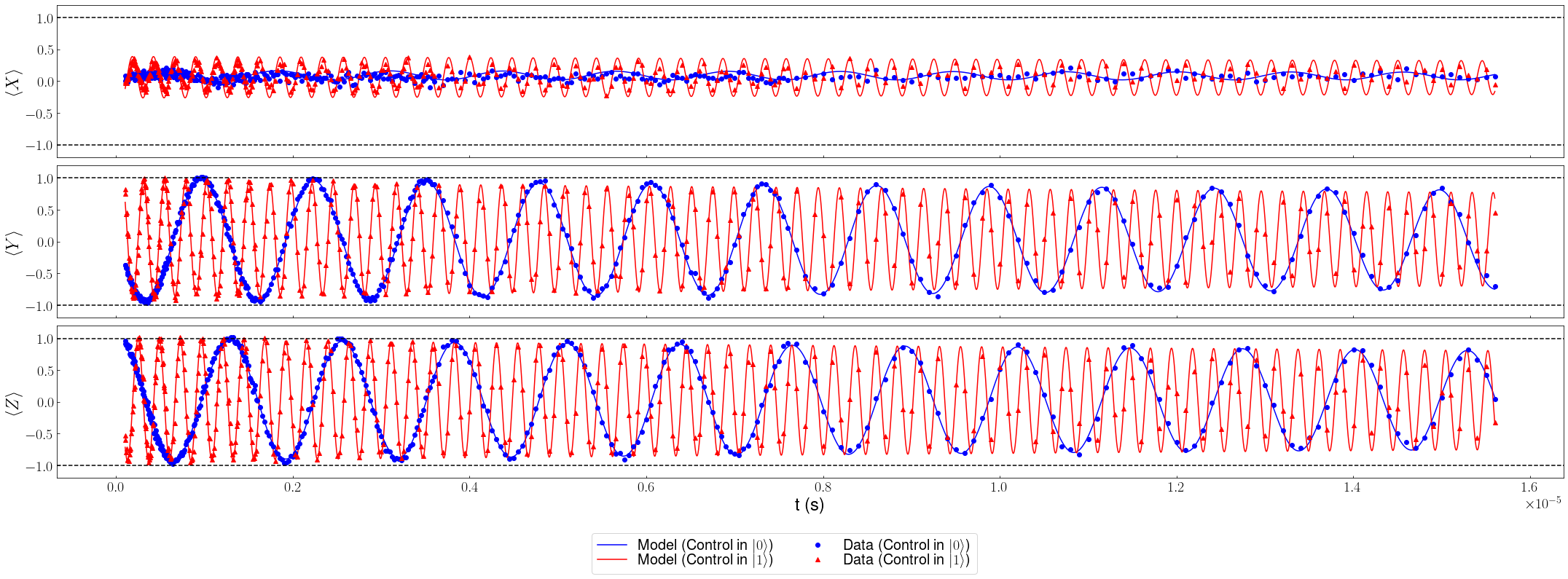}
    \caption{Rabi oscillations obtained experimentally through queries to a CR Hamiltonian on the IBM Quantum device (\textit{ibmq\_boeblingen}) for different measurement operators (rows), preparation operators (markers) and evolution times $t$ (x-axis). The set of measurement and preparation operators will be described in Section~\ref{sec:cr_query_space}. The experimental data is indicated by markers and the model fit to the data by lines. The model fit was generated by learning the CR Hamiltonian (Eq.~\ref{eq:cr_hamiltonian_block_form}) from experimental data using MLE and predicting values of Rabi oscillations for the query space using the learned Hamiltonian.}
    \label{fig:rabi_oscillations_illustration}
\end{figure}

To obtain the physical meaning of the parameter vector $\bm{\Lambda}$, we consider Rabi oscillations. For different measurement operators $M \in \measSpace$ and preparation operators $U \in \prepSpace$, the corresponding Rabi oscillation is the difference in probability densities of the ground state and excited state of the target qubit with time $t \in \timeSpace$. A typical example of the Rabi oscillations from querying the CR Hamiltonian on a noisy quantum system is shown in Figure~\ref{fig:rabi_oscillations_illustration}. We remark that $\omega_{0,1}$ defines the frequency of the Rabi oscillations for the two different preparation operators we consider. The parameters $\delta_{0,1}$ and $\phi_{0,1}$ define the offsets, amplitudes and phase shifts of the Rabi oscillations. In Figure~\ref{fig:rabi_oscillations_illustration}, we can see the effects of different noise sources such as readout and depolarization which will be discussed in Section~\ref{sec:noise_sources}.
\subsection{Experimental Setup} \label{sec:experimental_setup}
In this section, we give a quick overview of the different IBM Quantum devices that we employ for our application of Hamiltonian learning. This is followed by a description of the query space considered for our application and how the queries to the CR gate on the IBM Quantum devices are made in experiment.

\subsubsection{Quantum Devices} \label{sec:quantum_devices}
We will present data and results for four different IBM Quantum devices which we call A, B, C, and D. All of these devices are based on superconducting architectures requiring only microwave control \cite{rigetti2010fully,chow2011simple} and consist of fixed-frequency transmon qubits with shared quantum buses \cite{chow2014implementing}. Device A is a two-qubit device. Device B is a four-qubit device on which we will query only one of the CR gates that can be implemented between two qubits. Device C is a five-qubit device with a bow-tie layout. Device D is the 20-qubit \textit{ibmq\_boeblingen} which was accessed via the IBM Quantum cloud computing service. See Figure~\ref{fig:quantum_device_connectivity_maps} in Appendix~\ref{app_sec:quantum_device_description} for the connectivity maps of all the devices. We give a summary of the properties of the pairs of qubits involved in the cross-resonance Hamiltonians we considered on these devices in Table~\ref{tab:quantum_device_parameters} in Appendix~\ref{app_sec:quantum_device_description}.

\subsubsection{Experimental Implementation and Query Space} \label{sec:cr_query_space}
Queries to the CR Hamiltonians between different qubit pairs on the IBM Quantum devices are made through appropriate pulse sequences. These pulse sequences are constructed and executed on the hardware using \texttt{Qiskit-Pulse} \cite{alexander2020qiskit}, which is a pulse programming module within \texttt{Qiskit} \cite{abraham2019qiskit} and serves as a front-end implementation of the \texttt{OpenPulse} interface \cite{mckay2018qiskit}. Each \texttt{Qiskit-Pulse} \cite{alexander2020qiskit} program consists of pulses, channels and instructions. Here, we describe a \texttt{Qiskit-Pulse} program and describe how a query to a CR gate on a IBM Quantum device is specified.

A pulse is a time-series of complex-valued amplitudes with maximum unit norm and which we denote as $a_k$ where $k \in [n-1]$ corresponds to the time stamp. The difference between these time-stamps is considered to be $dt$ which is typically the sample rate of the waveform generator. The output signal thus has an amplitude of
\begin{equation}
    A_k = \text{Re}\left[ e^{i2\pi f k dt + \gamma} a_k \right]
    \label{eq:pulse_signal_amplitude}
\end{equation}
at time $k dt$ where $f$ and $\gamma$ are a modulation frequency and phase. A pulse is specified in \texttt{Qiskit-Pulse} by specifying the individual amplitudes $a_k$ and the phase $\phi$. Alternatively, one can use parametric pulse shapes that are offered by the library such as \texttt{Gaussian}, \texttt{GaussianSquare}, etc. These pulses are then implemented on the hardware via channels which label signal lines used for transmitting and receiving signals between the control electronics, and the hardware. In particular, these are implemented on the \texttt{PulseChannel} and are used to control the system Hamiltonian to implement different gates. 

\begin{figure}[ht!]
	\centering
    \includegraphics[scale=0.6]{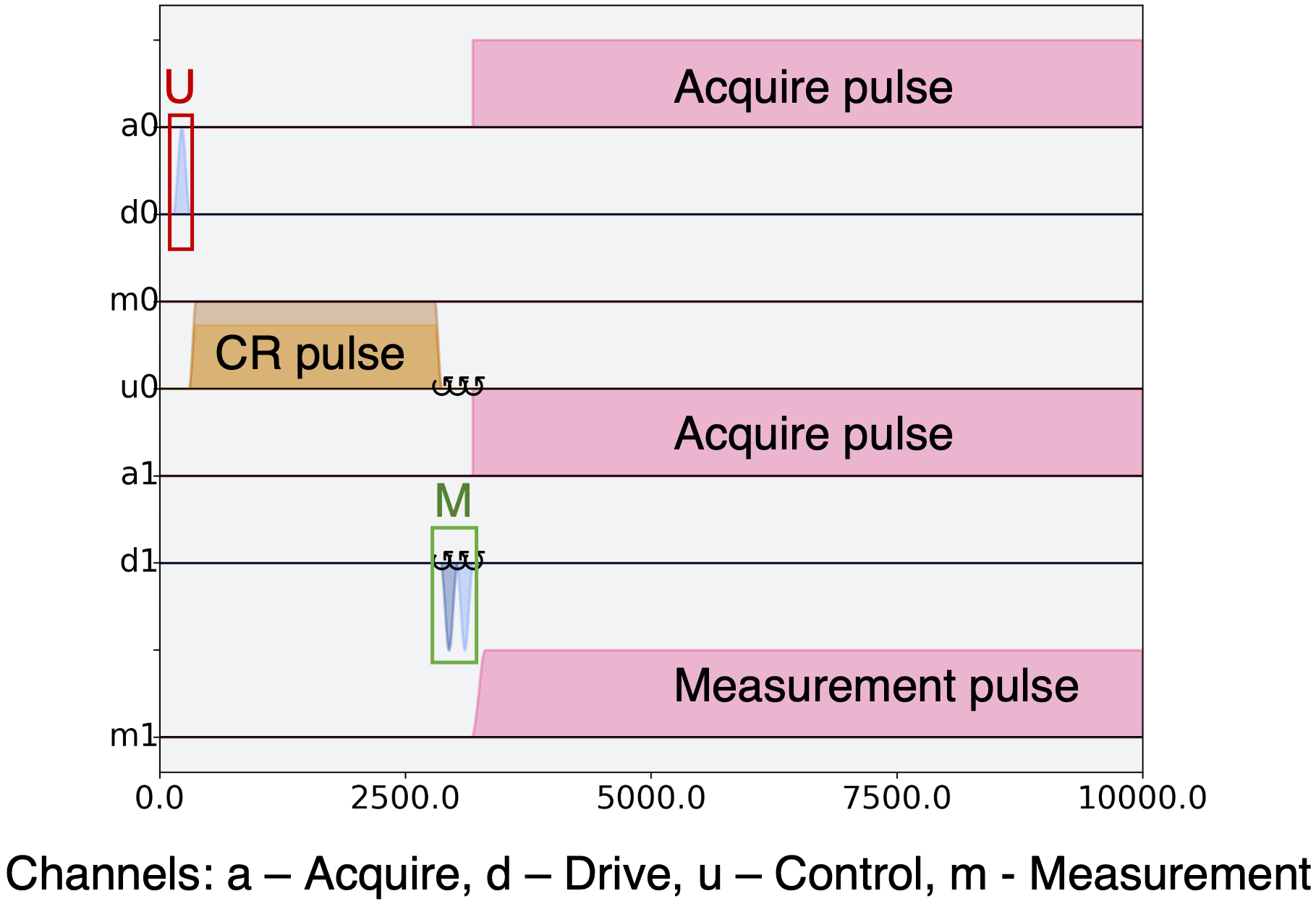}
    \caption{An example of a CR pulse schedule on the IBM Quantum device \textit{ibmq\_boeblingen} considering to the query of $x=(M,U,t)$ where $M= \sigma_I \otimes \exp \left(-i \frac{\pi}{4} \sigma_X \right)$, $U= \sigma_X \otimes \sigma_I$, and time duration $t=6 \times 10^{-7}$s. The x-axis corresponds to time normalized by $dt=2.22\times 10^{-10}$ (Eq.~\ref{eq:pulse_signal_amplitude}). The different channels corresponding to each qubit (y-axis) are written as the type of channel (see plot legend) followed by qubit number. Qubit $0$ is set to be the control qubit and $1$ to be the target qubit. The envelope of the different pulses are shown in each channel. The rotations on the drive or control channels indicate virtual Z gates. An equivalent representation of the quantum circuit is shown in Fig.~\ref{fig:query_specification}.}
    \label{fig:CR_expt_pulse_sequence}
\end{figure}

To implement a query to a quantum device, we need an equivalent description of the quantum circuit shown in Fig.~\ref{fig:query_specification} in the form of a pulse schedule. We discuss this here in parallel with a description of the query space $\querySpace$ considered for learning CR Hamiltonians. To prepare the initial state, we consider the set of preparation operators $\prepSpace = \{\sigma_I \otimes \sigma_I,\sigma_X \otimes \sigma_I \}$ applied to the pure state $\ket{00}$. The single-qubit gates are implemented as a sequence of Gaussian pulses of the appropriate amplitude and duration \cite{alexander2020qiskit}. Assuming the first (left) qubit is the control and the second (right) qubit is the target, the effect of the preparation operators is to place the control in $\ket{0}$ and $\ket{1}$ respectively. We evolve the initial state $\ket{\psi(0)}$ for time $t\in \timeSpace$ which we will specify when discussing the results of our application of Hamiltonian learning in Section~\ref{sec:Results}. This is done by switching the CR interaction on for time duration $t$ which is done in practice by implementing a \texttt{GaussianSquare} pulse of duration $t$. A \texttt{GaussianSquare} pulse is a square pulse with truncated Gaussian-shaped rising and falling edges. We discuss later in Section~\ref{sec:noise_sources} how using this pulse may introduce non-idealities in the system evolution. Finally after obtaining the final state $\ket{\psi(t)}$, we apply the measurement operators in $\measSpace= \{\sigma_I \otimes \exp \left(i \frac{\pi}{4} \sigma_Y \right), \sigma_I \otimes \exp \left(-i \frac{\pi}{4} \sigma_X \right), \sigma_I\otimes \sigma_I \}$ and measure only the second qubit which we have chosen as the target qubit. The query space is then $\querySpace = \measSpace \times \prepSpace \times \timeSpace$. An example of a pulse schedule is shown in Fig.~\ref{fig:CR_expt_pulse_sequence} highlighting the different parts of the query. 
Moreover, in our experimental setup, we obtain measurements of the single-shot signal (integrated cavity amplitude) $\readoutSignal$ which is a function of the measurement outcomes $\mathbf{y}$ which we have described earlier. In the following Section~\ref{sec:noise_sources}, we will discuss how to model the different noise sources that affect our system and how they are determined through experiments.

\subsection{Estimates of Noise and Nonidealities for Experimental System} \label{sec:noise_sources}
Using the noise models presented in Section~\ref{sec:hlproblem_noise}, we give the specific relevant models for readout noise, imperfect pulse shaping, and decoherence for the different IBM Quantum devices which we study.

Let us introduce some notation that we will use in the following discussions. We use the index $k$ for the queries. The $k$th query is given by $x^{(k)} = (M^{(k)},U^{(k)},t^{(k)})$, and the corresponding measurement outcome of the target qubit as $y^{(k)}$. The measurement outcome $y^{(k)}$ is not directly observed but inferred from the corresponding signal $\readoutSignal^{(k)}$. We denote the inferred value as $\hat{y}^{(k)}$.

\subsubsection{Readout Noise} \label{sec:readout_noise}
\begin{figure}[ht!]
    \centering
    \xincludegraphics[scale=0.35, label=\textbf{(a)}]{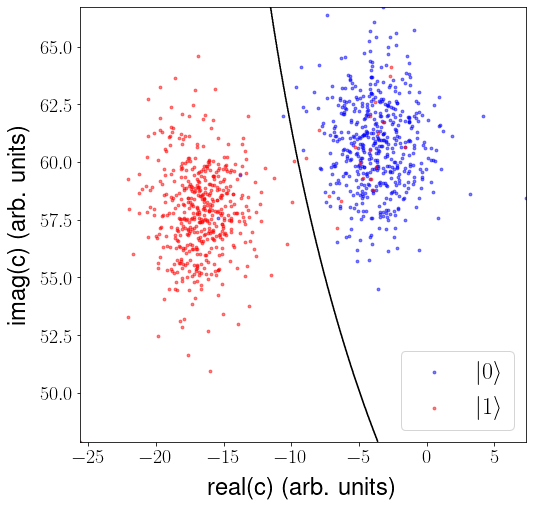}
    \hspace{4em}
    \xincludegraphics[scale=0.35, label=\textbf{(b)}]{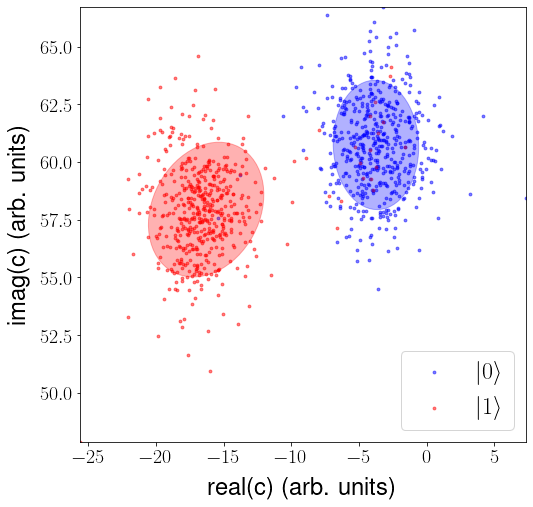}
	\caption{Characterization of readout noise from calibration data of single qubit readouts considering a bit-flip channel model using (a) a trained Bayesian naive classifier, and (b) fitting Gaussian distributions. In (a)-(b), the experimental data points of the complex readout signal $\mathbf{c}$ are shown as markers. In (a), the decision boundary is shown as a line. In (b), the contours indicate the single standard deviation.}
	\label{fig:readout_models}
\end{figure}

As discussed in Section~\ref{sec:hlproblem_noise}, we assume the measurement noise model to be a bit-flip channel with the input of unobserved measurement outcomes $y$ and the output of readout $\tilde{y}$ observed through signal $\readoutSignal$. For this inference task, we use calibration data of single qubits initially prepared in states $\ket{0}$ or $\ket{1}$, and subsequently measured in the usual computational basis. In Figure \ref{fig:readout_models}(a), we plot different realizations $\{y^{(k)},\readoutSignal^{(k)}\}_{k=1}^{100}$ from IBM Quantum device D $ibmq\_boeblingen$ that were used for training a binary classifier. The classifier provides us with the ability to predict $y$ given $\readoutSignal$. Moreover, the misclassification errors $p_{\hat{\mathrm{y}}|\mathrm{y}}(1|0)$ and $p_{\hat{\mathrm{y}}|\mathrm{y}}(0|1)$ can be used to approximate the properties of the bit-flip channel, in particular the conditional probabilities of a bit-flip $p_{\tilde{\mathrm{y}}|\mathrm{y}}(1|0)$ and $p_{\tilde{\mathrm{y}}|\mathrm{y}}(0|1)$ respectively. As these are independent of values of $\readoutSignal$ here, we denote the conditional probabilities of a bit-flip as $r_{0}=p_{\tilde{\mathrm{y}}|\mathrm{y}}(1|0)$ and $r_{1}=p_{\tilde{\mathrm{y}}|\mathrm{y}}(0|1)$. The MLE of the parameters incorporating this noise model is given by
\begin{align}
    \hat{\boldsymbol{\theta}} = \arg \min_{\boldsymbol{\theta}} - \frac{1}{N} \sum_{k=1}^N & \log \left[ p_{\tilde{\mathrm{y}}|\mathrm{x}}(\tilde{y}^{(k)}|x^{(k)};\boldsymbol{\theta}) \right] \\
    = \arg \min_{\boldsymbol{\theta}} - \frac{1}{N} \sum_{k=1}^N & \log \Big[ (1-\hat{y}^{(k)})\left( (1-r_0)p_{\mathrm{y}|\mathrm{x}}(y^{(k)}=0|x^{(k)};\bm{\theta}) + r_1(1-p_{\mathrm{y}|\mathrm{x}}(y^{(k)}=1|x^{(k)};\bm{\theta})) \right) + \\
    & \hat{y}^{(k)} \left( (1-r_1)(1-p_{\mathrm{y}|\mathrm{x}}(y^{(k)}=1|x^{(k)};\bm{\theta})) + r_0 p_{\mathrm{y}|\mathrm{x}}(y^{(k)}=0|x^{(k)};\bm{\theta}) \right) \Big] 
\end{align}
where $p_{\mathrm{y}|\mathrm{x}}(y^{(k)}|x^{(k)};\bm{\theta})$ is given by Eq.~\ref{eq:borns_rule}. Alternately, instead of assigning a deterministic result $\hat{y}^{(k)}$ for each $c^{(k)}$, we can incorporate $p_{\mathrm{c}|y}$ directly into the log-likelihood function which could yield a more accurate model. This could be done through our choice of binary classifier or by fitting a distribution to the training data. Noting that single qubit measurement outcomes correspond to their energy levels, we fit Gaussian distributions to the training data and hence obtain a parameteric form of $p_{\mathrm{c}|y}$ in Figure \ref{fig:readout_models}(b). The MLE is now
\begin{equation}
    \hat{\boldsymbol{\theta}} = \arg \min_{\boldsymbol{\theta}} - \frac{1}{N} \sum_{k=1}^N \log \left[ p_{\mathrm{c}|\mathrm{y}}(c^{(k)}|0) p_{\mathrm{y}|\mathrm{x}}(y^{(k)}=0|x^{(k)};\bm{\theta}) + p_{\mathrm{c}|\mathrm{y}}(c^{(k)}|1) (1-p_{\mathrm{y}|\mathrm{x}}(y^{(k)}=0|x^{(k)};\bm{\theta})) \right] 
\end{equation}
A useful tool for calibration and diagnostics is Rabi oscillations which we denote by $p_{\mathrm{rabi}}(x)$. Rabi oscillations are obtained from evaluating the difference in the population densities of the ground state and excited state of the target qubit or $p_{\mathrm{rabi}}(x) = p_{\mathrm{y}|\mathrm{x}}(0|x) - p_{\mathrm{y}|\mathrm{x}}(1|x)$. We can compute Rabi oscillations in two different ways, either through binary classification or through fitting Gaussians. Using the misclassification errors from the binary classifier, we can then write
\begin{equation}
    \hat{p}_{\mathrm{rabi}}(x) = p_{\hat{\mathrm{y}}|\mathrm{x}}(0|x) \left( \frac{1-r_0+r_1}{1-r_0-r_1} \right) - p_{\hat{\mathrm{y}}|\mathrm{x}}(1|x) \left( \frac{1+r_0-r_1}{1-r_0-r_1} \right) 
\end{equation}
where we have denoted the computed Rabi oscillations as $\hat{p}_{\mathrm{rabi}}(x)$. While $p_{\mathrm{y}|\mathrm{x}}(0|x)$ and $p_{\mathrm{y}|\mathrm{x}}(1|x)$ are guaranteed to be valid probability distributions, the above estimation does not ensure that $\hat{p}_{\mathrm{rabi}}(x)$ is bounded by $-1$ and $1$. To obtain more accurate estimates of the Rabi oscillations, we can solve the following MLE problem
\begin{equation}
    \hat{p}_{\mathrm{rabi}}(x) = \arg \min_{q \in [-1,1]} \left( - \sum_k \mathds{1}\{x^{(k)} =x\} \log\left[(1+q)p_{\mathrm{c}|\mathrm{y}}(c^{(k)}|0) + (1-q)p_{\mathrm{c}|\mathrm{y}}(c^{(k)}|1) \right] \right)
    \label{eq:rabi_oscillations_mle}
\end{equation}
where we use the estimated conditional distributions $p_{\mathrm{c}|y}$ from the Gaussian fits. The indicator function $\mathds{1}\{x^{(k)}=x\}$ is used to ensure that the summation is only over measurement outcomes of given query $x$. The analytical expressions for $p_{\mathrm{rabi}}(x)$ are given by
\begin{align}
    M_{\braketLR{X}}: \, & p_{\mathrm{rabi}}(x) = \sin \delta_j \cos \delta_j \cos \phi_j + \cos \delta_j \sqrt{1-\cos^2 \delta_j \cos^2 \phi_j} \cos(2\omega_j t + \alpha_j) \\
    M_{\braketLR{Y}}: \, & p_{\mathrm{rabi}}(x) = \sin \delta_j \cos \delta_j \sin \phi_j + \cos \delta_j \sqrt{1-\cos^2 \delta_j \sin^2 \phi_j} \cos(2\omega_j t + \gamma_j) \\
    M_{\braketLR{Z}}: \, & p_{\mathrm{rabi}}(x) = 1 - 2 \cos^2 \delta_j \sin^2 (\omega_j t) = \sin^2 \delta_j + \cos^2 \delta_j \cos(2\omega_j t)
    \label{eq:rabi_oscillations_cr_gate}
\end{align}
where we denote $\alpha_j = \arg( (-\sin \delta_j \cos \phi_j) + i(-\sin \phi_j))$ and $\gamma_j = \arg((-\sin \delta_j \sin \phi_j) + i \cos \phi_j)$ in the above expressions. The subscript $j=0$ is used to denote the preparation operator $U=\sigma_I \sigma_I$ and $j=1$ to denote $U= \sigma_X \sigma_I$. 

\begin{figure}[ht!]
	\centering
    \xincludegraphics[scale=0.31, label=\textbf{(a)}]{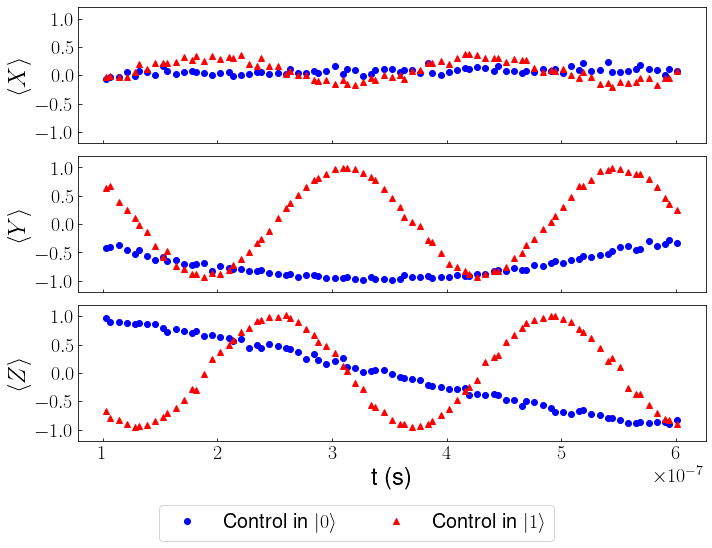}
    \qquad
    \xincludegraphics[scale=0.31, label=\textbf{(b)}]{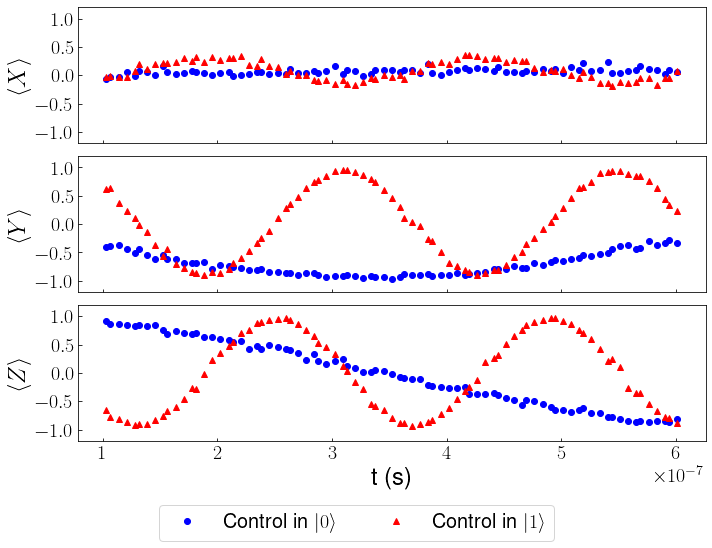}
    \caption{Examples of Rabi oscillations computed from experimental data collected from IBM Quantum device D \textit{ibmq\_boeblingen} for different measurement operators $M$ (rows), preparation operators $U$ (markers) and evolution times $t$ (x-axis). These were computed (a) using a binary classifier and (b) using Gaussian distribution fits. The difference between the two approaches is negligible when the misclassification error of the binary classifier is low ($r_0,r_1<0.01$) as is the case here.} \label{fig:rabi_oscillations}
\end{figure}

In Figure~\ref{fig:rabi_oscillations}, we plot the Rabi oscillations for the different $M \in \measSpace$ and $U \in \prepSpace$ over the time range of $t \in \timeSpace$ using the above two methods. We observe that the Rabi oscillations are not bounded between $-1$ and $+1$ for the case of $M_{\braketLR{Y}}$ in Figure~\ref{fig:rabi_oscillations}(a) when using misclassification error to compensate for the readout noise. The estimated Rabi oscillations in Figure~\ref{fig:rabi_oscillations}(b) do not suffer from the same issue.

As we will see later in Section~\ref{sec:estimation_procedure}, Rabi oscillations will also be used in our estimation procedure for obtaining an initial guess for the parameter estimate $\hat{\boldsymbol{\theta}}$ that is used as an input to the optimizer for solving the MLE problem. It can also be used as a quantitative tool for ascertaining how well the model fits the data. This will become apparent in the next few sections where we discuss other noise models.
\subsubsection{Imperfect Pulse Shaping} \label{sec:imperfect_pulse_shaping}
Another nonideality is introduced through the pulses used to control the Hamiltonian and implement different unitary operators. It is convenient to think of cross-resonance control pulses as rectangular pulses that modulate the sinusoidal control pulse. The modulated signal results in unitary operators of the form ${U(t_1)=\exp{(-iH t_1)}}$ which we would want to implement in a quantum circuit.  However, in practice, rectangular pulses cause significant amounts of signal energy to be distributed above and below the frequency of the control sinusoid. This distribution of energy can potentially excite higher energy states of the superconducting transmon being used as a qubit (i.e., $\ket{2}$ and above) as well as excite neighboring spectator transmon qubits. To minimize such effects, pulse shaping is employed to reduce this energy spread by smoothing the rising and falling edges of the pulse, which has the effect of reducing the magnitudes of the frequency artifacts above and below the frequency of the control sinusoid.  

\begin{figure}[ht!]
	\centering
	\includegraphics[scale=0.36]{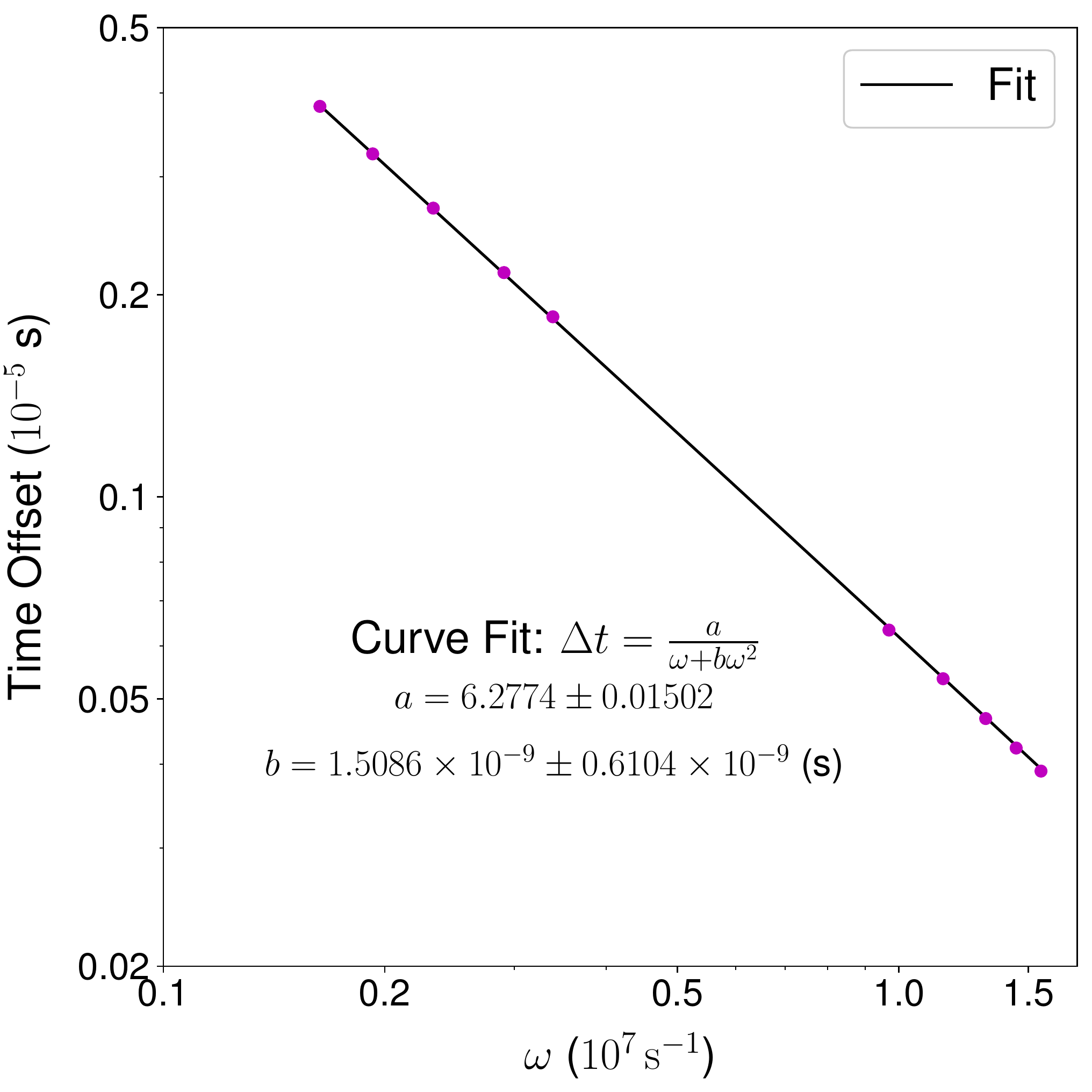}
	\caption{Dependence of the time offset $\Delta t$ on parameters $\omega$ for IBM Quantum device D \textit{ibmq\_boeblingen}. The plotted data points correspond to driving the device under different conditions and hence different cross-resonance Hamiltonians. The imperfect pulse shaping model extracted from these experimental data points is shown by a fit and this is later used in the MLE.}
	\label{fig:data_driven_model_imperfect_pulses_ibmq_boeb}
\end{figure}

Pulse shaping is accomplished by taking a Gaussian-shaped pulse, splitting it in half, and then inserting a rectangular pulse between the halves. This results in the \texttt{GaussianSquare} pulse, previously described in Section~\ref{sec:cr_query_space}. Thus, we actually implement operators of the form ${\tilde{U}(t_1)=\exp{(-i \mathbb{T} \int_{0}^{t_1} \tilde{H}(t) dt)}}$ where $\mathbb{T}$ is the time ordering operator, $\tilde{H}$ is the Hamiltonian at any particular time given by $\tilde{H}(t)=H(t, v(t))$ with $H$ as the cross-resonance Hamiltonian and $v(t)$ as a function of the cross-resonance pulse amplitude.  Up to a first-order approximation in $t$, we can however model this as (see Appendix~\ref{app_sec:imperfect_pulse_shaping} for details)
\begin{equation}
    \tilde{H}(t) = v(t)H
\end{equation}
We denote $\Delta t_r$ and $\Delta t_f$ as the time durations of the rising and falling edges of the shaped pulse. The central portion of $v(t')$ is a rectangular function; i.e., for $\Delta t_r \le t' \le t-\Delta t_f$, $v(t') = \mathbbm{1}_{t' \in [\Delta t_r, t-\Delta t_f]}$ where $t$ is the duration of the pulse.  We then have
\begin{align}
    \tilde{U}(t) &= \exp{\left( -i \mathbb{T} \int_{0}^{t} \tilde{H}(t') dt' \right)} \\
        &= \exp{\left( -i H \int_{0}^{t} v(t') dt' \right)} \\
        &= \exp{\left( -i H \left[ (t-\Delta t_f - \Delta t_r) + \underbrace{\int_{0}^{\Delta t_r} v(t') dt' + \int_{t - \Delta t_f}^{\Delta t_f} v(t') dt'}_{\Delta t_{\text{eff}}}  \right] \right)} \\
        &= \exp{\left( -i H ( t_{\text{expt}} + \Delta t_{\text{eff}})  \right)}
\end{align}
where in the last step, we set $t_{\text{expt}}=t-\Delta t_f - \Delta t_r$ which is the evolution time that is reported in our experiments. It should be noted that this can change from one experimental setup to another. The value of $\Delta t_{\text{eff}}$ can be interpreted as an effective total edge duration that takes the shapes of the rising and falling edges into account. This first-order pulse-shaping model introduces another model parameter, namely $\Delta t_{\text{eff}}$, that we need to estimate. We can do this directly in our MLE:
\begin{align}
    [\hat{\boldsymbol{\theta}}, \Delta \hat{t}_{\text{eff}}] = \arg \min_{\boldsymbol{\theta}, \Delta t_{\text{eff}}} - \frac{1}{N} \sum_{k=1}^N \log \left[ \sum_{y \in \{0,1\}} \left( p_{\mathrm{c}|\mathrm{y}}(c^{(k)}|y) \sum_{z \in \{0,1\}} \left|\braketExp{yz}{M^{(k)} e^{-iH(\boldsymbol{\theta})(t^{(k)} + \Delta t_{\text{eff}}) }U^{(k)}}{00}\right|^2 \right) \right] 
\end{align}
While this can be certainly done while learning the Hamiltonian, one can also determine the dependence of $\Delta t_{\text{eff}}$ on the different Hamiltonian parameters using prior calibration data. We determined that $\Delta t_{\text{eff}}$ depends only on the parameters of $\omega_{0,1}$ as $\Delta \hat{t}_{\text{eff}}(\bm{\theta}) = a/(\omega + b \omega^2)$. In Figure~\ref{fig:data_driven_model_imperfect_pulses_ibmq_boeb}, we plot the dependence of $\Delta t$ (short for $\Delta t_{\text{eff}}$) on $\omega$ for the IBM Quantum device D \textit{ibmq\_boeblingen}. The results for the other IBM Quantum devices A, B and C, are shown in Figure~\ref{fig:data_driven_model_imperfect_pulses} (Appendix~\ref{app_sec:details_results_HL}). For the IBM Quantum device D (\textit{ibmq\_boeblingen}), the values are $a=6.2774 \pm 0.01502$ and $b = 1.5086 \times 10^{-9} \pm 0.6104 \times 10^{-9}$s. How we arrived at this dependence is discussed later in Section~\ref{sec:Results}. Using this data-driven model allows us to use reduce the number of parameters in the estimation and hence reduce the associated computational cost of the optimization. In the following sections, we will denote this model by $\Delta \hat{t}_{\text{eff}}(\bm{\theta})$.

\subsubsection{Decoherence} \label{sec:noise_sources_decoherence}
Recalling our discussion in Sec.\ref{sec:hlproblem_noise}, we model decoherence as a depolarization channel acting on the quantum state $\rho(t)=\exp(-iHt) \rho(0)$ produced as a result of Hamiltonian evolution for a time duration $t$. The resulting state (from Eq.~\ref{eq:depolarization_error}) was given by 
\begin{equation}
    \mathcal{E}(\rho(t)) = (1-p_d(t))\rho(t) + p_d(t) \frac{I}{2^n}
    \label{eq:depolarization_error_repeated}
\end{equation}
One approach to obtain a description of $p_d(t)$ is to assume the functional form $1-\exp(- (t-t_0)/\mu)$, resulting from a Poisson process with rate $\mu$ as described in Sec.~\ref{sec:hlproblem_noise} and then estimate $\mu$ from the training examples after incorporating this into the MLE. Here, we describe a model for $p_d(t)$ using the measured device properties of $T_1$ and $T_2$ times of each qubit.

We consider an independent noise model on each of the two qubits used for implementing a cross-resonance gate. Let us denote the amplitude damping and phase damping of $k$th qubit as $\mathcal{E}_{a,k}$ and $\mathcal{E}_{p,k}$ respectively. They have the following Kraus operators
\begin{align}
    \mathcal{E}_{a,k} : & \left\{ \begin{bmatrix} 1 & 0 \\ 0 & \sqrt{1 - \gamma_{a,k}} \end{bmatrix},  \begin{bmatrix} 0 & \sqrt{\gamma_{a,k}} \\ 0 & 0 \end{bmatrix} \right\} \\
    \mathcal{E}_{p,k} : & \left\{ \begin{bmatrix} 1 & 0 \\ 0 & \sqrt{1 - \gamma_{p,k}} \end{bmatrix},  \begin{bmatrix} 0 & 0 \\ 0 & \sqrt{\gamma_{p,k}} \end{bmatrix} \right\}
\end{align}
where
\begin{align}
    \gamma_{a,k} &= 1 - \exp \left(- \Gamma_{a,k} t \right) \\
    \gamma_{p,k} &= 1 - \exp \left(- \Gamma_{p,k} t \right)
\end{align}
with 
\begin{equation}
    \Gamma_{a,k} := \frac{1}{2 \mathrm{T}_{1,k}}, \, \Gamma_{p,k} := \frac{1}{\mathrm{T}_{\phi,k}}
\end{equation}
where $\mathrm{T}_{1,k}$ is the $\mathrm{T}_1$ time of the $k$th qubit and $\mathrm{T}_{\phi,k}$ is the pure dephasing rate of the $k$th qubit related to the $\mathrm{T}_2$ time of the $k$th qubit as 
\begin{equation}
    \frac{1}{\mathrm{T}_{\phi,k}} = \frac{1}{\mathrm{T}_{2,k}} - \frac{1}{2 \mathrm{T}_{1,k}}
\end{equation}
The overall noise operator acting on the two qubits is then given by \cite{sundaresan2020reducing}
\begin{equation}
    \mathcal{E} = \otimes_{k=1}^2 (\mathcal{E}_{a,k} \circ \mathcal{E}_{p,k})
\end{equation}
where $\circ$ indicates taking a composition of the two noise operators. The probability $p_d(t)$ can be based on the unitarity \cite{wallman2015estimating} of the noise operator $\mathcal{E}$ which is a completely positive linear map quantifying the coherence of the noise operator. The probability $p_d(t)$ is then given by
\begin{align}
    1-p_d(t) = 1 + \frac{1}{15} \big( & \gamma_{a,1}^2 (3 \gamma_{a,2}(\gamma_{p,2} - 2) - 4\gamma_{p,2} + 7) \\ \nonumber
    &+ 4 \gamma_{a,1} (\gamma_{p,1} - 2)(\gamma_{a,2}^2 + \gamma_{a,2}(\gamma_{p,2} - 2) - \gamma_{p,2} + 2) \\ \nonumber
    &+ \gamma_{a,2}^2 (7-4\gamma_{p,1}) - 4 \gamma_{a,2} (\gamma_{p,1} - 2) (\gamma_{p,2} - 2) + 4 \gamma_{p,1} \gamma_{p,2} - 8 \gamma_{p,1} - 8 \gamma_{p,2} \big)
\end{align}  

This can be further generalized to $n$-qubit system (see Appendix G of \cite{sundaresan2020reducing} for details). Let the measurement of Hamiltonian evolution for time $t$ followed by this noisy depolarization channel be $\tilde{y}$ and the corresponding Rabi oscillation $\tilde{p}_{\mathrm{rabi}}$. Note that the Rabi oscillations in this case are related to the noiseless case (from Eq.~\ref{eq:rabi_oscillations_cr_gate}) as
\begin{equation}
    \tilde{p}_{\mathrm{rabi}} = (1-p_d(t)) p_{\mathrm{rabi}}
\end{equation}

In Table~\ref{tab:decoherence_rabi_fits_summary}, we give the RMSE betweeen Rabi oscillations obtained using different decoherence models assuming we know the true Hamiltonian parameters and Rabi oscillations inferred from data. For the \textit{single parameter model}, we consider a prefactor of $(1-p_d(t)) = \exp(-(t-t_0)/\mu$ with the single parameter $\mu$ on the Rabi oscillations obtained using no decoherence model. This parameter $\mu$ is then estimated from the data and is found to be $(7.75 \pm 0.91) \times 10^{-5}$s. In the \textit{two parameter model}, this is allowed to vary with the state preparation operator $U$ being applied. For $U=\sigma_I \otimes \sigma_I$, we have $\mu = (5.52 \pm 0.82) \times 10^{-5}$s and for $U=\sigma_X \otimes \sigma_I$, $\mu = (2.51 \pm 0.59) \times 10^{-5}$s. It is advantageous to use such models due to the low number of parameters present and when the $\mathrm{T}_1$ or $\mathrm{T}_2$ times of the different qubits are not available. However, if these times are available, one can use the two-qubit decoherence model or the two parameter model as we note from the values of RMSE in Table~\ref{tab:decoherence_rabi_fits_summary}.

\begin{table}[ht!]
\begin{tabular}{|l|c|c|}
\hline
\textbf{Decoherence Model}     & \textbf{Root Mean Squared Error} & \textbf{KL Divergence} \\ \hline
No decoherence                 & $0.109321$ & $0.012370$ \\ \hline
Single-qubit decoherence model & $0.095903$ & $0.007408$ \\ \hline
Single Parameter Model         & $0.091644$ & $0.006515$ \\ \hline
Two-qubit decoherence model    & $0.090953$ & $0.006392$ \\ \hline
Two Parameter Model         & $0.084289$ & $0.005376$ \\ \hline
\end{tabular}
\caption{\label{tab:decoherence_rabi_fits_summary} Comparison of different decoherence models in fitting Rabi oscillations inferred from experimental data collected from IBM Quantum Device D. Kullback-Leibler divergence (KL divergence) is computed as $D_{KL}(p_\mathrm{data} || p_\mathrm{model})$ where $p_\mathrm{data}$ is the probability inferred from data and $p_\mathrm{model}$ is that predicted from the model.}
\end{table}

\subsection{Procedures for Experiments evaluating HAL Algorithm}
Having described the cross-resonance Hamiltonian and the different noise sources or non-idealities on the IBM Quantum devices that we consider for our application, we are now in a position to describe the HAL algorithm for this particular application. We firstly discuss the details of the HAL-FI algorithm implemented for learning the CR Hamiltonian on the IBM Quantum devices. As the implementation of HAL-FIR for this application is very similar, it is omitted. We then describe our estimation procedure for solving the MLE problem for estimating the Hamiltonian parameters from the training data generated during learning.

\subsubsection{Implementation of HAL-FI Algorithm for Learning CR Hamiltonians}\label{sec:HAL-FI_CR_Gate}
Let us review the details of the implementation of the HAL-FI algorithm in the context of our experiments on learning the CR Hamiltonian. We consider the following inputs. The initial query space which may be changed during the course of the HAl-FI algorithm during training is $\querySpace^{(0)}=\measSpace\times \prepSpace \times \timeSpace^{(0)}$ with $\measSpace$ and $\prepSpace$ as described in Section~\ref{sec:cr_query_space}. Here, we explicitly denote the superscript on $\timeSpace$ which is initially set to $\timeSpace^{(0)}$ but may change during training if an adaptive query space strategy is employed. We set $\timeSpace^{(0)}$ to be the $81$ equispaced times in the interval $[10^{-7},6 \times 10^{-7}]s$. We consider the initial number of queries as $N_{\text{tot}}^{(0)}=2430$ (or five times the number of different queries in the initial query space $\querySpace^{(0)}$), a constant batch size $N_b=486$ (or the number of queries in the initial query space $\querySpace^{(0)}$), and the initial query distribution $q^{(0)}$ as the uniform random distribution over $\querySpace^{(0)}$.

The initial set of training examples $(X^{(0)},Y^{(0)})$ are obtained by sampling $X^{(0)}$ from $\querySpace^{(0)}$ with respect to $q^{(0)}$ in Line 1 of Algorithm~\ref{algo:AL_hamiltonians} and collecting the corresponding set of measurement outcomes $Y^{(0)}$ through queries to the CR Hamiltonian on the IBM Quantum device in Line 2. The set of training examples $(X^{(i)},Y^{(i)})$ is progressively increased (Line 10) during learning by adding $N_b$ queries $X_q^{(i)}$ sampled from query distribution $q^{(i)}$ chosen by HAL-FI in Line 6 and collecting the corresponding measurement outcomes $Y_q^{(i)}$ in Line 9. The learning is continued until our query budget is expended or the desired learning error is achieved.

To determine the initial parameter estimate $\hat{\bm{\theta}}^{(0)}$ from the initial set of training examples (Line 4) and subsequent $\hat{\bm{\theta}}^{(i)}$ from $(X^{(i)},Y^{(i)})$ (Line 11), we solve the MLE (Eq.~\ref{eq:general_MLE}) for the CR gate incorporating the Hamiltonian model description and presence of different noise sources. The parameter estimate $\hat{\bm{\theta}}$ is then given by
\begin{align}
    \hat{\boldsymbol{\theta}} = \arg \min_{\boldsymbol{\theta}} - \frac{1}{N} \sum_{k=1}^N \log \Bigg[ \sum_{y \in \{0,1\}} \Bigg( p_{\tilde{\mathrm{y}}|\mathrm{y}}(\tilde{y}^{(k)}|y) \sum_{z \in \{0,1\}} & \Bigg[ \left(1-p_d\left(t^{(k)}\right)\right) \left|\braketExp{yz}{M^{(k)} e^{-iH(\boldsymbol{\theta})(t^{(k)} + \Delta \hat{t}_{\text{eff}}(\bm{\theta}) ) }U^{(k)}}{00}\right|^2 + \nonumber \\ & \frac{1}{4}p_d\left(t^{(k)}\right) \Bigg] \Bigg) \Bigg]
    \label{eq:mle_hl_noise}
\end{align}
or
\begin{align}
    \hat{\boldsymbol{\theta}} = \arg \min_{\boldsymbol{\theta}} - \frac{1}{N} \sum_{k=1}^N \log \Bigg[\sum_{y \in \{0,1\}} \Bigg( p_{\mathrm{c}|\mathrm{y}}(c^{(k)}|y) \sum_{z \in \{0,1\}} & \Bigg[ \left(1-p_d\left(t^{(k)}\right)\right) \left|\braketExp{yz}{M^{(k)} e^{-iH(\boldsymbol{\theta})(t^{(k)} + \Delta \hat{t}_{\text{eff}}(\bm{\theta}) ) }U^{(k)}}{00}\right|^2 + \nonumber \\ & \frac{1}{4}p_d\left(t^{(k)}\right) \Bigg] \Bigg) \Bigg]
    \label{eq:mle_hl_noise_GMM_readout}
\end{align}
depending on how the readout noise is modeled. The imperfect pulse-shaping model $\Delta \hat{t}_{\text{eff}}(\bm{\theta})$ was discussed in Section~\ref{sec:imperfect_pulse_shaping} and $p_d(t)$ is the probability of depolarization associated with the two-qubit decoherence model (Section~\ref{sec:noise_sources_decoherence}). The choice of the MLE (Eq.~\ref{eq:mle_hl_noise} or Eq.~\ref{eq:mle_hl_noise_GMM_readout}) for each IBM Quantum device will be specified in Section~\ref{sec:Results}. 

The parameter estimates $\hat{\bm{\theta}}$ obtained after solving the MLE problem are used by HAL-FI to construct the Fisher information matrices based on the model and obtain the query distribution $q^{(i)}$ by solving the SDP program of Eq.~\ref{eq:SDP_query_optimization} in Line 6. The expressions for the Fisher information matrices for different queries (Section~\ref{sec:cr_query_space}) considering the CR Hamiltonian and noise models are given in Appendix~\ref{app_sec:details_cr_hamiltonians}. The query space $\querySpace^{(i)}$ used in Lines 6 and 7 depend on the querying strategy and hence the learning scenario we consider.

As described in Section~\ref{sec:HAL-FI_algo_description}, we can define four different learning scenarios based on the presence of an active learner and how the query space is adaptively changed during learning. In passive learning, an active learner is not present and we set the query distribution to the uniformly random distribution over $\querySpace$. In the case of active learning with fixed query space, the query space remains fixed during training i.e., $\querySpace^{(i)}=\querySpace \forall i$, and the query distribution is determined by solving Eq.~\ref{eq:SDP_query_optimization} using the current estimate of the parameters $\hat{\boldsymbol{\theta}}$ over this fixed query space. When considering active learning with an adaptively growing query space, we consider two different situations on the basis of how the query space is changed between batches during learning. We consider two cases: (i) $\querySpace^{(i)}$ grows linearly by linearly increasing the $\timeSpace^{(i)}$ between batches and (ii) $\querySpace^{(i)}$ grows exponentially by doubling the allowed set of system interaction time $\timeSpace^{(i)}$. The query distribution is then determined by solving the corresponding SDP problem of Eq.~\ref{eq:SDP_query_optimization} over the query space $\querySpace^{(i)}$ corresponding to the $i$th batch. These different learning scenarios for HAL-FI are summarized in Table~\ref{tab:learning_scenarios_summary}.

\begin{table}[ht!]
\centering
\begin{tabular}{|c|c|c|}
\hline
\textbf{Learning Scenario} & \textbf{Query Space} & \textbf{Query Distribution}  \\ \hline
Passive Learning & Fixed & uniformly random \\ \hline
Active Learning with Fixed Query Space  & Fixed & $q$ through Eq.~\ref{eq:query_optimization_variance_of_params} \\ \hline
Active Learning with Adaptive Query Space I  & Linearly growing $\timeSpace$ & $q$ through Eq.~\ref{eq:query_optimization_variance_of_params} \\ \hline
Active Learning with Adaptive Query Space II  & Exponentially growing $\timeSpace$ & $q$ through Eq.~\ref{eq:query_optimization_variance_of_params} \\ \hline
\end{tabular}
\caption{\label{tab:learning_scenarios_summary} Summary of different learning scenarios}
\end{table}

\subsubsection{Estimation Procedure}\label{sec:estimation_procedure}
The MLE problem (Eq.~\ref{eq:mle_hl_noise} and Eq.~\ref{eq:mle_hl_noise_GMM_readout}) for the two different parameterizations of $\mathbf{J}$ and $\bm{\Lambda}$ is nonlinear and non-convex. An example of the energy landscape of the log-likelihood loss function for the IBM Quantum device \textit{ibmq\_boeblingen} is shown in Figure~\ref{fig:results_energy_landscapes} of Appendix~\ref{app_sec:est_procedure_cr_ham}. The presence of multiple local minima make the MLE problem in general challenging to solve. To ensure that we converge to the global minimum and do not get stuck in a local minimum during estimation, we divide the estimation into multiple stages. In the first stage, we obtain an initial crude estimate which is refined over subsequent stages. 

The estimation procedure summarized in Algorithm~\ref{algo:MLE_Estimation_Procedure} is divided into two main stages where in the first stage we use Rabi oscillations (Eq.~\ref{eq:rabi_oscillations_cr_gate}) inferred from the experimental data to come up with an initial estimate for the MLE solve in the second stage. It proceeds as follows. Let us consider the parameterization of $\bm{\Lambda}$. This is more useful to work with as the frequencies of oscillation in $e^{-iHt}$ are determined by $\omega_{0}$ and $\omega_{1}$ unlike $\mathbf{J}$ where it is determined by all the parameters. This provides a way to create initial estimates of $\omega_{0,1}$ independent of the other parameters.

We compute the Rabi oscillations $\hat{p}_\text{rabi}$ for each query $x$ made from the experimental data by solving the optimization problem of Eq.~\ref{eq:rabi_oscillations_mle}. This serves as an input to the initial estimation procedure. Initial estimates of the parameters $\omega_{0,1}$ are then obtained by applying a Discrete (Fast) Fourier Transform to the Rabi oscillations. These initial estimates are then refined by fitting nonlinear regression equations of the form $A\cos (\omega t) + B \sin (\omega t) + C$ to the Rabi oscillations, where the fit minimizes the total $L_2$ error, the coefficients $A$, $B$, and $C$ for each Rabi oscillation are estimated using linear least-squares regression, and a bracketed gradient-based search is performed to refine the estimates of $\omega_{0,1}$. The corresponding optimization problem can be framed as minimizing the following residual error
\begin{align}
    \min E(\mathbf{A},\omega_{0,1}) = \sum_{t \in \mathcal{T}} \left( p_{\mathrm{rabi}}(t) - \mathbf{A}\bm{\Omega}(\omega t) \right)^2
    \label{eq:nonlinear_regression_residual_error}
\end{align}
where the coefficients $\mathbf{A} = (A,B,C)$ are known functions of $\delta_{0,1},\phi_{0,1}$ through the analytical forms of the Rabi oscillations for the query space considered (see Eq.~\ref{eq:rabi_oscillations_cr_gate}) and $\Omega(\omega t)$ is a vector of cosines and sines (fully described in Appendix~\ref{app_sec:est_procedure_cr_ham}). Thus, we can then obtain initial estimates for $\delta_{0,1},\phi_{0,1}$ from the $A$, $B$, and $C$ coefficients of the nonlinear regression equations for each of the Rabi oscillations.  Finally, we obtain an initial estimate $\hat{\bm{\Lambda}}_0$ for $\bm{\Lambda}$ by fixing the values of $\omega_{0,1}$ and carrying out a gradient descent procedure using the same cost function. This is used as an initial condition to the MLE solve (Eq.~\ref{eq:mle_hl_noise}). We have denoted the negative log-likelihood loss function which appears in the optimization problem of Eq.~\ref{eq:mle_hl_noise} as $L$. We solve the MLE problem using the optimizer of stochastic gradient method of ADAM \cite{kingma2014adam} which encourages getting out of any local minima. The parameter estimate produced by ADAM is then refined using the second-order quasi-Newton method of L-BFGS-B \cite{lbfgsb:1997}. The computational complexity of our estimation procedure is dominated by ADAM. This motivates us to directly use L-BFGS-B after initial estimation for latter batches during learning. The full description, computational details and extensions of the estimation procedure is given in Appendix~\ref{app_sec:est_procedure_cr_ham}.

\begin{algorithm}[H]
	\caption{Estimation Procedure for Hamiltonian Learning}
	\small
	\textbf{Input}: Training examples of size $m$: $D=\{(X,Y)\}=\{(x_i,y_i)\}_{i \in [m]}$ \\
	\textbf{Output}: $\hat{\boldsymbol{\Lambda}}$
	\begin{algorithmic}[1]
		\State Obtain Rabi oscillations $p_{\mathrm{rabi}}$ from $D$ by solving Eq.~\ref{eq:rabi_oscillations_mle}
		\State $\omega_{0,1} \leftarrow \text{FFT}(p_{\mathrm{rabi}})$
		\State $(\delta_{0,1},\phi_{0,1}) \leftarrow p_{\mathrm{rabi}}(\Omega(\omega t))^{-1}$ \Comment{Regression on Rabi oscillations given $\omega_{0,1}$}
		\State Refine estimate of $\hat{\bm{\Lambda}}_0$ through gradient descent on $E(\mathbf{A},\omega_{0,1})$
		\State $\hat{\boldsymbol{\Lambda}} \leftarrow \arg \min \limits_{\boldsymbol{\Lambda}} L(\boldsymbol{\Lambda};X,Y)$ \Comment{using $\hat{\bm{\Lambda}}_0$ as a guess}
		\State \Return $\hat{\bm{\Lambda}}$
	\end{algorithmic}
	\label{algo:MLE_Estimation_Procedure}
\end{algorithm}

\section{Results} \label{sec:Results}
We now present the results of the experiments described in Section~\ref{sec:CR_Gate_Model_and_Setup} and show that they support the claims made in Section~\ref{sec:hlproblem}. In this section, we compare the active learner HAL introduced in Section~\ref{sec:HAL} against the passive learner (Section~\ref{sec:active_learner}) in learning the CR Hamiltonian through experiments on different IBM Quantum devices and finally analyze the results of these experiments to affirmatively answer the learning problems proposed in Section~\ref{sec:hlproblem}. 

In Section~\ref{sec:results_data}, we first describe the datasets used for Hamiltonian learning and the experimental protocol for evaluating the performance of the learners. In Section~\ref{sec:results_performance}, we show results of the learners on these different datasets. In Section~\ref{sec:results_analysis}, we describe how HAL-FI can be used to achieve Heisenberg (or super-Heisenberg) limited scaling and evaluate the query advantage of HAL-FI over the baseline considering different learning scenarios.

\subsection{Data and Experiment Protocol} \label{sec:results_data}
In this section, we describe the different kinds of datasets that were used in assessing the performance of the HAL-FI algorithm and how they were collected. We then summarize the parameters of the cross-resonance Hamiltonian and the noise sources discussed in Section~\ref{sec:noise_sources} for the different IBM Quantum devices (Section~\ref{sec:quantum_devices}).

\subsubsection{Datasets from IBM Quantum Devices} \label{sec:results_datasets_expt}
The different datasets that we use for Hamiltonian learning are a combination of experimental data collected from the IBM Quantum devices described in Section~\ref{sec:quantum_devices} and that collected from a \textit{simulator} which we will describe in Section~\ref{sec:results_datasets_sim}. 

Experimental data is collected from the different IBM Quantum devices according to the query space described in Section~\ref{sec:cr_query_space}. The set of evolution times $\timeSpace$ is set to $81$ equispaced times in the interval $\timeSpace=[10^{-7},6 \times 10^{-7}]s$. For IBM Quantum devices A, B, and C, there are $200$ measurement outcomes (or shots) for each query $\mathbf{x} \in \querySpace$. For IBM Quantum device D \textit{ibmq\_boeblingen}, there are $512$ measurement outcomes for each query. Recall from our discussion of the Hamiltonian learning framework in \ref{sec:hlproblem} and HAL algorithm in \ref{sec:comment_HS_scaling_HAL}, the outputs of our queries are not expectation values but rather single shot readouts of the target qubit. 

The experimental data is then collected and made available as an offline dataset that the active learner can query. Unlike deploying an active learner in real-time where a particular query can be made to the system unlimited number of times, using experimental datasets imposes the additional constraint of the number of times a query can be made by the active learner due to the limitation on the number of measurement outcomes available for each query. In Figure~Appendix~\ref{app_sec:query_opt}, we discuss how we handle this constraint during query optimization.

\subsubsection{Parameters of the CR Hamiltonian and Noise Sources for IBM Quantum Devices} \label{sec:results_params_ham_noise_ibmq}
Considering the entire collected experimental datasets for each IBM Quantum device (Section~\ref{sec:quantum_devices}) as training data, we compute the Hamiltonian parameters $\mathbf{J}$/$\bm{\Lambda}$, and that of the different noise sources using the estimation procedure specified in Section~\ref{sec:estimation_procedure} for the MLE. We solve the MLE of Eq.~\ref{eq:mle_hl_noise} for IBM Quantum device D which has very low readout noise and use MLE of Eq.~\ref{eq:mle_hl_noise_GMM_readout} for the other IBM Quantum devices. A summary of the estimated parameters of the CR Hamiltonian and noise sources on IBM Quantum device D $ibmq\_boeblingen$ is shown in Figure~\ref{tab:ibmq_boeb_ham_params_summary} under different drive configurations. We summarize the estimated parameters for the other devices in Appendix~\ref{app_sec:details_results_HL}. 
\begin{table}[ht!]
\centering
\small
\begin{tabular}{|l|l|ll|ll|}
\hline
\multicolumn{1}{|c|}{\multirow{2}{*}{\begin{tabular}[c]{@{}c@{}}\textbf{Drive} \\ \textbf{Config.}\end{tabular}}} & 
\multicolumn{1}{c|}{\multirow{2}{*}{\begin{tabular}[c]{@{}c@{}}\textbf{CR Amp.} \\ \textbf{(arb. units)}\end{tabular}}} & 
\multicolumn{2}{c|}{\textbf{Hamiltonian Parameters [$\mathbf{\times10^6s^{-1}}$]}} & 
\multicolumn{2}{c|}{\textbf{Noise Sources}} \\ \cline{3-6} 
\multicolumn{1}{|c|}{} & \multicolumn{1}{c|}{} & \multicolumn{1}{l|}{$\mathbf{J}=(J_{IX},J_{IY},J_{IZ},J_{ZX},J_{ZY},J_{ZZ})$} & $(\omega_0,\omega_1)$ &
\multicolumn{1}{l|}{\textbf{Readout $(r_0, r_1)$}} & 
\textbf{Time Offset $(\Delta t_{\text{eff},0},\Delta t_{\text{eff},1})$ [ns]} \\ \hline
1 & 0.24 & \multicolumn{1}{l|}{(-3.88, -1.08,  -0.24, 5.44,  1.07,   0.21)} & (1.57, 9.58) & \multicolumn{1}{l|}{$(0.012, 0.025)$} & $(1965, 289)$ \\ \hline
2 & 0.30 & \multicolumn{1}{l|}{(-4.57, -1.47, -0.29, 6.50, 1.39, 0.41)} & (1.94, 11.45) & \multicolumn{1}{l|}{$(0.0078, 0.033)$} &  $(1581, 226)$\\ \hline
3 & 0.36 & \multicolumn{1}{l|}{(-5.12, -1.65, -0.23, 7.52, 1.66, 0.33)} & (2.40, 13.07) & \multicolumn{1}{l|}{$(0.0078, 0.035)$} & $(1267, 203)$ \\ \hline
4 & 0.42 & \multicolumn{1}{l|}{(-5.42, -1.95, 0.37, 8.38,  1.90, 0.07)} & (2.97, 14.33) & \multicolumn{1}{l|}{$(0.0078, 0.039)$} & $(1016, 182)$ \\ \hline
5 & 0.48 & \multicolumn{1}{l|}{(-5.72, -2.13, 0.03, 9.20,  2.15, 0.11)} & (3.48, 15.51) & \multicolumn{1}{l|}{$(0.0078, 0.023)$} &  $(862, 166)$\\ \hline
\end{tabular}
\caption{\label{tab:ibmq_boeb_ham_params_summary} Summary of estimated CR Hamiltonian parameters for the IBM Quantum device D $ibmq\_boeblingen$ with different drive configurations (Config.) corresponding to amplitude (Amp.) of CR pulse. We give the Hamiltonian parameters in the parameterization $\mathbf{J}$ and the physically relevant frequency components in $\bm{\Lambda}$. The readout noise is defined by the parameters of $r_0$ and $r_1$ which are the conditional probabilities of bit flip given the measurement outcomes are $y=0$ and $y=1$ respectively (see Section~\ref{sec:readout_noise}).} \end{table}
\subsubsection{Datasets from Simulation} \label{sec:results_datasets_sim}
The experimental datasets in Section~\ref{sec:results_datasets_expt} contain noise sources other than those modeled even if negligible and may not span a long enough time range over $\timeSpace$ for testing different learning scenarios. In order to understand the behavior of the HAL-FI and HAL-FIR algorithms considering all the noise sources are known, we set up a simulator. The advantage of using a simulator over experimental data is that it allows us to assess the limits of the performance of the HAL-FI algorithm in the presence or absence of different noise sources such as decoherence. Studies carried out on the simulator also allow us to test the robustness of the active learner in the presence or absence of different noise sources.

The simulator is an \textit{oracle} which imitates the different quantum devices but where all the different noise sources are known and perfectly modeled, and which we can query. The Hamiltonian of the simulator is set to that learned from the full set of training examples contained in an experimental dataset collected from a particular quantum device. Thus, we can have simulators for each of the IBM Quantum devices A, B, C, and D, under different drive configurations. All the modeled noise sources of readout noise, imperfect pulse-shaping, and decoherence as described in Section~\ref{sec:noise_sources} are included. We apply HAL-FI in real-time on the simulator as there is no limitation on the number of times we select a particular query $\mathbf{x} \in \querySpace$.

\begin{figure}[ht!]
	\centering
	\xincludegraphics[scale=0.22, label=\textbf{(a)}]{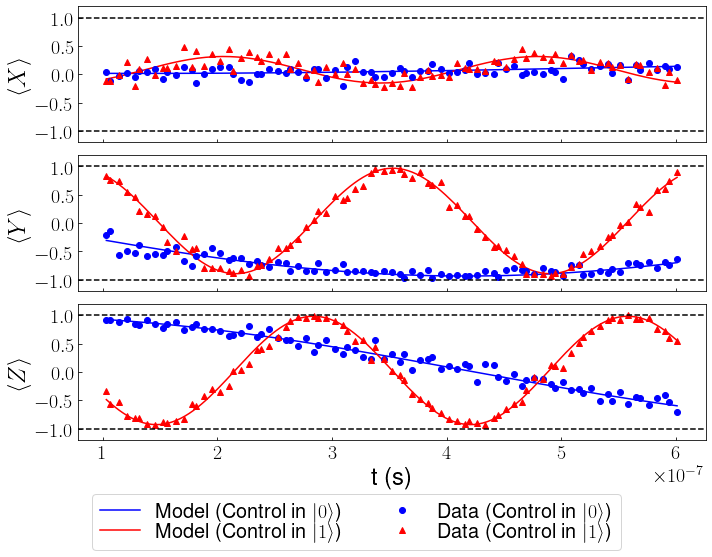}
	\quad
    \xincludegraphics[scale=0.22,label=\textbf{(b)}]{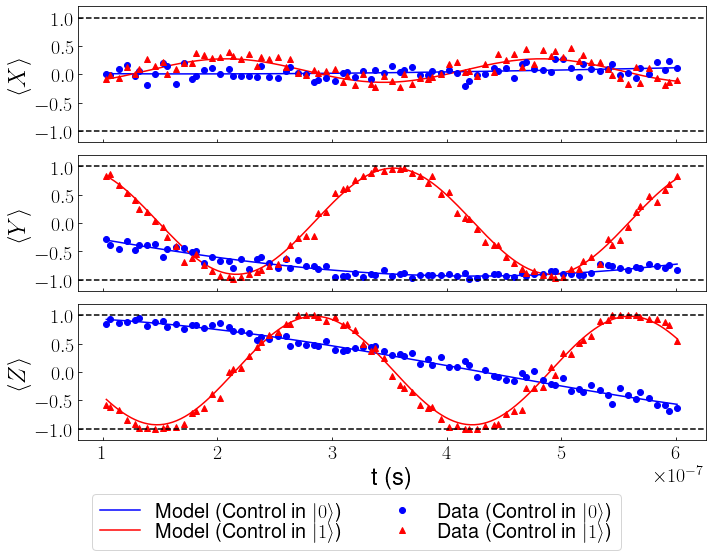}
    \quad
    \xincludegraphics[scale=0.22,label=\textbf{(c)}]{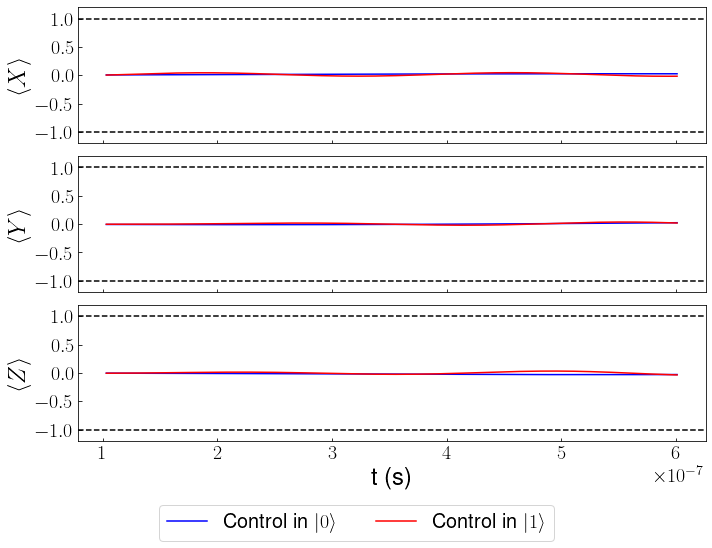}
    \caption{Comparison of Rabi oscillations computed from data (46800 queries) of IBM Quantum device D \textit{ibmq\_boeblingen} from the two different oracles of (a) experimental data, and (b) simulator. In (c), we plot the difference in the model Rabi oscillations predicted from the Hamiltonians learned in (a) and (b). In the subplots of (a)-(c), we plot Rabi oscillations (or difference) for different measurement operators $M$ (rows), preparation operators $U$ (colors), and evolution times $t$ (x-axis), corresponding to the query space described in Section~\ref{sec:cr_query_space}.}
	\label{fig:comparison_rabi_oracles}
\end{figure}

In Figure~\ref{fig:comparison_rabi_oracles}, we compare Rabi oscillations from the two different oracles based on IBM Quantum device D $ibmq\_boeblingen$ under drive configuration $2$. A set of $46800$ training examples are generated from both oracles assuming a uniform query distribution over the query space. This is used to learn Hamiltonian parameters in each case. The small difference in the predicted model Rabi oscillations from the Hamiltonian parameters learned on the two oracles, further indicates that the simulator faithfully represents the collected experimental datasets.
\subsubsection{Protocol for Comparing Performance of Hamiltonian Learning Methods}\label{sec:protocol_comparing_HL_methods}
Query complexity is used for comparing the performance of different learners on the simulator or oracle with access to an experimental dataset. In particular, our main goal is to extract the scaling of the query complexity with respect to the root mean square error (RMSE) of Hamiltonian parameters $\mathbf{J}$. We compute the empirical RMSE at each round of learning as follows:
\begin{equation}
    \text{RMSE} = \sum_{i=1}^m \Expectation \left[ \left( \frac{\hat{\theta}_i}{\xi_i} - \frac{\theta^\star_i}{\xi_i} \right)^2 \right]^{1/2} = \sum_{i=1}^m \sum \limits_{k=1}^{N_{\text{runs}}} \left[ \left( \frac{\hat{\theta}^{(k)}_i}{\xi_i} - \frac{\theta^\star_i}{\xi_i} \right)^2 \right]^{1/2}
\end{equation}
where we approximate the expectation by an average over parameter estimates from $N_{\text{runs}}$ runs and the true parameter values $\bm{\theta}^\star$ by the mean of these runs. The normalization factors $\xi_i$ are selected to be $10^6\,\mathrm{s}^{-1}$ for all $i$ as the components of $\mathbf{J}$ with highest magnitude are expected to the order of $10^{6 \pm 1}$ \cite{magesan2018effective} for these IBM Quantum devices under these drive configurations. We implement the following experimental protocol. For each of the quantum devices described in Section~\ref{sec:quantum_devices}, we compute the empirical RMSE for the learners from $200$ independent runs of the simulator and $500$ independent runs on the experimental dataset. These number of runs on each oracle were required to obtain accurate scalings of trends in RMSE with number of queries and ensure the uncertainty (or two standard deviations) of each scaling was at most $10\%$. In each run, we carry out the Hamiltonian learning algorithm for the different learning scenarios as detailed in Section~\ref{sec:HAL-FI_CR_Gate} and summarized in Table~\ref{tab:learning_scenarios_summary}. Additionally, we track the testing error of the learner with number of queries $N$. The so obtained trends are used to comment on the robustness of the estimation procedure used for MLE (Section~\ref{sec:estimation_procedure}) and the benefits of using the active learner HAL-FIR for making predictions of queries to the Hamiltonian (Problem~\ref{prob:HL_prediction_testing}) over a baseline. The testing error is computed empirically as well on a testing dataset collected from the simulator or experimental dataset using $\ptest$.

\subsection{Performance of Hamiltonian Learning Methods} \label{sec:results_performance}
In this section, we assess the performance of the different Hamiltonian learning methods introduced so far using the protocol described in Section~\ref{sec:protocol_comparing_HL_methods} in tackling the learning problems posed in Section~\ref{sec:hlproblem}. We report results of the different Hamiltonian learning scenarios (Table~\ref{tab:learning_scenarios_summary}): (i) baseline of passive learner with estimation based on FFT and regression, (ii) passive learner with an estimation procedure to solve the MLE problem, (iii) active learner in fixed query space, and (iv) active learner in an adaptively growing (linearly/exponentially) query space. We use the HAL-FI algorithm which was discussed in Section~\ref{sec:activelearning} in the latter active learning approaches.

We present the scaling behavior of each algorithm during learning under different regimes. For each scenario, we show trends of learning error (RMSE) with number of queries. These trends indicate the performance of each learning algorithm, culminating in evidence of query advantage. For brevity of presentation, we focus on the results obtained from IBM Quantum Device D (\textit{ibmq\_boeblingen}) under the drive configuration $2$.

\subsubsection{Scaling Behavior and Convergence Studies}
In this section, we assess the trends of learning error with number of queries for our baseline, passive learner, and active learner (HAL-FI). We consider HAL-FI in a fixed query space and a linearly growing query space. Results of the exponentially growing query space are postponed to Section~\ref{sec:results_analysis_HS} where we comment on the achievability of Heisenberg limited scaling. 

We focus on the Hamiltonian learning task of Problem~\ref{prob:HL_model_inference} and touch upon Problem~\ref{prob:HL_prediction_testing} to show that HAL is a general framework for tackling both problems when equipped with the appropriate query optimization. We show results on the simulator for two different cases of time range $\mathcal{T}$ of the query space $\mathcal{Q}$. For successful frequency detection using FFT \cite{cohen1989time,parhizkar2015sequences}, the time range must be sufficiently long to see a single cycle of the sinsusoid corresponding to the lowest frequency. We call this as the minimum-frequency criteria (MFC). We thus show results on the simulator where $\timeSpace$ is such that the MFC for frequency estimation (using FFT) is satisfied and when MFC is not necessarily satisfied. The Minimum-Frequency criteria is not to be confused with Nyquist criteria that is a comment on the sampling rate required. This is a buildup to our comparison on the experimental dataset where $\mathcal{T}$ is such that MFC is not necessarily satisfied. In each case, we plot the empirical learning error (RMSE) versus number of queries made on a log-log scale so that the slope $s$ of the plotted lines can be directly interpreted as the scaling of learning error with complexity $\epsilon \sim N^{s}$.

\begin{figure}[ht!]
    \centering
	\xincludegraphics[scale=0.31, label=\textbf{(a)}]{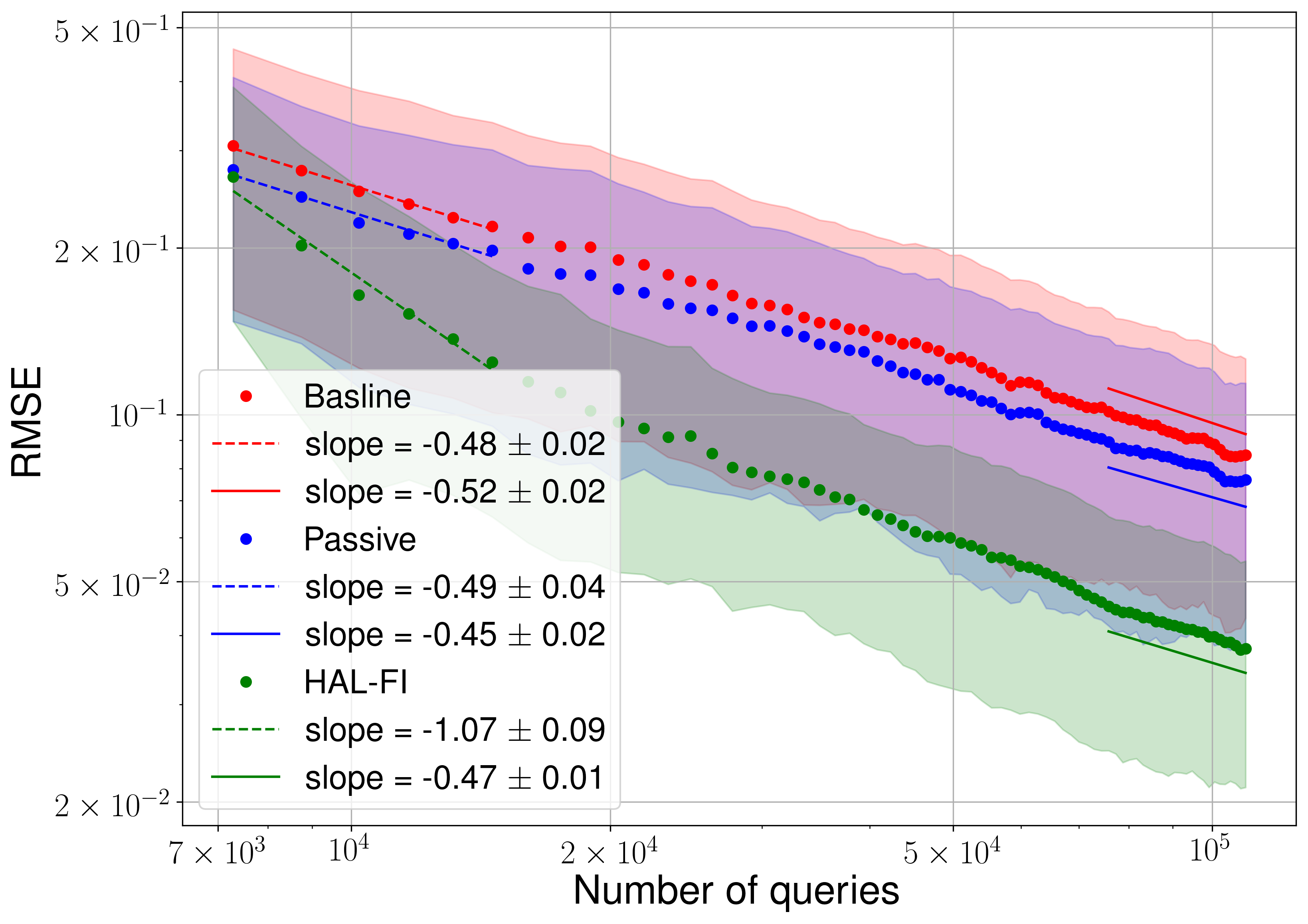}
    \quad
    \xincludegraphics[scale=0.31,label=\textbf{(b)}]{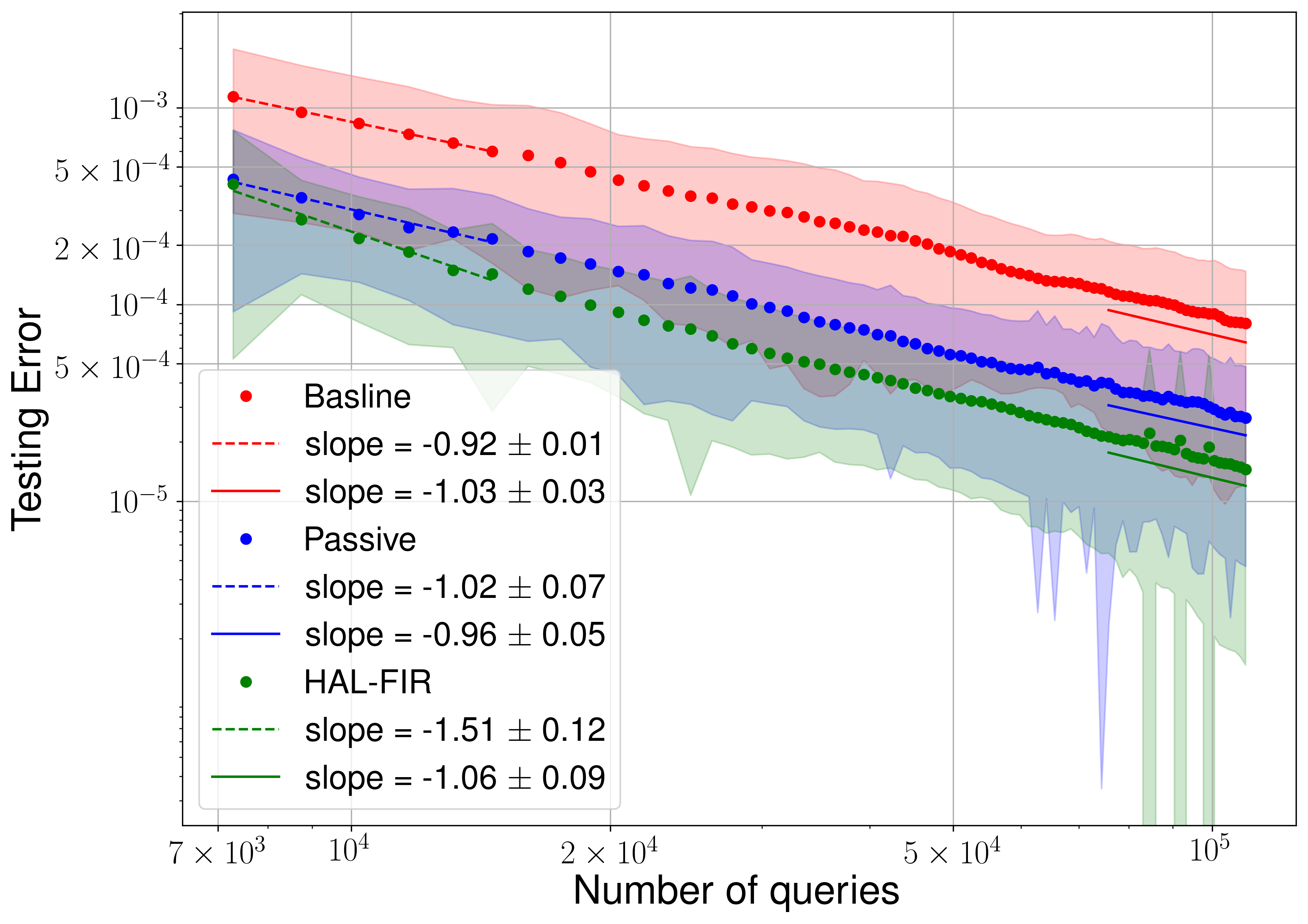}
    \caption{Scaling of learning error with number of queries for different learners on a simulator when MFC is satisfied. In (a), we compare the trends in RMSE of HAL-FI against those of the baseline and passive learner. In (b), we compare the trends in testing error of HAL-FIR against those of baseline and passive learner. Slopes indicate the scaling of learning error (RMSE/Testing Error) $\epsilon$ with number of queries $N$ in the finite sample and asymptotic sample regimes. Filled in areas indicate the respective errors on trends for each learner.}
    \label{fig:CR_Error_Scalings_Simulator_Nyquist}
\end{figure} 

\paragraph{Simulator and Minimum-Frequency criteria is satisfied} \label{sec:noisy_simulator_nyquist}
We consider the query space as defined in Section~\ref{sec:cr_query_space} with $\timeSpace$ set to be the $243$ equispaced times in the interval of $[10^{-7},18\times 10^{-7}]$s. A comparison of different learners for Hamiltonian learning considering this query space is shown in Figure~\ref{fig:CR_Error_Scalings_Simulator_Nyquist}(a). For both the baseline and the passive learner, we observe a scaling of $\epsilon \sim 1/\sqrt{N}$ or $N \sim \epsilon^{-2}$ in RMSE with number of queries. This is in agreement with the SQL. The approximately constant gap between the baseline and the passive learner corresponds to a constant query reduction when using an estimation procedure based on solving the MLE over a procedure based on FFT and regression. Thus, a query advantage can be obtained by changing estimation procedures for Hamiltonian learning.

We observe two different scalings for the HAL-FI algorithm, an initial scaling which is higher than SQL and similar to Heisenberg limited scaling, and a scaling of SQL in the asymptotic query regime. This shows that depending on the desired learning error, we can expect to see a higher rate of convergence. However, asymptotically the number of queries $N$ required by the active learner HAL-FI is a constant fraction of that required by the passive learner.

Additionally, we compare the baseline, passive learner, and active learner HAL-FIR for tackling Problem~\ref{prob:HL_prediction_testing} in Figure~\ref{fig:CR_Error_Scalings_Simulator_Nyquist}(b). The testing error is computed using Eq.~\ref{eq:query_optimization_testing_error} on a testing dataset of $10^5$ i.i.d. samples. The testing distribution $\ptest$ is considered to be known and set to be the uniform distribution over the query space $\querySpace$. It is assumed that HAL-FIR is given access to this testing distribution. We observe a scaling of $\epsilon \sim 1/N$ for the baseline and the passive learner which is expected as the log-likelihood loss function associated with $\hat{\bm{\theta}}$ is divergence-free in the asymptotic case \cite{sourati2017asymptotic}. As in the case of HAL-FI for Hamiltonian learning, we observe a higher initial scaling for HAL-FIR and a a scaling of $\epsilon \sim 1/N$ in the asymptotic regime, consistent with that observed for the passive learner. 

As discussed in Section~\ref{sec:hlproblem}, it is not typical for the testing distribution in Problem~\ref{prob:HL_prediction_testing} to be known and the result here can viewed as a validation of the Hamiltonian model learned and hence the learners on a set of queries sampled using the testing distribution. 

We note sudden peaks in uncertainty associated with the passive learner and HAL-FIR for higher values of samples. This might be indicative of traveling between multiple local minima in a larger convex hull when solving the MLE (Eq.~\ref{eq:mle_hl_noise}). The trends in testing error are not severely impacted by this in expectation indicating that these are rare events and our learners (with their estimation procedure) are robust.

Having made a case for the robustness of the learners and estimation procedures being used in this work on the simulator considering a query case $\mathcal{Q}$ which satisfies MFC, we are now in a position to compare the performance of the different learners when the $\querySpace$ is not guaranteed or known to satisfy such a criteria. In the results that follow, the observations made here are used as a basis for the behaviour to expect among the learners.

\paragraph{Simulator and Minimum-Frequency criteria is not necessarily satisfied} \label{sec:noisy_simulator}
In practice, the range of system evolution times $\timeSpace$ corresponding to the query space $\querySpace$ cannot be known apriori to satisfy MFC for frequency estimation using FFT on Rabi oscillation data. It is then crucial for learners equipped with estimation procedures to either: (i) succeed at Hamiltonian learning given this query space or (ii) alert the user that a longer $\timeSpace$ is required upon failure to learn a Hamiltonian model. Here, we ensure the former by modifying the standard FFT routine as discussed earlier in Section~\ref{app_sec:est_procedure_cr_ham}. Further details are also provided in Appendix~\ref{sec:estimation_procedure}. Note that this solve is also carried out as the first step in our estimation procedure for obtaining initial conditions to the MLE solve.

\begin{figure}[ht!]
    \centering
    \xincludegraphics[scale=0.31,label=\textbf{(a)}]{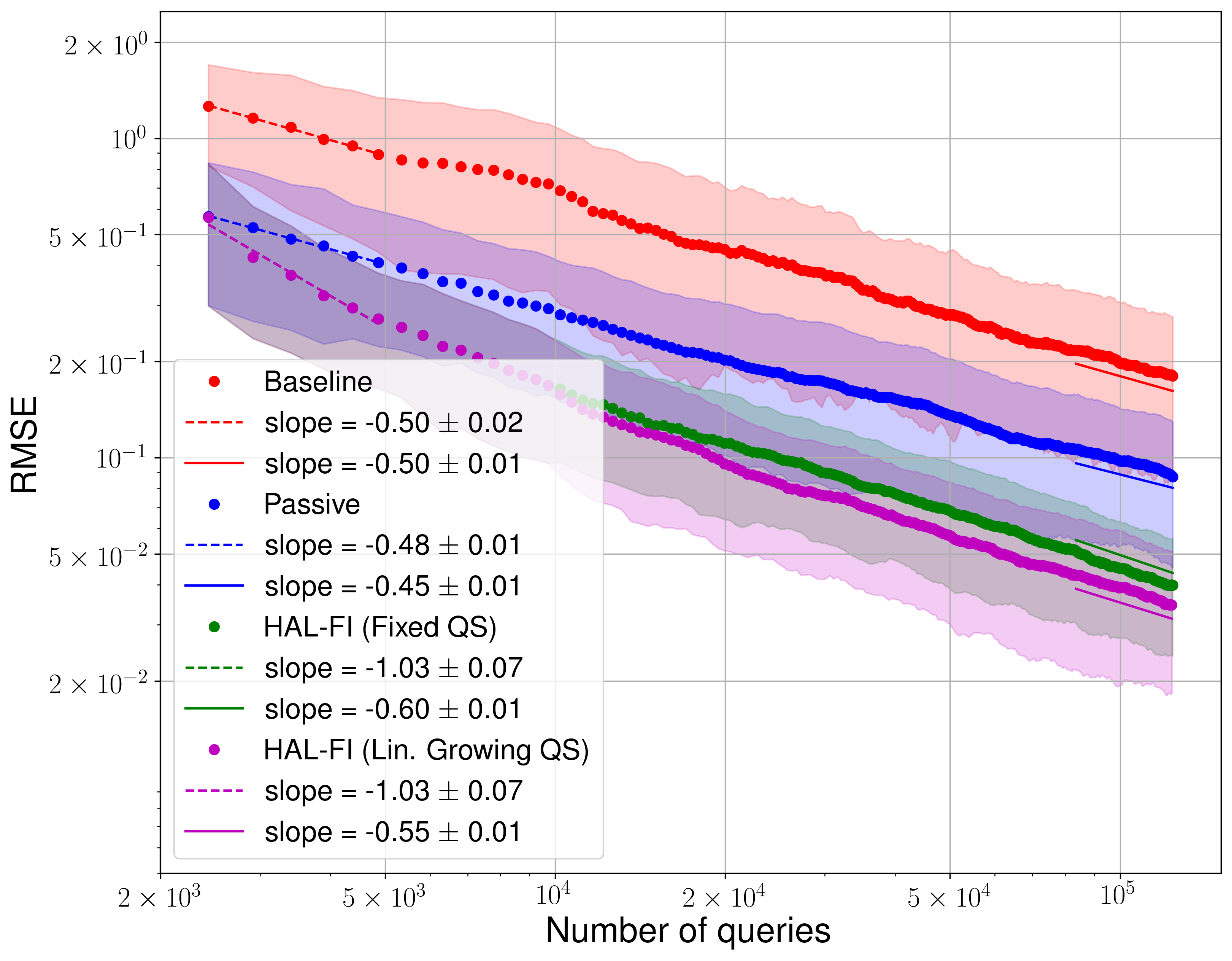}
    \quad
    \xincludegraphics[scale=0.31,label=\textbf{(b)}]{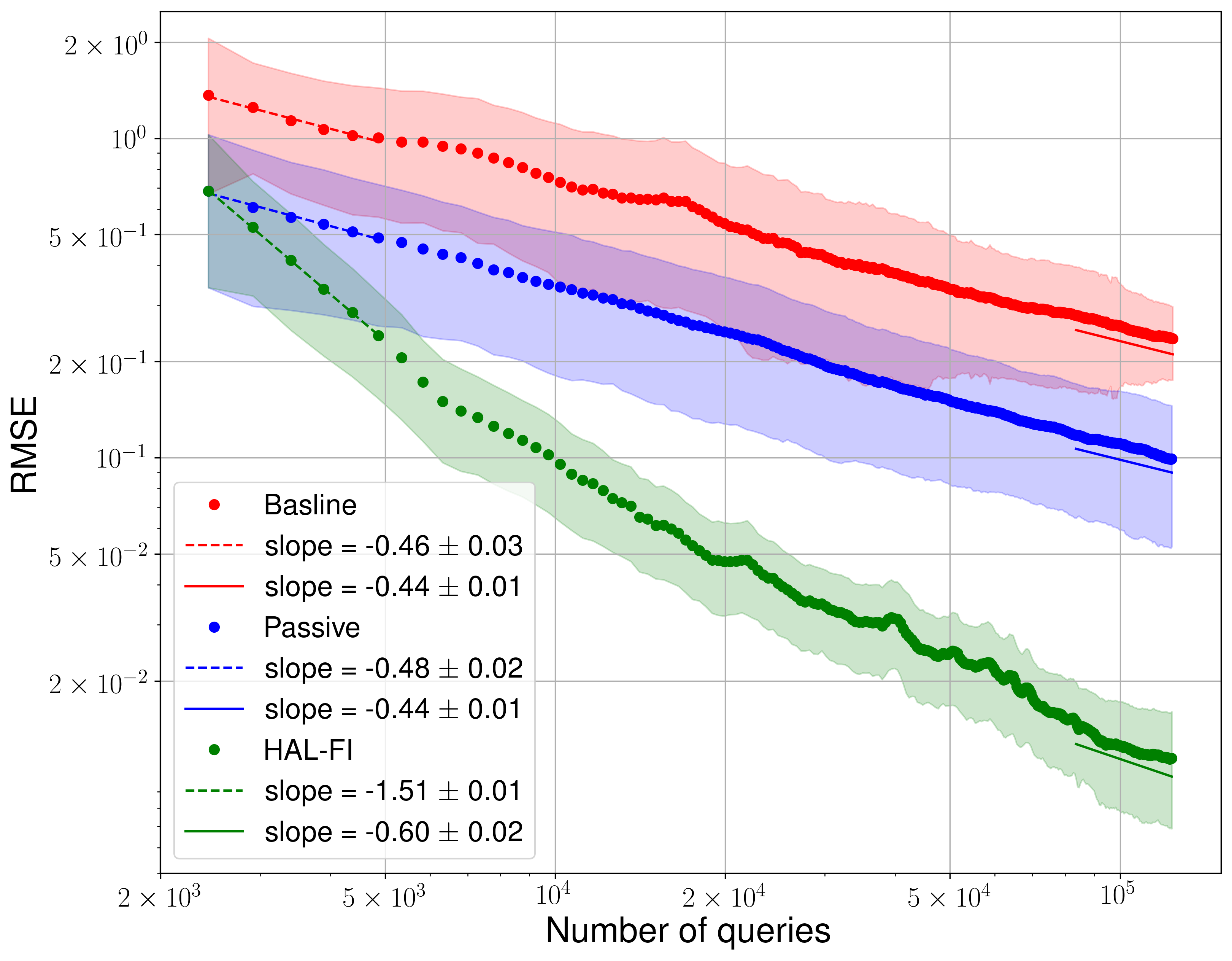}
    \caption{Scaling of RMSE with number of queries for different learners on (a) a simulator, and (b) oracle with access to experimental data. In (a)-(b), the performance of HAL-FI is compared against the passive learner and baseline. The time range $\timeSpace$ of the query space is such that Minimum-Frequency criteria is not necessarily satisfied for frequency estimation using FFT. Slopes indicate the scaling of learning error $\epsilon$ with number of queries $N$ in the low and high sample regimes. Filled in areas indicate the respective errors on trends for each learner.}
    \label{fig:CR_Error_Scalings_Simulator_Expt}
\end{figure}

A comparison of the different learners considering a $\timeSpace$ that does not satisfy Minimum-Frequency criteria on the simulator is shown in Figure~\ref{fig:CR_Error_Scalings_Simulator_Expt}(a). We set $\timeSpace$ to be the set of $81$ equispaced times in the interval of $10^{-7}, 6 \times 10^{-7}]$s. As obtained earlier on the simulator with a longer $\timeSpace$, we observe a SQL scaling of $\epsilon \sim 1/\sqrt{N}$ or ($N \sim \epsilon^{-2}$) in RMSE with number of queries for both the baseline and passive learner. However, there is now a noticeably wider gap in between the trends corresponding to around $80.6\%$ reduction in queries when using the passive learner over the baseline.

For the HAL-FI algorithm, we consider cases of when the query space is fixed and when it is adaptively grown by linearly growing the $\timeSpace$. For both cases, we see an initial scaling which is higher than SQL and similar to Heisenberg limited scaling, and a scaling of SQL in the asymptotic query regime. This behaviour is expected from our observations on the simulator earlier. We note that using HAL-FI combined with a linearly growing query space does not show significant improvement over HAL-FI in the fixed query space. This is due to the low rate at which we adaptively grow the query query space during learning and the fact that growing the query space is only advantageous for learning a subset of the Hamiltonian parameters. We discuss this further in Section~\ref{sec:results_analysis_HS}. 

\paragraph{Experimental Dataset} \label{sec:HL_experimental_data}
We show a comparison of the performance of the different learners on the $\textit{oracle}$ with access to experimental data in Figure~\ref{fig:CR_Error_Scalings_Simulator_Expt}(b). As expected from our observations on the simulator, we observe SQL like scalings in RMSE with $N$ for both the baseline and passive learner. We observe around $82.2\%$ query reduction when using the passive learner over the baseline, similar to that previously observed on the simulator. This query reduction was computed by fixing the RMSE value at $0.2$, and comparing the number of queries required by the passive learner to achieve this RMSE value versus the baseline. The trends themselves are remarkably similar to those obtained on the simulator, supporting the fact that the main noise sources affecting the quantum device were identified and the simulator is a good representation of the real quantum hardware. 

For HAL-FI, we only show results for fixed query space as the results for the linearly growing query space are very similar, as was also observed on the simulator. We see an initial scaling of $\epsilon \sim 1/N^{3/2}$ (or $N \sim \epsilon^{-2/3}$) in RMSE with number of queries which is higher than SQL and Heisenberg limited scaling, and a scaling of SQL in the asymptotic query regime. The performance of HAL-FI on the experimental data is surprisingly better than that on the simulator. This behaviour is consistent across different drive configurations on \textit{ibmq\_boeblingen} and is not particular for this \textit{oracle}. The initially accelerated learning where we observe super-Heisenberg limited scaling in HAL-FI and lower values of RMSE achieved for a smaller number of $N$ than on the simulator might be due to: (i) noise sources not included in our model that do not contribute significantly to the noise affecting the quantum device but encourage exploration by HAL-FI, and (ii) correlations between the samples collected in the experimental data (e.g., due to thermal fluctuations). 

We have already seen benefits of using an active learner over a baseline strategy or a passive learner through the lower values of RMSE that can be achieved. This is analyzed in terms of query advantage in Section~\ref{sec:results_analysis_QA}.

\subsection{Analysis}\label{sec:results_analysis}
In this section, we analyze the results of the performance of the different learners from Section~\ref{sec:results_performance}. We firstly comment on the achievability of the Heisenberg limited scaling by HAL-FI with an adaptively growing query space as claimed in Section~\ref{sec:hlproblem_noise}. In the process, we consider a different learning scenario motivated by recalibrations of quantum devices. We then describe the query advantage of the active learner under different conditions over the baseline strategy.

\subsubsection{Heisenberg limited scaling} \label{sec:results_analysis_HS}
In Section~\ref{sec:results_performance}, we did not observe Heisenberg limited scaling for HAL-FI (even with an adaptively growing query space). This is due to the fact that the query space is not rich enough to achieve Heisenberg limited scaling i..e., there is no sequence of queries even in the adaptively growing query space to achieve Heisenberg limited scaling. We discuss when Heisenberg limited scaling is achievable for Hamiltonians based on the CR Hamiltonian in Appendix~\ref{app_sec:HLS_Discovery}. We show that the behavior of the learners observed so far is expected through another set of experiments in Appendix~\ref{app_sec:details_results_HL}.

It should however be possible to achieve Heisenberg limited scaling for a subset of Hamiltonian parameters given the query space (see Section~\ref{sec:cr_query_space}) when the task is to learn this subset of Hamiltonian parameters and we are given access to information about the other Hamiltonian parameters. This is exactly the setting of a recalibration where prior information about the Hamiltonian parameters is available from previous calibrations and the goal is to learn the subset of parameters which drift significantly with time. Motivated by this, we consider the following learning scenario on $ibm\_boeblingen$ under drive configuration $3$ (see Table~\ref{tab:ibmq_boeb_ham_params_summary}).

\begin{figure}[h!]
	\centering
    \xincludegraphics[scale=0.28, label=\textbf{(a)}]{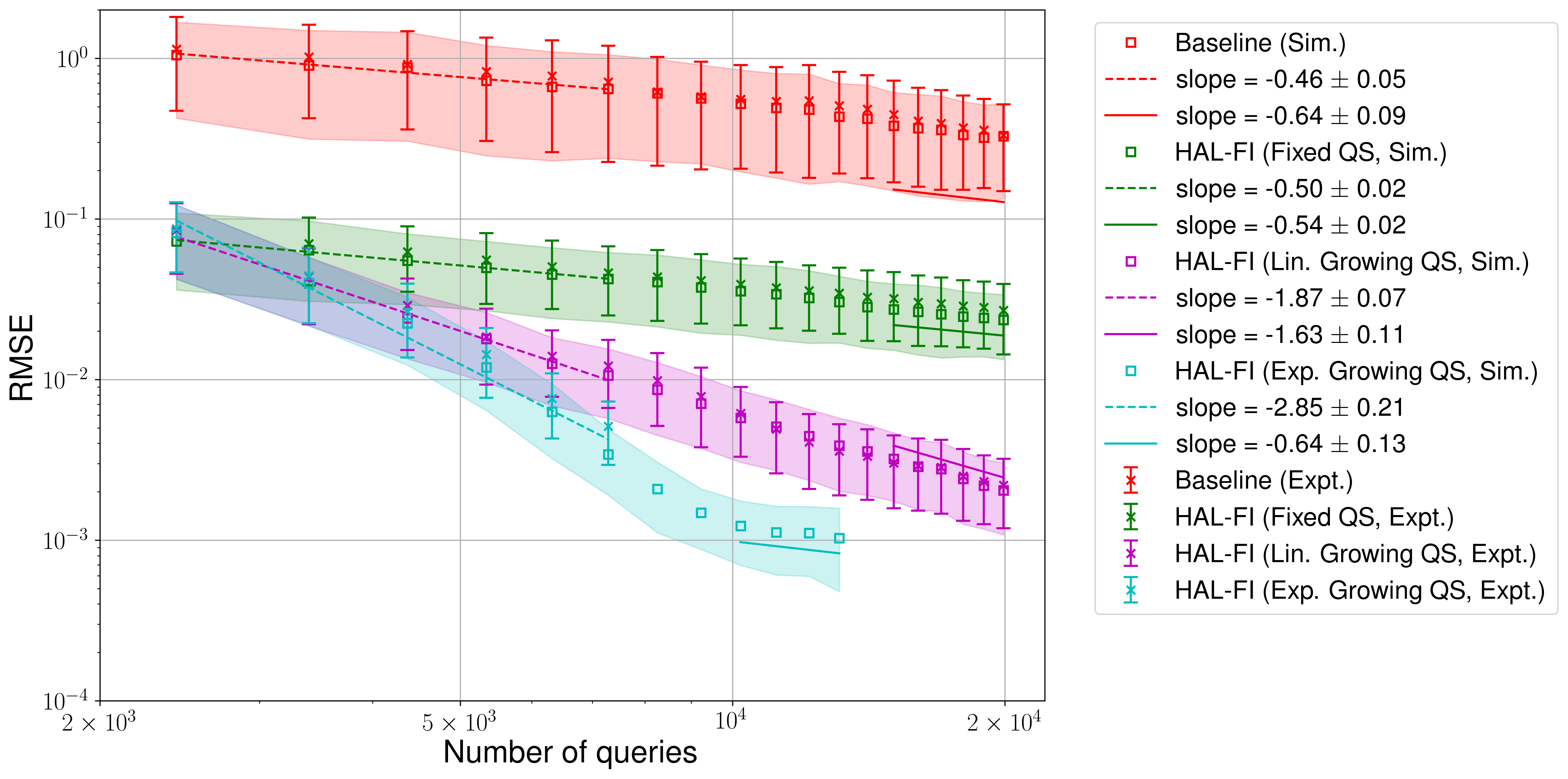}
    \quad
    \xincludegraphics[scale=0.31, label=\textbf{(b)}]{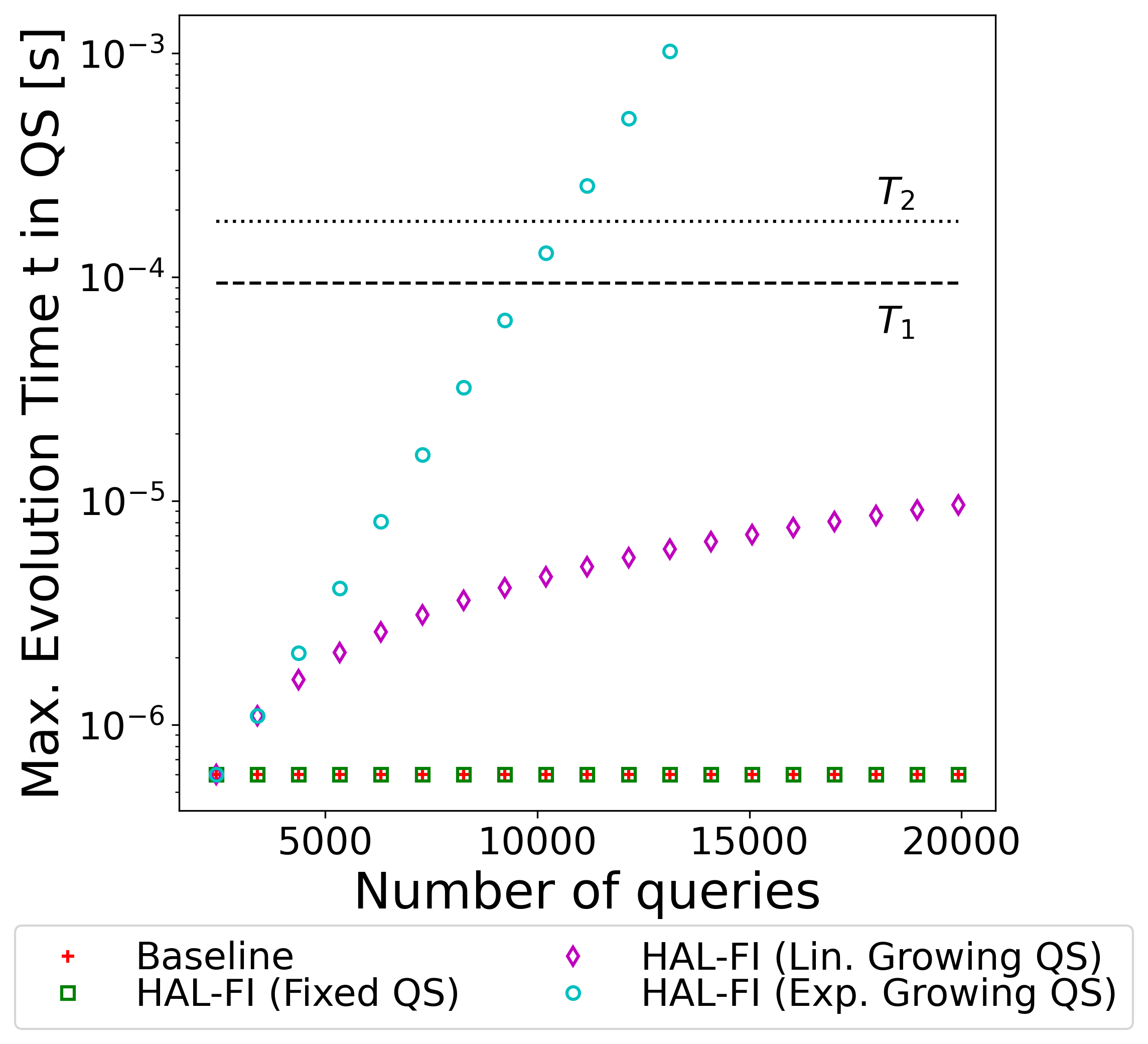}
    \caption{(a) Trends of RMSE with number of queries for different learners in fixed or growing query spaces (QS) on experimental data (Expt.) compared against simulator (Sim.). Slopes indicate the scaling of RMSE $\epsilon$ with number of queries $N$ in the low query/sample and high query/sample regimes. Filled in areas indicate the respective errors on trends for each learner on the simulator. Passive learner which has similar behavior to HAL-FI in fixed query space is not shown for brevity. (b) Trends of the maximum evolution time associated with query spaces of different learners.}   
	\label{fig:reduced_HL_sim_expt_data}
\end{figure}

We have access to an estimate of the Hamiltonian parameters $\hat{\bm{\theta}}$ from a previous calibration during which Hamiltonian learning was run on a uniformly sampled set of queries from $\mathcal{Q}$ (of size $N = 2430$). The goal is to then learn the parameters $\omega_{0,1}$ using the different learners at our disposal. We plot comparisons of different learners (using $N_b = 972$) on this recalibration task in Figure~\ref{fig:reduced_HL_sim_expt_data} considering the oracles of the simulator and experimental data. 

We observe the SQL scaling in the baseline, passive learner and HAL-FI in the fixed query space. There is nearly a constant gap between the baseline strategy and HAL-FI in the fixed query space, indicating a constant query reduction in achieving a desired learning error. We observe a super-Heisenberg limited scaling of RMSE $\epsilon \sim N^{-3/2}$ in number of queries in HAL-FI with a linearly growing query space. This is also observed for HAL-FI with an exponentially growing query space in the low sample regime. The deterioration in the scaling of HAL-FI with an exponentially growing query space to the SQL is due to the maximum evolution time corresponding to the growing query space eventually exceeding $T_1$ and $T_2$. This is shown in Figure~\ref{fig:reduced_HL_sim_expt_data}(b). In fact, the bend in the trend of RMSE versus $N$ for HAL-FI with an exponentially growing query space occurs immediately after the maximum evolution time in $\timeSpace$ exceeds $T_1$. As $\timeSpace$ for HAL-FI with a linearly growing query space is grown much more slowly, effects of decoherence are not yet felt and super-Heisenberg limited scaling convergence rate in learning error is achieved.

Qualitatively, it is clear that much lower values of learning error can be achieved with a given budget of queries using HAL-FI with an adaptively growing query space over the baseline for recalibration. This is quantified in terms of query advantage in Section~\ref{sec:results_analysis_QA}.

\subsubsection{Query Advantage} \label{sec:results_analysis_QA}
We have so far compared the trends and obtained the scalings of RMSE $\epsilon$ with number of queries $N$ for different learners for Hamiltonian learning (Section~\ref{sec:results_performance}) and with prior information (Section~\ref{sec:results_analysis_HS}). To quantify the benefits of using HAL-FI over a baseline strategy or a passive learner, we now show the corresponding query advantage (defined in Section~\ref{sec:def_query_advantage}), a performance metric that summarizes the reduction in resources required to achieve a desired learning error.

\begin{figure}[h!]
	\centering
    \xincludegraphics[scale=0.29, label=\textbf{(a)}]{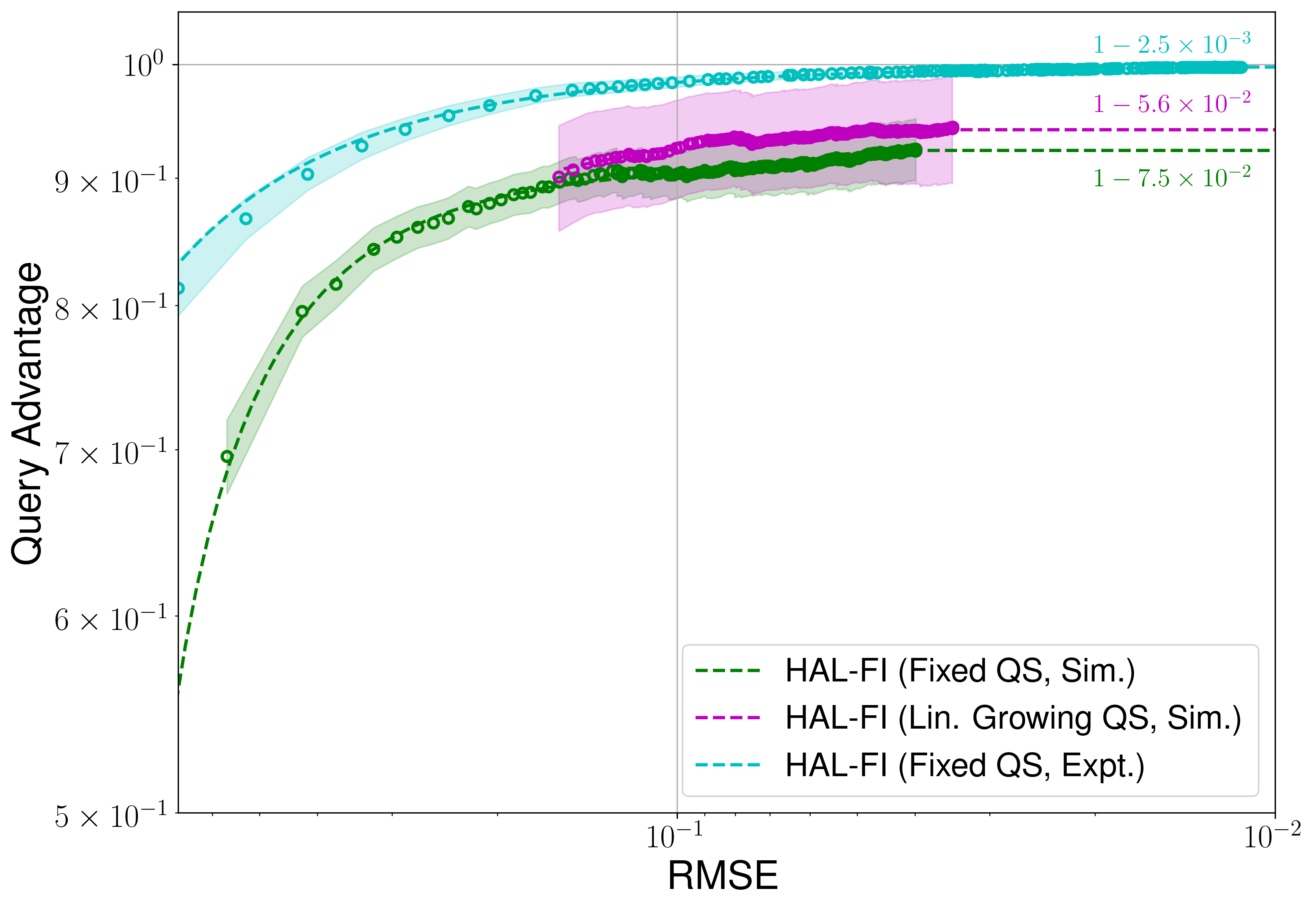}
    \quad
    \xincludegraphics[scale=0.29, label=\textbf{(b)}]{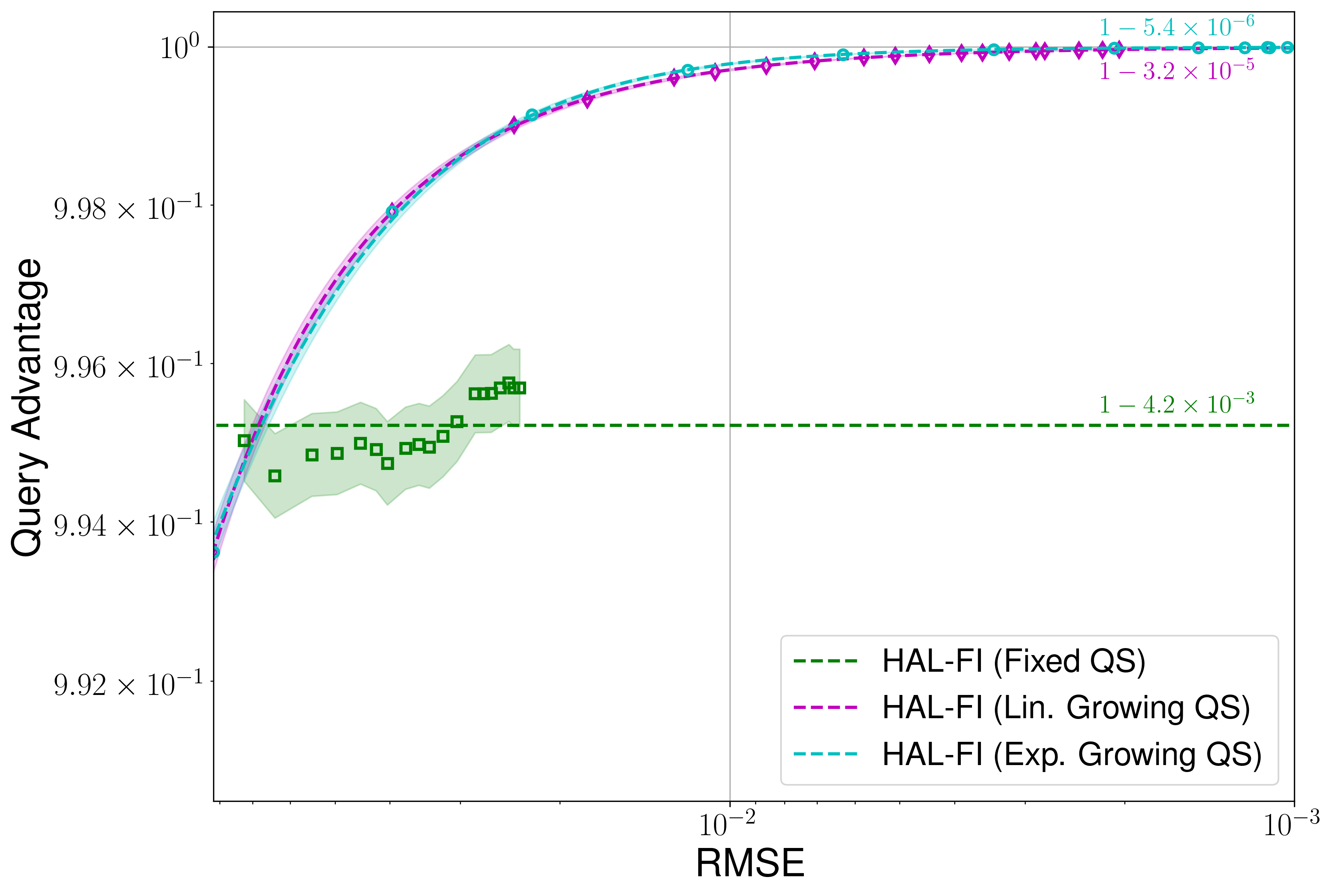}
    \caption{Query advantage (QA) of different learners over the baseline considering the problem of (a) Hamiltonian learning without prior information and (b) Hamiltonian learning with prior information on subset of Hamiltonian parameters from previous calibrations. In (a), we plot the QA of HAL-FI over the baseline as observed on the simulator (Sim.) and experimental data (Expt.). In (b), we plot the QA of HAL-FI over the baseline on the simulator for fixed and adaptively growing query spaces (QS). In (b), the QAs obtained on the simulator closely match those obtained on the experimental data and hence the latter is not shown. In (a)-(b), the data points correspond to values of QA computed from the data points of Figures~\ref{fig:CR_Error_Scalings_Simulator_Expt},\ref{fig:reduced_HL_sim_expt_data} and the lines are the fits to these values of QA. The annotated text for the different trends indicate a lower bound on the QA for the corresponding learner over the baseline for the lowest reported value of RMSE. Filled in areas indicate the uncertainty of each trend.}
	\label{fig:query_advantage_HL}
\end{figure}

We plot the query advantage of HAL-FI over the baseline strategy for Hamiltonian learning on the simulator and experimental data in Figure~\ref{fig:query_advantage_HL}(a). When computing the query advantage of a learner over the baseline, we consider that they see the same oracle i.e., the simulator or experimental data. We observe a query advantage of around $80\%$ in HAL-FI on the simulator and experimental data over the baseline, for high values of RMSE. The initial accelerated learning observed in Figure~\ref{fig:CR_Error_Scalings_Simulator_Expt} for HAL-FI on the simulator and experimental data translates to an accelerated query advantage for high values of RMSE. The trend in query advantage for HAL-FI on the simulator flattens to an asymptotic value of at least $95.1\%$ when the query space is fixed during learning and at least $96.3\%$ when the query space is grown linearly. The corresponding value for HAL-FI on the experimental data in the fixed query space is $99.8\%$ for low values of RMSE. While not shown, the query advantage of HAL-FI on the experimental data over a passive learner is at least $99.1\%$ for low values of RMSE.

Similarly for the learning scenario of recalibration, we plot the query advantage of HAL-FI over the baseline strategy for Hamiltonian learning when prior information is available in Figure~\ref{fig:query_advantage_HL}(b). The trends shown here correspond to that of the simulator as similar behavior is observed for experimental data. The query advantage of HAL-FI in a fixed query space over the baseline is constant with RMSE and around $~99.6\%$ i.e., HAL-FI requires only $0.4\%$ of the queries required by the baseline. For HAL-FI with a linearly growing query space, the query advantage for low values of reported RMSE is around $(100 - 3.2 \times 10^{-3})\%$. Further, we note that the scaling of query advantage QA of HAL-FI in a linearly growing query space with RMSE $\epsilon$ scales as $\text{QA} \sim (1 - O(\epsilon^{-1/6}))$ until decoherence starts effecting the scaling. For HAL-FI with an exponentially growing query space, the query advantage flattens to around $(100 - 5.4 \times 10^{-4})\%$ for low values of RMSE. Under this learning setting at low values of RMSE around $10^{-3}$, HAL-FI has a query space with evolution times far exceeding $T_1$ or $T_2$ and thus this reported query advantage is expected for even lower values of RMSE.

From the analysis of query advantage above, we observe that two orders of magnitude reduction in queries can be obtained over a baseline strategy by adopting HAL-FI. Moreover, during recalibrations, we observe three orders of magnitude reduction when HAL-FI is used with a linearly growing query space and five orders of magnitude reduction in queries when HAL-FI is used with an exponentially growing query space. 

Adopting HAL-FI not only allows us to achieve query reduction but it also allows us to save wall clock time taken to calibrate a quantum device. We again consider the query space of \ref{sec:cr_query_space} where a query on average takes $2400ns$ to run (accounting for time duration of implementing measurement pulses). The repetition rate of current IBM Quantum devices for executing circuits is $10kHz$. We can then reduce the duration of Hamiltonian learning of all CR Hamiltonians of directly connected qubit pairs on a $20$-qubit IBM Quantum device of \textit{ibmq\_boeblingen} to reach a RMSE of $5 \times 10^{-2}$ from around $10$ minutes to $5$ seconds by using HAL-FI instead of the baseline strategy. These timings only take into account the time on the quantum hardware and not additional latencies in classical electronics interfacing with the hardware.

\section{Conclusion} \label{sec:conclusion}
In this paper, we proposed the active learning algorithms of HAL-FI for Hamiltonian learning and HAL-FIR for predictions of queries to a Hamiltonian, sampled from a testing distribution. The performance of HAL-FI/HAL-FIR was compared against different learners for learning a CR Hamiltonian on the $20$-qubit IBM Quantum device \textit{ibmq\_boeblingen} on a simulator and an oracle with access to experimental data. We showed that HAL-FI can achieve a query advantage of around $~99.8\%$ compared to a baseline strategy and $~99.1\%$ over a passive learner for low values of learning error on experimental data. During recalibration when learning the Hamiltonian with access to information from previous calibrations, we observed that HAL-FI can achieve query advantages of $~99.5\%$ over a baseline strategy. Further, we showed that we achieve Heisenberg limited rate of convergence where possible when an active learner is used in conjunction with an adaptive query space during learning before the evolution time of queries exceed qubit $T_1$ or $T_2$ and decoherence starts deteriorating information content available in queries. 

Overall, using the active learner HAL-FI can yield in reduction of resources of up to two orders of magnitude during calibration and five orders of magnitude during recalibration for low values of learning error. This improvement in query complexity has multiple practical consequences besides accelerating Hamiltonian learning during calibration and recalibrations of quantum computers. 

Another important calibration step is determining controls to implement desired single and multi-qubit quantum gates. This often relies on building a Hamiltonian model of the true environment i.e., the quantum computer. This can be accomplished by HAL to ensure minimal queries are used. Gates, once implemented, are characterized through quantum process tomography which is not query efficient but could be accelerated with an active learner. For the specific application of learning the CR Hamiltonian, more noise sources can be included as they become relevant e.g., leakage errors which become pronounced under strong driving. It would also be interesting to see if one can achieve asymptotic Heisenberg limited scaling even in the presence of noise such as decoherence by using appropriate quantum error correction protocols. It should be noted that Hamiltonian learning and for that matter quantum process tomography both suffer from an exponential scaling in the size of the quantum device $n$ and using an active learner only ensures a better scaling in $\epsilon$ or query advantage. One could get a better scaling in $n$ if additional information such as the structure of the Hamiltonian was known to be a $k$-local Hamiltonian which can be learned with a query complexity that scales as $O(poly(n))$ \cite{anshu2021sample}.

There are other calibration steps where Hamiltonian learning may not be required but the concept of active learner can be introduced. This would particularly be advantageous where a variety of experiments can be carried out but it is not clear which of them are more informative for the learning task. For example, a query efficient method is desired for learning cross-talk on a superconducting quantum device \cite{abrams2019methods,dai2021calibration}. 

Additionally, there is much room for improvement and extension in the algorithm itself. Currently, expert knowledge is required to specify a complete query space to HAL to ensure all the Hamiltonian parameters can be learned and possibly with Heisenberg limited scaling. It is desirable to remove this expert and replace them with a method to synthesize queries. The current active learning strategy is also based on the query criterion of Fisher information but this could be modified to incorporate cost of different queries and increase exploration. Moreover, a more general query criteria could possible be learned through reinforcement learning as illustrated in recent work on classical applications \cite{pang2018meta}.

\begin{acknowledgments}
AD was supported in part by the MIT-IBM Watson AI Lab.  AD, JS, and ILC were supported in part by the U.S. Department of Energy, Office of Science, National Quantum Information Science Research Centers, Co-Design Center for Quantum Advantage under contract DE-SC0012704. AD acknowledges the MIT SuperCloud and Lincoln Laboratory Supercomputing Center for providing computational resources.
\end{acknowledgments}

\clearpage

\appendix


\section{Notation} \label{app_sec:notation}
In this section, we summarize the notation used in the paper. In Table~\ref{tab:notation_summary_HL}, we describe the notation used in general for Hamiltonian learning throughout this paper. In Table~\ref{tab:notation_summary_HAL}, we give the notation used specifically for different rounds of the HAL algorithms (HAL-FI and HAL-FIR). Note that the superscript of $(i)$ is used in two different ways below. It is used to refer to the $i$th example of a training set $D=\{(x^{(i)},y^{(i)}\}_{i \in [N]}$ as well as the $i$th round of the HAL algorithm. The usage in the paper will be clear from context.

\begin{table}[ht!]
\centering
\begin{tabular}{r|l}
\hline
\textbf{Notation} & \textbf{Definition} \\ \hline
$H$ & Hamiltonian of a quantum system \\
$H^\star$ & \textit{True} Hamiltonian of an unknown quantum system \\
$\hat{H}$ & \textit{Estimated} Hamiltonian of an unknown quantum system \\
$n$ & Number of qubits in a quantum system \\
$\bm{\theta}$ & Hamiltonian model parameters \\
$\bm{\theta}^\star$ & True Hamiltonian model parameters of the unknown quantum system \\
$\hat{\bm{\theta}}$ & Estimated Hamiltonian model parameters \\
$\bm{\xi}$ & Normalization vector of $\bm{\theta}$ \\
$\mathrm{x}$ & Random variable describing query to the quantum system (See Sec.~\ref{sec:hlproblem} for details) \\
$\querySpace$ & Query space or alphabet of $\mathrm{x}$ \\
$x$ & An example of the random variable $\mathrm{x}$ \\
$\mathrm{y}$ & Random variable describing the measurement outcomes obtained after querying the quantum system \\
$n_r$ & Number of qubits being readout from a quantum system, $n_r \leq n$ \\
$\mathcal{Y}$ & Alphabet of $\mathrm{y}$, $\mathcal{Y} = \{0,1\}^{n_r}$ \\
$y$ & An example of the random variable $\mathrm{y}$ \\
$(x^{(i)}, y^{(i)})$ & $i$th training example of $(\mathrm{x}, \mathrm{y})$ \\
$N$ & Query complexity or number of training examples of $\{(x,y)\}$ \\
$\tilde{\mathrm{y}}$ & Noisy observation of the measurement outcome $\mathrm{y}$ \\
$\tilde{y}$ & Example of $\tilde{\mathrm{y}}$ \\
$q$ & Query distribution or arbitrary distribution associated with $\mathrm{x}$ \\
$q^\star$ & Optimal query distribution for a given learning problem \\
$\ptest$ & Testing distribution \\
$p_{\mathrm{y}|\mathrm{x}} \left(y|x;\bm{\theta} \right)$ & Likelihood of $y$ given $x$ and considering model parameters $\bm{\theta}$\\
$L(y|x;\theta)$ & Shorthand notation of negative log-likelihood i.e., $- \log p_{\mathrm{y}|\mathrm{x}} \left(y|x;\bm{\theta} \right)$ \\
$\mathrm{M}$ & Random variable describing the measurement operator in a query \\
$\mathcal{M}$ & Measurement space or alphabet of $\mathrm{M}$ \\
$\mathrm{U}$ & Random variable describing the state preparation operator in a query \\
$\mathcal{U}$ & Preparation operator space or alphabet of $\mathrm{U}$ \\
$\mathrm{t}$ & Random variable describing the evolution time in a query \\
$\mathcal{T}$ & Range of evolution times or alphabet of $\mathrm{t}$ \\
$(M,U,t)$ & Example of the tuple $(\mathrm{M},\mathrm{U},\mathrm{t})$ and query $\mathrm{x}$ \\
$p_{\mathrm{rabi}}$ & Rabi oscillations $p_{\mathrm{y}|\mathrm{x}}(0|x) - p_{\mathrm{y}|\mathrm{x}}(1|x)$ \\
$(r_0,r_1)$ & Probability of bit-flip $p_{\tilde{\mathrm{y}}|\mathrm{y}}(1|0)$ and $p_{\tilde{\mathrm{y}}|\mathrm{y}}(0|1)$ due to readout noise \\
$\fisherInfo_\mathrm{x}(\bm{\theta})$ & Fisher information matrix corresponding to query $\mathrm{x}$ considering model parameters $\bm{\theta}$\\
$\fisherInfo_q(\bm{\theta})$ & Fisher information matrix corresponding to query distribution $q$ considering model parameters $\bm{\theta}$\\
$\mathbf{J}$ & Parameter vector of coefficients of Pauli product terms of the cross-resonance Hamiltonian (Eq.~\ref{eq:cr_hamiltonian_block_form}) \\
$\mathbf{\Lambda}$ & Alternate parameter vector of cross-resonance Hamiltonian (Eq.~\ref{eq:cr_hamiltonian_block_form}) based on its eigendecomposition \\
$\{\sigma_I,\sigma_X,\sigma_Y,\sigma_Z\}$ & Single-qubit Pauli operators \\
\hline
\end{tabular}
\caption{\label{tab:notation_summary_HL} Notation used in Hamiltonian learning}
\end{table}

\begin{table}[ht!]
\centering
\begin{tabular}{r|l}
\hline
\textbf{Notation} & \textbf{Definition} \\ \hline
$N_{\mathrm{tot}}^{(i)}$ & Number of queries at the $i$th round \\
$q^{(i)}$ & Query distribution at the $i$th round \\
$\querySpace^{(i)}$ & Query space at the $i$th round \\
$N_b$ & Batch size \\
$i_\mathrm{max}$ & Maximum number of iterations\\
$\hat{\bm{\theta}}^{(i)}$ & Estimate of the Hamiltonian model parameters at the $i$th round \\
$X^{(i)}$ & Set of all queries made up till the $i$th round (inclusive) \\
$Y^{(i)}$ & Set of all measurement outcomes up till the $i$th round (inclusive) \\
$X_q$ & Set of queries obtained from sampling $\querySpace$ according to query distribution $q$ \\
$Y_q$ & Set of measurement outcomes obtained from quantum system after making queries in $X_q$ \\
\hline
\end{tabular}
\caption{\label{tab:notation_summary_HAL} Notation specific to HAL Algorithms (HAL-FI/HAL-FIR)}
\end{table}

\section{Details of Cross-Resonance Hamiltonian} \label{app_sec:details_cr_hamiltonians}
In this section of the Appendix, we give details of the different IBM Quantum devices employed for assessing the performance of the HAL algorithms (HAL-FI and HAL-FIR) proposed in Section~\ref{sec:HAL} for learning cross-resonance (CR) Hamiltonians (Eq.~\ref{eq:cr_hamiltonian_block_form}). In Section~\ref{sec:imperfect_pulse_shaping}, we stated a model for the noise source of imperfect pulse-shaping. Here, we describe how this model was obtained. This is then followed by relevant analytical expressions of likelihood and Fisher information considering the query space in Section~\ref{sec:cr_query_space}.

\subsection{Description of IBM Quantum Devices} \label{app_sec:quantum_device_description}
\begin{figure}[H]
    \centering
    \includegraphics[scale=0.35]{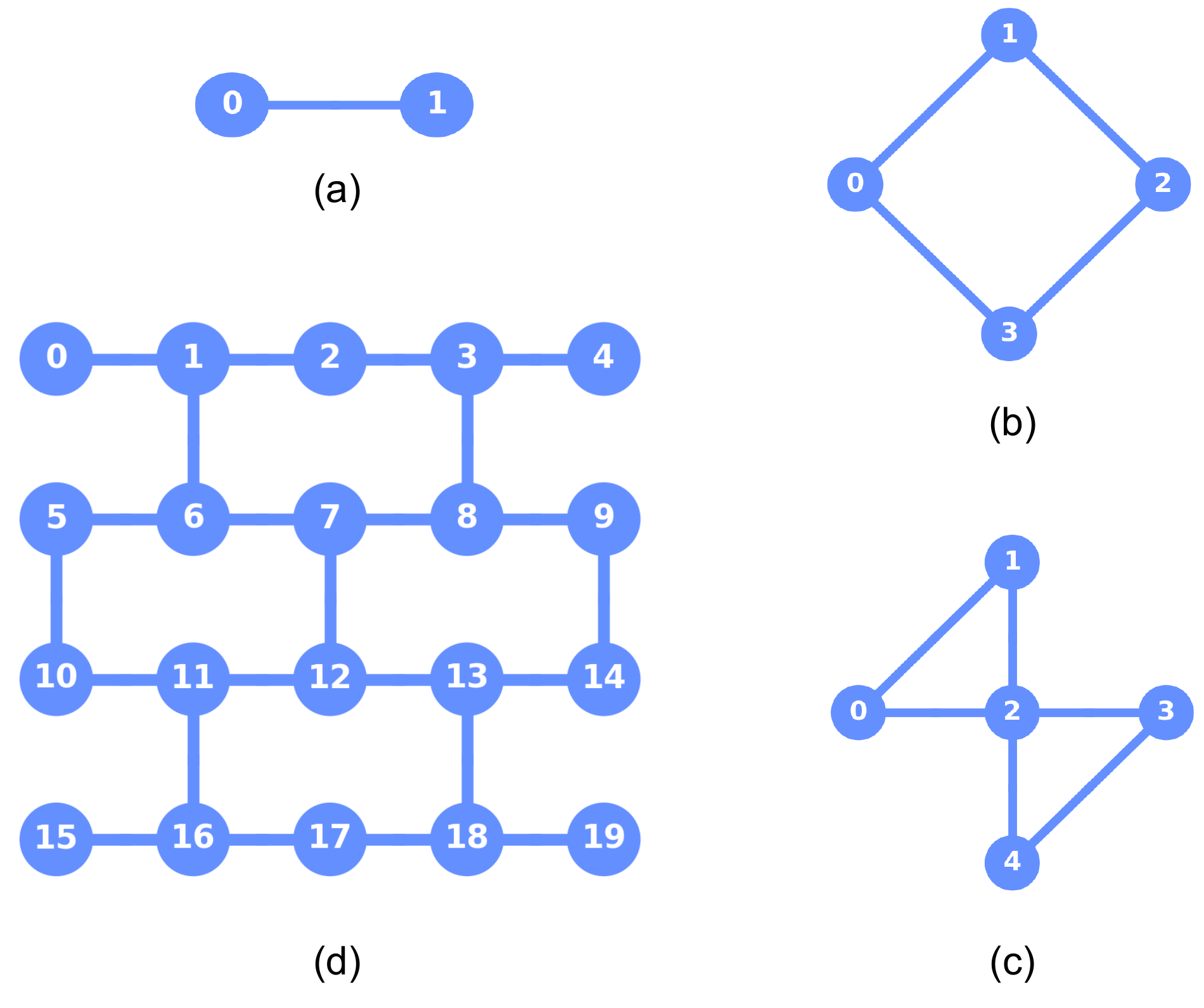}
    \caption{Connectivity maps for (a) IBM Quantum device A, (b) IBM Quantum device B, (c) IBM Quantum device C, and (d) IBM Quantum device D \textit{ibmq\_boeblingen}. Each node represents a physical qubit on the chip and the presence of an edge between two nodes in the connectivity map indicates that a CR gate can be applied between these two nodes.}
    \label{fig:quantum_device_connectivity_maps}
\end{figure}

We consider CR Hamiltonains on the four different IBM Quantum devices described in Section~\ref{sec:quantum_devices}. The connectivity maps of these devices are shown in Figure~\ref{fig:quantum_device_connectivity_maps}. We consider CR gates on particular qubit pairs on each device which are summarized in Table~\ref{tab:quantum_device_parameters}. In Table~\ref{tab:quantum_device_parameters}, we describe the properties of each qubit involved in the CR gate including their $T_1$ or $T_2$ times and the average infidelity of single-qubit gates.

\begin{table}[H]
\small
\centering
\begin{tabular}{|c|c|c|c|c|c|c|c|c|c|c|}
\hline
\multirow{2}{*}{\textbf{Device}} & \multicolumn{2}{c|}{\textbf{Qubit}} & \multicolumn{2}{c|}{\textbf{Qubit Freq. (GHz)}} & \multicolumn{2}{c|}{$\mathbf{T_1}$ ($\mathrm{\mu}\text{s}$)} & \multicolumn{2}{c|}{$\mathbf{T_2}$ ($\mathrm{\mu}\text{s}$)}  & \multicolumn{2}{c|}{\textbf{Error per gate} ($\times 10^{-4}$)} \\ \cline{2-11}

& \multicolumn{1}{c|}{\textbf{Control}} & \multicolumn{1}{c|}{\textbf{Target}}
& \multicolumn{1}{c|}{\textbf{Control}} & \multicolumn{1}{c|}{\textbf{Target}}
& \multicolumn{1}{c|}{\textbf{Control}} & \multicolumn{1}{c|}{\textbf{Target}}
& \multicolumn{1}{c|}{\textbf{Control}} & \multicolumn{1}{c|}{\textbf{Target}}
& \multicolumn{1}{c|}{\textbf{Control}} & \multicolumn{1}{c|}{\textbf{Target}} \\ \hline

A &	1 &	0	& $5.0593$ & $4.8441$ & $65.4 \pm 10.8$ & $32.9 \pm 9.5$ & $71.1 \pm 5.9$ &	$57.7 \pm 9.2$ & $5.65 \pm 0.13 $	& $8.85 \pm 0.46$ \\ \hline
B &	3 &	0	& $5.1482$ & $4.9273$ & $63.2 \pm 11.7$ & $78.1 \pm 24.9$ & $73.9 \pm 9.9$ & $124.3 \pm 21.5$ &	$5.92 \pm 0.30$	& $6.29 \pm 0.23$ \\ \hline
\multirow{3}{*}{C} & 0 &	1	& $5.3613$ & $5.2910$ & $34.2 \pm 2.5$ & $45.0 \pm 19.1$ &	$39.2 \pm 3.3$ & $63.1 \pm 7.6$ & $18.3 \pm 1.0$ & $20.6 \pm 1.3$ \\ \cline{2-11}
	& 0 &	2	& $5.3613$ & $5.2543$ & $34.2 \pm 2.5$ & $45.0 \pm 19.1$ & $35.8 \pm 0.7$ & $52.4 \pm 4.2$ &	$18.3 \pm 1.0$ & $9.23 \pm 0.26$ \\ \cline{2-11}
	& 1 &	2	& $5.2910$ & $5.2543$ & $39.2 \pm 3.3$ & $63.1 \pm 37.6$ & $35.8 \pm 0.7$ & $52.4 \pm 4.2$ & $20.6 \pm 1.3$ & $9.23 \pm 0.26$ \\ \hline
D &	0 &	1	& $5.0466$ & $4.8468$ & $94.0 \pm 6.0$ & $75.7 \pm 17.0$ & $177.2 \pm 44.8$ &	$128.1 \pm 29.7$ & $2.39 \pm 0.12$	& $3.12 \pm 0.11$ \\ \hline
\end{tabular}
\caption{\label{tab:quantum_device_parameters} Relevant parameters of IBM Quantum devices. The qubit used as the control or target qubit is indicated by its number in the device connectivity map shown in Fig~\ref{fig:quantum_device_connectivity_maps}. Error per gate refers to the average infidelity of single-qubit gates implemented on that qubit.}
\end{table}

\subsection{Modeling Pulse Shapes}\label{app_sec:imperfect_pulse_shaping}
We now describe how the imperfect pulse-shaping model stated in Section~\ref{sec:imperfect_pulse_shaping} was obtained. We consider cross-resonance control pulses (also called \texttt{GaussianSquare}) whose time-varying amplitudes are rectangular-shaped envelopes with tapered rising and falling edges, where the tapering is designed to minimize the signal energy that falls above and below the frequency of the sinsoid that is being modulated by the pulse envelope. The resulting unitary operators thus have the form ${\tilde{U}(t)=\exp{(-i \mathbb{T} \int_{0}^{t} \tilde{H}(t') dt')}}$, where $\mathbb{T}$ is the time ordering operator, $\tilde{H}$ is the Hamiltonian at any particular time given by $\tilde{H}(t')=H(t', v(t'))$, $H$ is the cross-resonance Hamiltonian, and $v(t')$ is the time-varying pulse envelope.  

Let $\Delta t_r$ and $\Delta t_f$ be, respectively, the durations of the rising and falling edges of the shaped pulse envelope.  The central portion of $v(t')$ is then a rectangular function such that, for $t' \in [\Delta t_r, t-\Delta t_f]$, $v(t') = V_{\text{max}}\mathbbm{1}_{t' \in [\Delta t_r, t-\Delta t_f]}$ where $t$ is the total duration of the pulse, and $V_{\text{max}}$ is the amplitude of this central rectangular portion.  We thus have
\begin{align}
  \tilde{U}(t) &= \exp{\left( -i \mathbb{T} \int_{0}^{t} \tilde{H}(t') dt' \right)} \\
  &= \exp{\left( -i \mathbb{T} \int_{t-\Delta t_f}^{t} H(t',v(t')) dt' \right)}
    \exp{\left( -i \mathbb{T} \int_{\Delta t_r}^{t-\Delta t_f} H(t',V_{\text{max}}) dt' \right)}
    \exp{\left( -i \mathbb{T} \int_{0}^{\Delta t_r} H(t',v(t')) dt' \right)}
  \\
  &= \exp{\left( -i \mathbb{T} \int_{t-\Delta t_f}^{t} H(t',v(t')) dt' \right)}
    \exp{\left( -i H_{\text{max}} \int_{\Delta t_r}^{t-\Delta t_f} dt' \right)}
    \exp{\left( -i \mathbb{T} \int_{0}^{\Delta t_r} H(t',v(t')) dt' \right)}
  \\
  &= \exp{\left( -i \mathbb{T} \int_{t-\Delta t_f}^{t} H(t',v(t')) dt' \right)}
    \exp{\left( -i H_{\text{max}} t_{\text{expt}} \right)}
    \exp{\left( -i \mathbb{T} \int_{0}^{\Delta t_r} H(t',v(t')) dt' \right)}
\end{align}
where $t_{\text{expt}}=t-\Delta t_f - \Delta t_r$ and $H_{\text{max}} = H(t',V_{\text{max}})$ which is constant assuming that any signal distortions that are introduced by the control electronics and/or along the signal path to the quantum device are negligible.

\begin{figure}[H]
    \centering
    \includegraphics[scale=0.2]{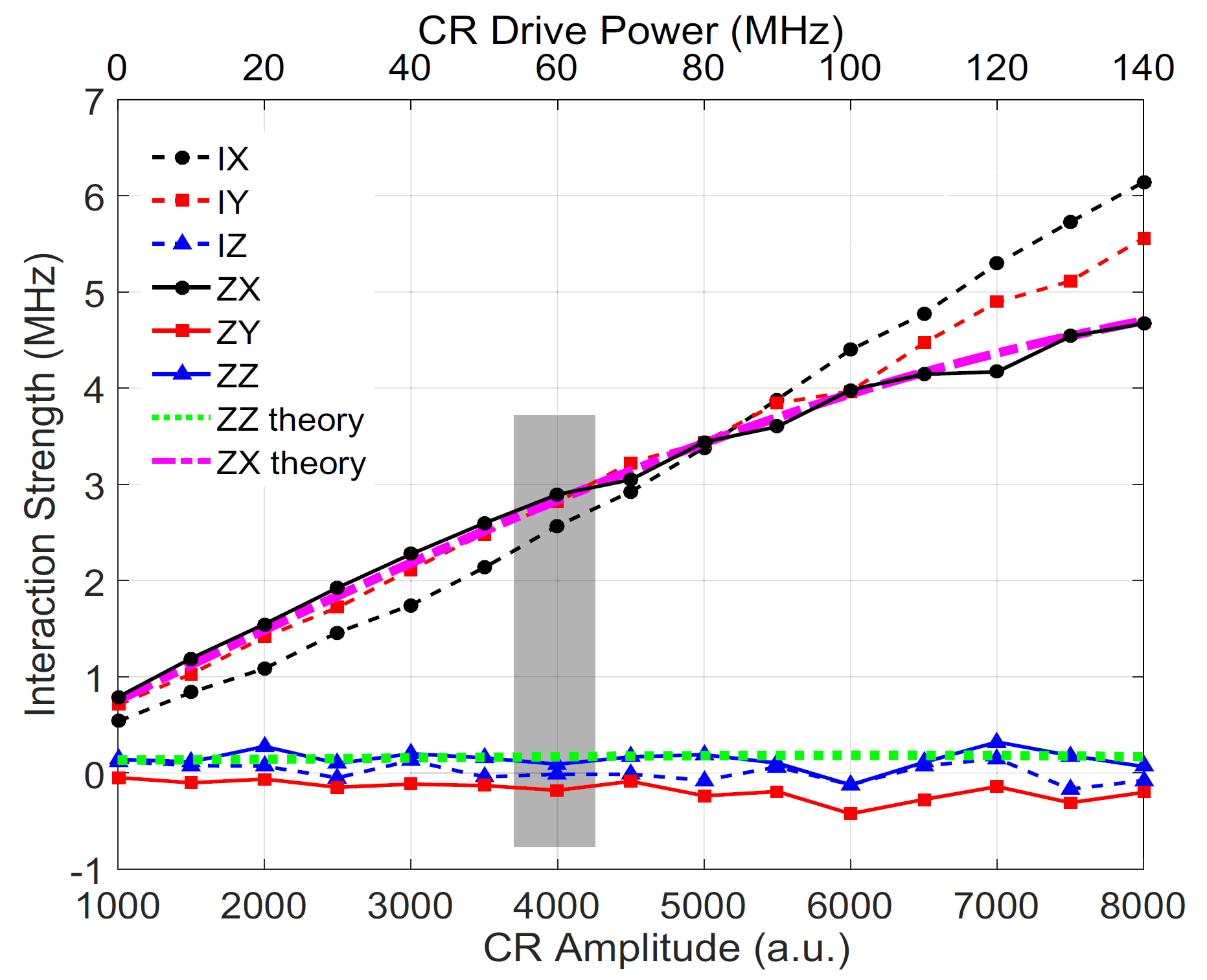}
    \caption{Hamiltonian parameters $\mathbf{J}$ as a function of the amplitude of the control pulse or drive as reported in Sheldon {\it et al}. \cite{sheldon2016procedure}}
    \label{fig:j_params_vs_amplitude}
\end{figure}

The above equation decomposes $\tilde{U}(t)$ into the time evolution of the Hamiltonian that corresponds to the central rectangular-pulse portion of the control pulse with pre- and post-rotations that are determined by the tapered rising and falling edges of the pulse.  In general, there is a nonlinear relationship between the shapes of these edges and the resulting pre- and post-rotations.  However, based on the results reported in \cite{sheldon2016procedure} and shown in Figure~\ref{fig:j_params_vs_amplitude}, the cross-resonanace Hamiltonian parameters tend to vary fairly linearly with respect to the overall pulse amplitude.  The pre- and post-rotations can thus be approximated by assuming a first-order model for the time-varing Hamiltonian parameters given by
\begin{equation}
  \mathbf{J}(t') \approx {v(t') \over V_{\text{max}}} \mathbf{J}_{\text{max}}
\end{equation}
where $\mathbf{J}_{\text{max}}$ is the vector of parameters for the Hamiltonian $H_{\text{max}}$ of the central recetangular portion of the pulse envelope.  The time-varying Hamiltonian is then approximated by  
\begin{equation}
  H(t',v(t')) \approx {v(t') \over V_{\text{max}}} H_{\text{max}} \; .
\end{equation}
The overall unitary operator $\tilde{U}(t)$ is then approximated by
\begin{align}
  \tilde{U}(t) 
  &\approx \exp{\left( -i H_{\text{max}} {1\over V_{\text{max}}} \int_{t-\Delta t_f}^{t} v(t') dt' \right)}
    \exp{\left( -i H_{\text{max}} t_{\text{expt}} \right)}
    \exp{\left( -i H_{\text{max}} {1\over V_{\text{max}}} \int_{0}^{\Delta t_r} v(t') dt' \right)}
  \\
  &= \exp{\left( -i H_{\text{max}} (t_{\text{expt}} + \Delta t_{\text{eff}}) \right)}
\end{align}
where
\begin{equation}
  \Delta t_{\text{eff}} = {1\over V_{\text{max}}} \left( \int_{0}^{\Delta t_r} v(t') dt' + \int_{t-\Delta t_f}^{t} v(t') dt' \right) \; .
\end{equation}

\subsection{Likelihood Function and Fisher Information Matrix for the CR Hamiltonian} \label{app_sec:fisher_info_cr}
In this section, we give the expressions for the likelihood function of $p_{\mathrm{y}|\mathrm{x}}(y|x;\boldsymbol{\theta)}$ and the Fisher information (FI) matrix. We consider the experimental setup as described in Section~\ref{sec:CR_Gate_Model_and_Setup} and query space $\querySpace$ as described in Section~\ref{sec:cr_query_space}. 

Recall from Section~\ref{sec:AL_query_criteria}, the FI matrix of a query $x$ is given by 
\begin{equation}
    \fisherInfo_x(\boldsymbol{\theta})[i,j] = \Expectation \left[ \frac{\partial \log p_{\mathrm{y}|\mathrm{x}}(y|x;\boldsymbol{\theta})}{\partial \theta_i} \frac{\partial \log p_{\mathrm{y}|\mathrm{x}}(y|x;\boldsymbol{\theta)}}{\partial \theta_j} \right]
\end{equation}
where $\log p(y|x;\boldsymbol{\theta})$ is the log-likelihood of the measurement outcome $y$ given the query $x$. In most cases in practice, the Fisher information matrix must be computed empirically in a Monte Carlo fashion. However, here we have a model of the CR Hamiltonian and models of the different noise sources affecting the quantum system available to us. We can thus evaluate the FI matrix analytically for different queries and in the presence or absence of noise.

\subsubsection{In Absence of Noise}
\paragraph{Likelihood} In the noiseless case, the likelihood function of different measurement outcomes $y \in \{0,1\}$ given query $x=(M,U,t) \in \querySpace$ is
\begin{equation}
    p_{\mathrm{y}|\mathrm{x}}(y|x;\boldsymbol{\theta}) =  \sum_{z \in \{0,1\}} \left|\braketExp{yz}{M e^{-iH(\boldsymbol{\theta}) t ) }U}{00}\right|^2 
\end{equation}
Evaluating this for the different queries in the query space as described in Section~\ref{sec:cr_query_space}, we obtain
\begin{align}
\small
    p_{\mathrm{y}|\mathrm{x}}(0|x;\boldsymbol{\theta}) = 
    \begin{cases}
    \frac{1}{2} ((\cos(\omega_j t) + \sin(\phi_j) \cos(\delta_j) \sin(\omega_j t))^2 + (\sin(\delta_j) \sin(\omega_j t) + \cos(\phi_j) \cos(\delta_j)\sin(\omega_j t))^2) &, x = (M_{\langle X \rangle}, U_j, t) \\
     \frac{1}{2} ((\cos(\omega_j t) - \cos(\phi_j) \cos( \delta_j) \sin(\omega_j t))^2 + (\sin(\delta_j) \sin(\omega_j t) + \sin(\phi_j) \cos(\delta_j) \sin(\omega_j t))^2) &, x = (M_{\langle Y \rangle}, U_j, t) \\
     1 - (\cos(\delta_j) \sin(\omega_j t))^2 &, x = (M_{\langle Z \rangle}, U_j, t)
    \end{cases}
\label{eq:expressions_likelihood_cr_noiseless}
\end{align}
where we have used the index $j \in \{0,1\}$ to refer to the different preparation operators $U_0 = \sigma_I \sigma_I$ and $U_1 = \sigma_X \sigma_I$. The measurement operators are: $M_{\langle X \rangle} = \sigma_I \otimes \exp \left(i \frac{\pi}{4} \sigma_Y \right)$, $M_{\langle Y \rangle}=\sigma_I \otimes \exp \left(-i \frac{\pi}{4} \sigma_X \right)$, and $M_{\langle Z \rangle}=\sigma_I\otimes \sigma_I$.

\paragraph{Fisher Information Matrix} Noting that the measurement outcome $y \in \{0,1\}$, we have
\begin{align}
    \fisherInfo_x(\boldsymbol{\theta})[i,j] &=  \sum_{y \in \{0,1\}} \frac{1}{p_{\mathrm{y}|\mathrm{x}}(y|x;\bm{\theta})} \frac{\partial p_{\mathrm{y}|\mathrm{x}}(y|x;\boldsymbol{\theta)}}{\partial \theta_i} \frac{ p_{\mathrm{y}|\mathrm{x}}(y|x;\boldsymbol{\theta)}}{\partial \theta_j} \\
    &= \frac{1}{p_{\mathrm{y}|\mathrm{x}}(0|x;\bm{\theta}) (1- p_{\mathrm{y}|\mathrm{x}}(0|x;\bm{\theta}))} \frac{\partial p_{\mathrm{y}|\mathrm{x}}(0|x;\boldsymbol{\theta)}}{\partial \theta_i} \frac{ p_{\mathrm{y}|\mathrm{x}}(0|x;\boldsymbol{\theta)}}{\partial \theta_j}
\end{align}
where in the second step, we have used the fact that $p_{\mathrm{y}|\mathrm{x}}(1|x;\bm{\theta})=1-p_{\mathrm{y}|\mathrm{x}}(0|x;\bm{\theta})$ and $\frac{\partial p_{\mathrm{y}|\mathrm{x}}(1|x;\boldsymbol{\theta)}}{\partial \theta_i} = - \frac{\partial p_{\mathrm{y}|\mathrm{x}}(0|x;\boldsymbol{\theta)}}{\partial \theta_i}$.

The FI matrix elements can also be expressed using the Rabi oscillation of a query $p_{\mathrm{rabi}}(x;\bm{\theta})$ as follows
\begin{equation}
    \fisherInfo_x(\boldsymbol{\theta})[i,j] = \frac{1}{1 - p_{\mathrm{rabi}}^2(x;\bm{\theta})} \frac{\partial p_{\mathrm{rabi}}(x;\bm{\theta})}{\partial \theta_i} \frac{ p_{\mathrm{rabi}}(x;\bm{\theta})}{\partial \theta_j}
\label{eq:fisher_info_rabi}    
\end{equation}

The FI matrices $\fisherInfo_x(\bm{\theta})$ for the different queries $x=(M,U,t)$ (Section~\ref{sec:cr_query_space}) depend on the parameterization of choice. We denote the FI matrix considering the parameterization of $\mathbf{\Lambda}$ as $\fisherInfo_x(\bm{\Lambda})$. Note that the Fisher information matrices $\fisherInfo_x(\mathbf{J})$ for the parameterization of $\mathbf{J}$ is related to the former through the jacobian of $\bm{\Lambda}$ with respect to $\mathbf{J}$.
\begin{equation}
    D_{\bm{\Lambda}, \mathbf{J}} = \begin{bmatrix} \frac{\mathrm{Re}(\beta_0)}{\omega_0} & -\frac{\mathrm{Re}(\beta_0)a_0}{|\beta_0|a_0} & -\frac{\mathrm{Im}(\beta_0)}{|\beta_0|^2} & \frac{\mathrm{Re}(\beta_1)}{\omega_1} & -\frac{\mathrm{Re}(\beta_1)a_1}{|\beta_1|a_1} & -\frac{\mathrm{Im}(\beta_1)}{|\beta_1|^2} \\ 
    \frac{\mathrm{Im}(\beta_0)}{\omega_0} & -\frac{\mathrm{Im}(\beta_0)a_0}{|\beta_0|a_0} & \frac{\mathrm{Re}(\beta_0)}{|\beta_0|^2} & \frac{\mathrm{Im}(\beta_1)}{\omega_1} & -\frac{\mathrm{Im}(\beta_1)a_1}{|\beta_1|a_1} & \frac{\mathrm{Re}(\beta_1)}{|\beta_1|^2} \\
    \frac{a_0}{\omega_0} & \frac{|\beta_0|}{\omega_0^2} & 0 & \frac{a_1}{\omega_1} & \frac{|\beta_1|}{\omega_1^2} \\
    \frac{\mathrm{Re}(\beta_0)}{\omega_0} & -\frac{\mathrm{Re}(\beta_0)a_0}{|\beta_0|a_0} & -\frac{\mathrm{Im}(\beta_0)}{|\beta_0|^2} & -\frac{\mathrm{Re}(\beta_1)}{\omega_1} & \frac{\mathrm{Re}(\beta_1)a_1}{|\beta_1|a_1} & \frac{\mathrm{Im}(\beta_1)}{|\beta_1|^2} \\ 
    \frac{\mathrm{Im}(\beta_0)}{\omega_0} & -\frac{\mathrm{Im}(\beta_0)a_0}{|\beta_0|a_0} & \frac{\mathrm{Re}(\beta_0)}{|\beta_0|^2} & -\frac{\mathrm{Im}(\beta_1)}{\omega_1} & \frac{\mathrm{Im}(\beta_1)a_1}{|\beta_1|a_1} & -\frac{\mathrm{Re}(\beta_1)}{|\beta_1|^2} \\
    \frac{a_0}{\omega_0} & \frac{|\beta_0|}{\omega_0^2} & 0 & -\frac{a_1}{\omega_1} & -\frac{|\beta_1|}{\omega_1^2} \\ \end{bmatrix} 
\end{equation}
where the $(i,j)$th element is given by $\partial \Lambda_j/\partial J_i$ with $a_{0,1}$ and $\beta_{0,1}$ as defined as in Eq.~\ref{eq:cr_hamiltonian_block_form}.

Note that $\fisherInfo_x(\bm{\Lambda})$ is of rank-$1$ for each query and takes a block form of $\begin{bmatrix} \fisherInfo_0 & 0 \\ 0 & 0 \end{bmatrix}$ for $U=\sigma_I \otimes \sigma_I$ and $\begin{bmatrix} 0 & 0 \\ 0 & \fisherInfo_1 \end{bmatrix}$ for $U=\sigma_X \otimes \sigma_I$. This indicates that queries involving $U=\sigma_I \otimes \sigma_I$ are informative about the Hamiltonian parameters $(\omega_0,\delta_0,\phi_0)$ and those involving $U=\sigma_X \otimes \sigma_I$ are informative about $(\omega_1,\delta_1,\phi_1)$.

\subsubsection{In Presence of Noise Sources and Nonidealities}
In Section~\ref{sec:readout_noise}, we modeled the effect of different noise sources and nonidealities on the quantum system. In particular, we discussed the effect of imperfections in control in Section~\ref{sec:imperfect_pulse_shaping}, effect of decoherence in Section~\ref{sec:noise_sources_decoherence} and how the observed measured outcome is subject to readout noise in Section~\ref{sec:readout_noise}. We consider the two-qubit decoherence model from Section~\ref{sec:noise_sources_decoherence}, and the readout noise model of a bit-flip channel based on binary classification from Section~\ref{sec:readout_noise}. We denote the noisy observed measurement outcome as $\tilde{y}$ and the hidden measurement outcome before the effect of the bit-flip channel as $y$. 

\paragraph{Likelihood} The likelihood function in the presence of the noise sources of readout noise, imperfect-pulse shaping, and decoherence, is then given by
\begin{align}
    p_{\tilde{\mathrm{y}}|\mathrm{x}}(\tilde{y}|x;\bm{\theta}) &= p_{\tilde{\mathrm{y}}|\mathrm{y}}(\tilde{y}|y) \sum_{z \in \{0,1\}} \left[ \left( 1 - p_d(t) \right)p_{\mathrm{yz}|\mathrm{x}}\left(yz| \left(M,U,t+\Delta t_{\text{eff}}(\bm{\theta})\right);\bm{\theta}\right) + \frac{1}{4}p_d(t)\right] \\
    &= \left(1-p_d(t)\right) \left[(1 - r_{1-\tilde{y}}) p_{\mathrm{y}|\mathrm{x}}(\tilde{y}| (M,U,t+\Delta t_{\text{eff}}(\bm{\theta}));\bm{\theta}) + r_{\tilde{y}} p_{\mathrm{y}|\mathrm{x}}(1-\tilde{y}| (M,U,t+\Delta t_{\text{eff}}(\bm{\theta}));\bm{\theta})\right] \nonumber \\ & + \frac{1}{2}p_d(t)\left(1 - r_{1-\tilde{y}} + r_{\tilde{y}} \right)
\end{align}
with the probability of the two-qubit string $yz$ given by
\begin{equation}
    p_{\mathrm{yz}|\mathrm{x}}\left(yz| \left(M,U,t)\right);\bm{\theta}\right) = \left|\braketExp{yz}{M e^{-iH(\boldsymbol{\theta}) t ) }U}{00}\right|^2 
\end{equation}
and the probability $p_{\mathrm{y}|\mathrm{x}}(y|x)$ given by Eq.~\ref{eq:borns_rule} (the noiseless case). In the expressions above, we have used the tuple representation of the query $x$, $p_d(t)$ which is the depolarization probability associated with the two-qubit decoherence model as discussed in Section~\ref{sec:noise_sources_decoherence} and $(r_0,r_1)$ which are the readout noise parameters as introduced in Section~\ref{sec:readout_noise}.

\paragraph{Fisher Information Matrix} The FI matrix for a given query $x$ is given by
\begin{align}
    \fisherInfo_x(\boldsymbol{\theta})[i,j] = \frac{1}{p_{\tilde{\mathrm{y}}|\mathrm{x}}(0|x;\bm{\theta}) (1- p_{\tilde{\mathrm{y}}|\mathrm{x}}(0|x;\bm{\theta}))} \frac{\partial p_{\tilde{\mathrm{y}}|\mathrm{x}}(0|x;\boldsymbol{\theta)}}{\partial \theta_i} \frac{ p_{\tilde{\mathrm{y}}|\mathrm{x}}(0|x;\boldsymbol{\theta)}}{\partial \theta_j}
\end{align}
where
\begin{equation}
    \frac{\partial p_{\tilde{\mathrm{y}}|\mathrm{x}}(0|x;\boldsymbol{\theta)}}{\partial \theta_i} = (1 - r_0 - r_1) \left( 1 - p_d(t) \right) \frac{\partial p_{\mathrm{y}|\mathrm{x}}(0|x;\boldsymbol{\theta)}}{\partial \theta_i}
\end{equation}
and
\begin{equation}
    \frac{\partial p_{\mathrm{y}|\mathrm{x}}(0|x;\boldsymbol{\theta)}}{\partial \theta_i} = 
    \begin{cases} 
        \frac{\partial  p(y=0| (M,U,t+\Delta t_{\text{eff}}(\bm{\theta}));\bm{\theta}) }{\partial \theta_i}, & \theta_i \notin \{\omega_0, \omega_1 \} \\ 
        \frac{\partial  p(y=0| (M,U,t+\Delta t_{\text{eff}}(\bm{\theta}));\bm{\theta}) }{\partial (\theta_i \Delta t_{\text{eff}}(\bm{\theta}))} \frac{\partial (\theta_i \Delta t_{\text{eff}}(\bm{\theta}))}{\partial \theta_i}, & \theta_i \in \{\omega_0, \omega_1 \}
    \end{cases}
\end{equation}
Note that special attention has to be given when taking the derivative with respect to $\omega_{0,1}$ as $\Delta t_{\text{eff}}(\bm{\theta})$ (Section~\ref{sec:imperfect_pulse_shaping}) actually only has a dependence on these two components in $\bm{\Lambda}$ and appears in the effective evolution time $t$ with a prefactor of $\omega_{0,1}$ in the likelihood (see Eq.~\ref{eq:expressions_likelihood_cr_noiseless}).

\section{Computational Details of Query Optimization} \label{app_sec:query_opt}
In Section~\ref{sec:results_datasets_expt}, we pointed out that the number of shots available for each query in the experimental datasets collected from the IBM Quantum devices are limited. In this section of the Appendix, we describe how the query optimizations for HAL-FI (Eq.~\ref{eq:query_optimization_variance_of_params}) and HAL-FIR (Eq.~\ref{eq:query_optimization_testing_error}) are solved under shot constraints for each query. We then describe how the computational cost of query optimization can be reduced through uncertainty filtering of the query space.

\subsection{Different Query Optimizations and Strategies for Handling Query Constraints}
We describe how constraints can be handled for the query optimization  (Eq.~\ref{eq:query_optimization_variance_of_params}) in HAL-FI (Algorithm~\ref{algo:QOA_FI}) but the same approach can also be used for HAL-FIR. Let us consider the $i$th round in active learning. The Hamiltonian parameter estimate in this round is $\bm{\hat{\theta}}^{(i)}$. Let us denote the number of shots available for each query $x \in \querySpace$ in the $i$th round as $N_\text{shots}^{(i)}(x)$. The number of shots available before learning has started is then $N_\text{shots}^{(0)}(x)$. Let the total number of shots that have already been made against query $x$ by the $i$th round (inclusive) be denoted as $N_\text{tot}^{(i)}(x)$. We will denote the total number of shots made over all queries inputted to the oracle by the $i$th round (inclusive) as $N_\text{tot}^{(i)}$.

We can frame the query optimization problem under shot constraints in two different ways, motivated by the asymptotic optimal query distribution associated with HAL-FI:
\begin{equation}
    q^\star = \arg \min_{q \in \mathcal{P}(\mathcal{Q})} \Tr(\fisherInfo_q^{-1}(\boldsymbol{\theta}))
    \label{app_eq:query_optimization}
\end{equation}
where $\mathcal{P}(\mathcal{Q})$ is the family of all probability distributions over the specified query space $\mathcal{Q}$. As discussed in Section~\ref{sec:activelearning}, we cannot solve for this distribution in practice as this requires us to have access to the true parameters $\bm{\theta}^\star$, and hence we solve for a sub-optimal query distribution $q^{(i)}(\hat{\bm{\theta}}^{(i)})$ given the current parameter estimates $\hat{\bm{\theta}}$. During active learning, we can solve for $q^{(i)}(\hat{\bm{\theta}}^{(i)})$ by viewing it as the query distribution over all the queries that have inputted to the oracle (i.e., including from previous rounds) or as the query distribution associated with the current  $i$th batch of queries that will be issued. These two viewpoints lead us to two different approaches of how the shot constraints are handled and how the queries to be inputted to the oracle are sampled from the query distribution. 

Firstly, consider the following query optimization problem
\begin{align}
    q^{(i)} &= \arg \min \limits_{q} \Tr(\fisherInfo_q^{-1}(\hat{\boldsymbol{\theta}}^{(i)})) \\ &\text{ subject to} \sum \limits_{x \in \querySpace^{(i)}} q(x) = 1, \text{ and} \frac{N_\text{tot}^{(i-1)}(x)}{N_\text{tot}^{(i)}} \leq q(x) \leq \min\left\{1,\frac{N_\text{shots}^{(i)}(x)}{N_\text{tot}^{(i)}}\right\} \forall x \in \querySpace^{(i)}
\label{eq:qopt_query_cosntraints_lower_bound}
\end{align}
with corresponding sampling of queries in that batch as:
\begin{equation}
    \text{Sample } N_{b} \text{ queries } X_q^{(i)} \text{ from } \querySpace^{(i)} \text{ w.p. } q_b^{(i)} 
\end{equation}
where $q_b^{(i)}(x) = q^{(i)}(x) - \frac{N_\text{tot}^{(i-1)}(x)}{N_\text{tot}^{(i)}}$. The query optimization in this case considers all the queries that have been sampled from earlier batches during active learning. The query distribution $q^{(i)}$ is then the distribution over all the queries made so far and $q_b^{(i)}$ is the query distribution for the batch to be issued. Further, the Fisher information matrix associated with $q^{(i)}$ is guaranteed to be invertible but that associated with $q_b^{(i)}$ may be non-invertible. This suggests that the outcomes of queries made in this round of the active learning procedure may not be informative about all the Hamiltonian parameters. 

In order to ensure that the query distributions for each batch are informative about all the Hamiltonian parameters, one may alternately solve the following query optimization problem
\begin{align}
    q^{(i)} &= \arg \min \limits_{q} \Tr(\fisherInfo_q^{-1}(\hat{\boldsymbol{\theta}}^{(i)})) \\ &\text{ subject to} \sum \limits_{x \in \querySpace^{(i)}} q(x) = 1, \text{ and } 0 \leq q(x) \leq \min\left\{\frac{N_\text{shots}^{(i)}(x)}{N_b^{(i)}},1 \right\}, \forall x \in \querySpace^{(i)}
    \label{eq:qopt_query_cosntraints_upper_bound}
\end{align}
where $N_b^{(i)}$ is the size of the batch of queries being issued to the oracle in the $i$th round. Queries of the issued batch are then sampled as 
\begin{equation}
    \text{Sample } N_{b} \text{ queries } X_q^{(i)} \text{ from } \querySpace^{(i)} \text{ w.p. } q^{(i)} 
\end{equation}
where $q^{(i)}$ is now the query distribution for each batch. This query optimization can be viewed as a greedy approach of query selection. The Fisher information matrix associated with $q^{(i)}$ is guaranteed to be invertible and hence informative about all the Hamiltonian parameters. We thus solve the query optimization considering shot constraints on each query according to Eq.~\ref{eq:qopt_query_cosntraints_upper_bound} for our application.

As we sample queries randomly according to $q^{(i)}$ and not proportional to $q^{(i)}$, the resulting $X_q$ might not satisfy query constraints exactly and an additional pruning step is required. Moreover, to reduce the computational cost of the query optimization solve, we consider a subset of queries $\mathcal{Q}^{(i)}_{\text{filtered}} \subset \mathcal{Q}^{(i)}$ for which shots are still available.

\begin{figure}[H]
	\centering
    \xincludegraphics[scale=0.35, label=\textbf{(a)}]{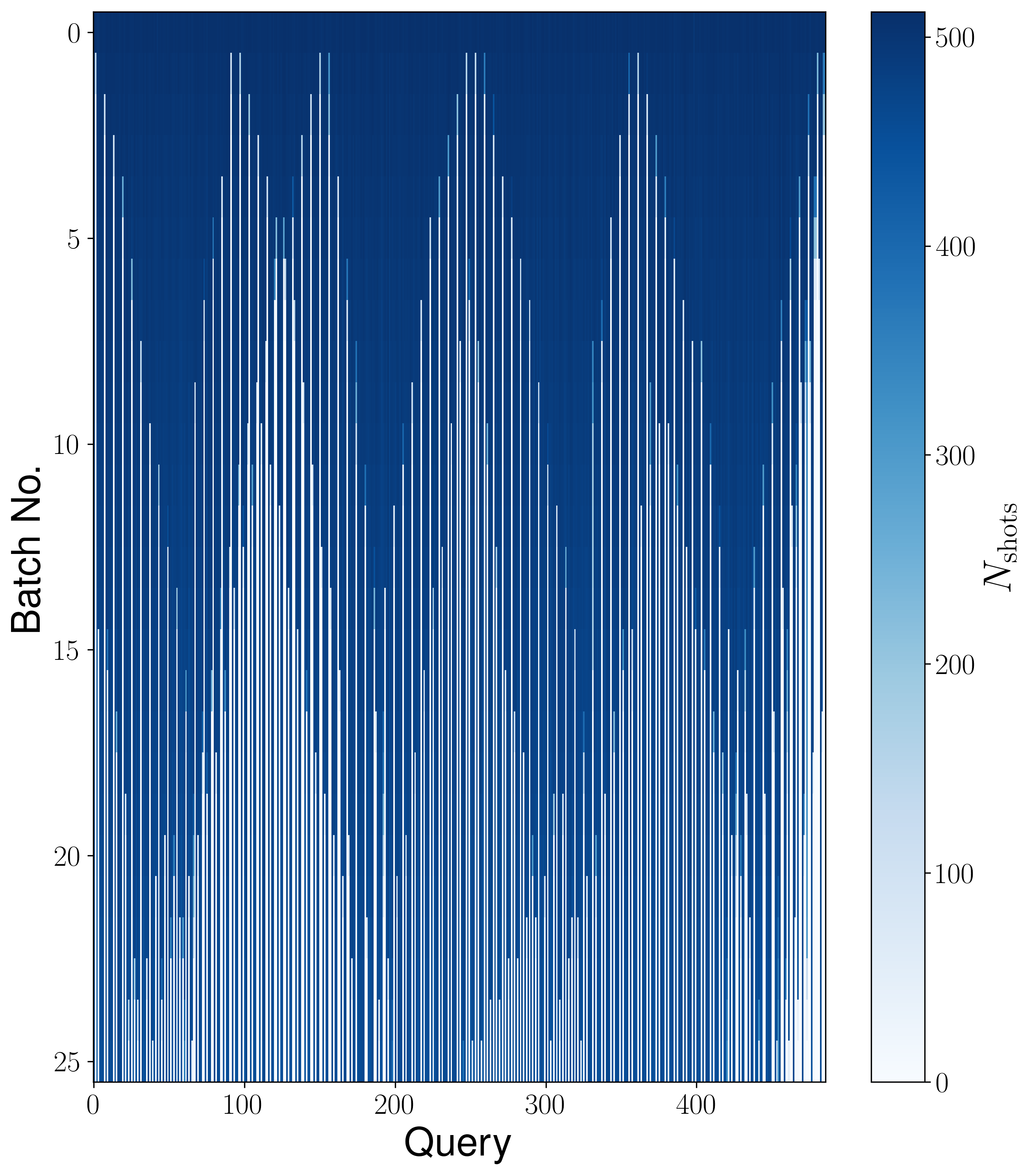}
    \hspace{4em}
    \xincludegraphics[scale=0.35, label=\textbf{(b)}]{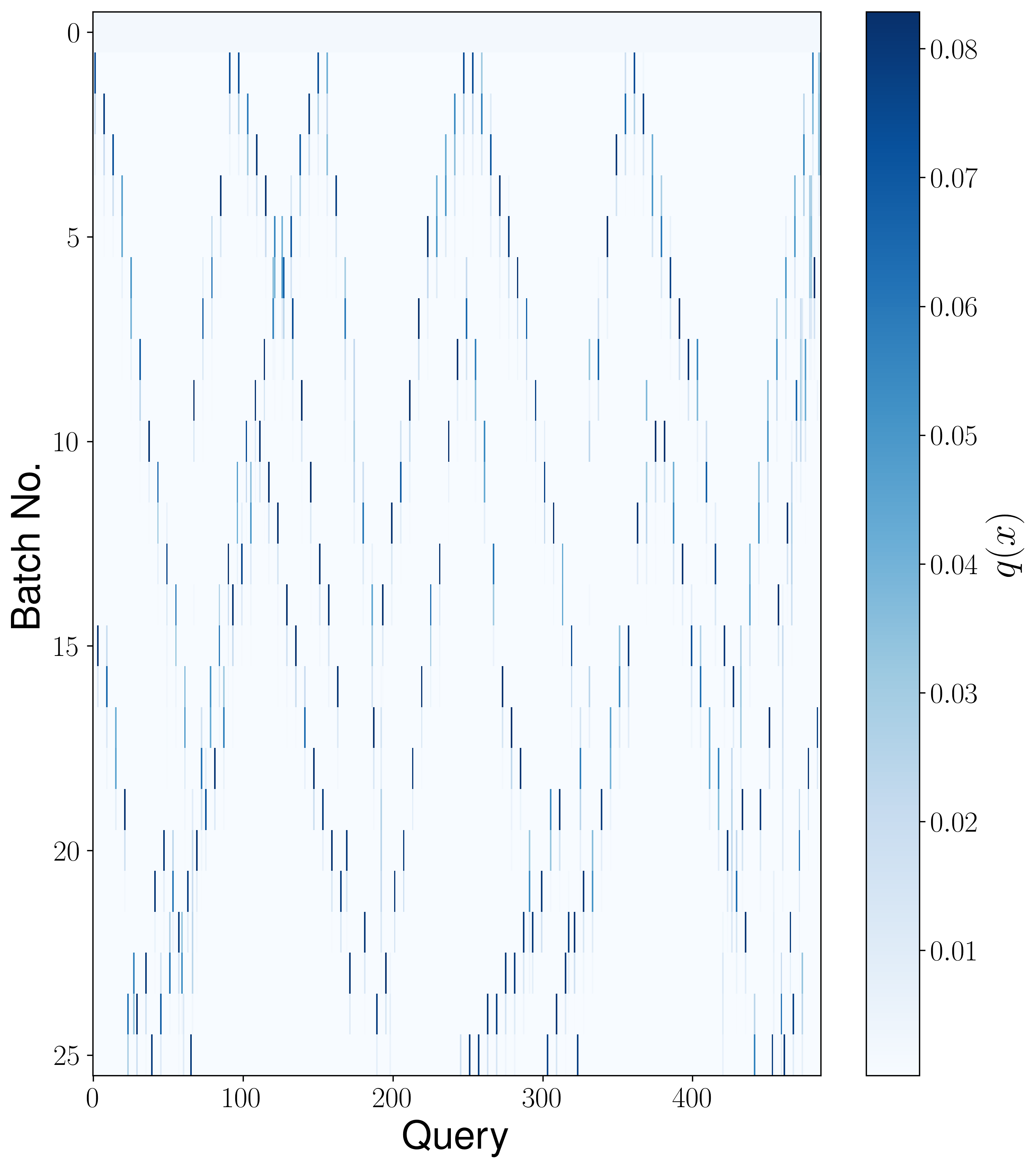}
    \caption{Visualization of queries being selected for each batch during a particular run of HAL-FI for $N_0=2000$ and $25$ batches of size $N_b=5000$. In (a), we plot the available number of shots for each query after batches of queries are made during active learning. In (b), we plot the query distribution for each batch during active learning. The number of shots available for each query before learning starts is $N_\text{shots}^{(0)}(x)=512\,\forall x$. Half the total number of shots available in this dataset is exhausted by the end of learning. Note that the different parameter values considered for HAL-FI are for stress testing the query constraints' handling procedure and are not tuned for the HAL-FI algorithm.}
	\label{fig:handling_query_constraints}
\end{figure}

In Algorithm~\ref{algo:algo_handling_query_constraints}, we summarize the steps taken to handle query constraints during query optimization for HAL-FI. It takes the inputs of the query space $\querySpace^{(i)}$ of the current $i$th round of active learning, number of shots available for each query and the size of the batch of queries to be issued. The inputted query space is first filtered by retaining only those queries for which shots are available. The resulting filtered query space is denoted by $\mathcal{Q}^{(i)}_{\text{filtered}}$, and we then compute the query distribution $q^{(i)}$ according to Eq.~\ref{eq:qopt_query_cosntraints_upper_bound}. We then sample queries for the batch $X_q$ according to this distribution after mixing and with incorporation of a pruning step. Note that lines 4 and 5 were also used in the original query optimization algorithms of HAL-FI (Algorithm~\ref{algo:QOA_FI}) and HAL-FIR (Algorithm~\ref{algo:QOA_FIR}) to encourage exploration. An illustration of the operation of the query optimization and handling of query constraints is shown in Figure~\ref{fig:handling_query_constraints} for a particular run of the HAL-FI learner.

\begin{algorithm}[H]
	\caption{Handling query constraints during query optimization in HAL-FI}
	\small
	\textbf{Input}: Current query space $\querySpace^{(i)}$, number of shots available for each query $N_{\text{shots}}^{(i)}(x)$, size of batch of queries requested $N_b^{(i)}$, total number of queries made so far $N^{(i-1)}_{tot}$  \\
	\textbf{Output}: Query set $X_q$ of size $N_b^{(i)}$
	\begin{algorithmic}[1]
		\State Filter $\mathcal{Q}^{(i)}$ by keeping only queries for which shots are available: $\mathcal{Q}^{(i)}_{\text{filtered}} = \{x \in \mathcal{Q}^{(i)} |  N_{\text{shots}}^{(i)}(x) > 0 \}$
		\State Obtain HAL-FI query distribution $q^{(i)}$ by solving the query optimization of Eq.~\ref{eq:qopt_query_cosntraints_upper_bound} with input of $\mathcal{Q}^{(i)}_{\text{filtered}}$
		\State Obtain uniform distribution over filtered query space: $p_U = 1/|\mathcal{Q}^{(i)}_{\text{filtered}}|$
		\State Set mixing coefficient: $\mu = 1 - 1/|N^{(i-1)}_{tot}|^{1/6}$
		\State Modify query distribution: $q^{(i)} = \mu q^{(i)} + (1 - \mu) p_U$
		\State Sample $X_q$ from $\mathcal{Q}^{(i)}_{\text{filtered}}$ according to $q^{(i)}$
		\State Prune queries from $X_q$ that cannot be made and randomly assign valid queries
		\State \Return $X_q$
	\end{algorithmic}
	\label{algo:algo_handling_query_constraints}
\end{algorithm}

\subsection{Uncertainty Filtering of Query Space} \label{app_sec:uncertainty_filtering}
\begin{figure}[H]
    \centering
    \includegraphics[scale=0.4]{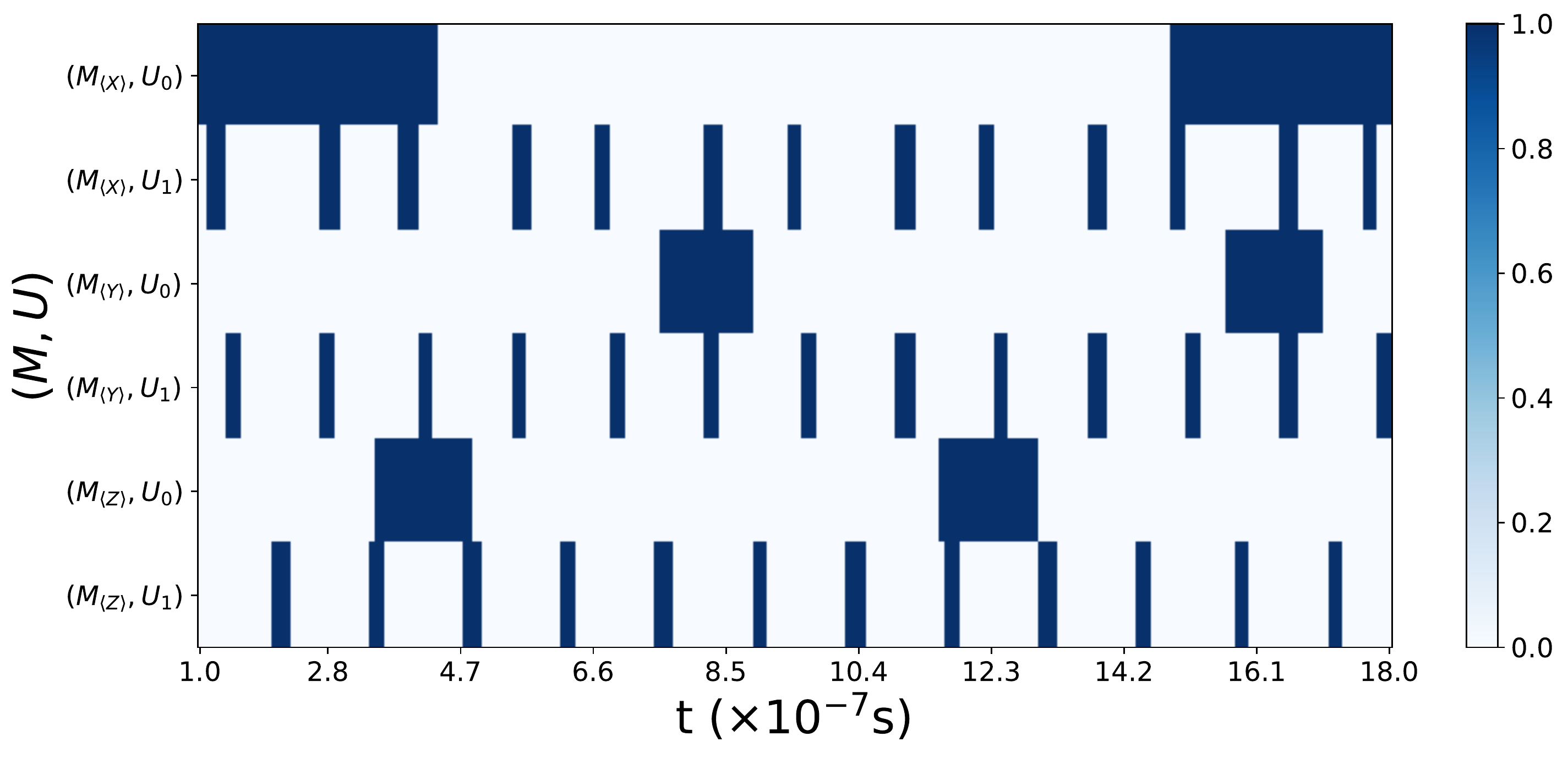}
    \caption{Uncertainty filtering of $\querySpace$ as defined in Section~\ref{sec:cr_query_space} with $\timeSpace$ set to be a sequence of $243$ linearly equispaced evolution times in $[10^{-7}, 18\times 10^{-7}]s$. The The x-axis corresponds to the system evolution times $t \in \timeSpace$. The y-axis indicates the different combinations of measurement operators and preparation operators available for each query in $\querySpace$. The different preparation operators are denoted as $U_0 = \sigma_I\sigma_I$ and $U_1=\sigma_X\sigma_I$. We consider the $\bm{\theta}^\star$ corresponding to IBM Quantum device D \textit{ibmq\_boeblingen} under drive configuration 2, with the different noise sources of readout noise, imperfect pulse-shaping, and decoherence being accounted for. Queries colored as dark blue are retained in the filtered query space and rest is filtered out.}
    \label{fig:uncertainty_filtering_ibmq_boeb}
\end{figure}

Uncertainty filtering becomes a crucial step during query optimization when we consider HAL-FI with an adaptively growing query space. We noted in Section~\ref{sec:active_learner} that the computational cost of query optimization (Eq.~\ref{eq:query_optimization_variance_of_params}) scales as $\mathcal{O}(n_{\mathcal{Q}}^2 m^3 + n_{\mathcal{Q}}m^4 + m^5)$ where $n_{\mathcal{Q}} = |\querySpace|$ is the number of queries in the query space, and $m$ is the length of the parameter vector $\bm{\theta}$. This computational cost is alleviated through filtering of the query space based on entropy $S(x)$. The entropy of the different queries can be computed from the model probability expressions available to us (see Section~\ref{app_sec:details_cr_hamiltonians}) given the current Hamiltonian parameter estimate $\hat{\bm{\theta}}$. The filtered query space based on entropy is determined as follows:
\begin{equation}
    \querySpace_{S} = \{ x | x \in \querySpace, S(x) > \tau \times \max_{x' \in \querySpace} S(x') \}
    \label{eq:maximum_entropy_query_space}
\end{equation}
where we set the threshold $\tau = 0.95$ i.e., we only consider queries with entropy that is at least $0.95$ times the highest entropy. This value of $\tau$ was chosen to ensure that at most only half of the queries are retained after uncertainty filtering of the query space. An illustration of uncertainty filtering of a query space is given in Figure~\ref{fig:uncertainty_filtering_ibmq_boeb}.

\section{Estimation Procedure for Learning Cross-Resonance Hamiltonians} \label{app_sec:est_procedure_cr_ham}
In this section of the Appendix, we discuss in detail the estimation procedure of Section~\ref{sec:estimation_procedure} that is used for solving the MLE problem of Eqs.~\ref{eq:mle_hl_noise},\ref{eq:mle_hl_noise_GMM_readout} for learning cross-resonance Hamiltonians (Eq.~\ref{eq:cr_hamiltonian_block_form}). This estimation procedure is used in conjunction with both the passive learner and HAL-FI/HAL-FIR. We consider the experimental setup discussed in Section~\ref{sec:CR_Gate_Model_and_Setup}. We also discuss how this estimation procedure can be improved or extended to other Hamiltonians. 

Let us recall the notation introduced in Section~\ref{sec:active_learner} and Section~\ref{algo:MLE_Estimation_Procedure}. The different rounds of active learning are indexed by $i \in [i_{\text{max}}]$. In each round of active learning, we use an estimation procedure divided into multiple steps. We index each of these fractional steps by $k$. The Hamiltonian parameter estimate at the $k$th fractional step in the $i$th round of AL will then be denoted by $\hat{\bm{\theta}}^{(i,k)}$ with the parameter estimate in the $i$th round at the end of the estimation procedure denoted simply by $\hat{\bm{\theta}}^{(i)}$. The training examples available at the $i$th round is given by $(X^{(i)},Y^{(i)})$.

We divide the estimation procedure into two main parts: (i) initial estimation and (ii) maximum-likelihood estimation (MLE).

\subsection{Initial Estimation}
As mentioned in Section~\ref{sec:estimation_procedure}, the first step of our estimation procedure involves frequency estimation followed by a nonlinear regression solve combined with a gradient descent procedure on the Rabi oscillations inferred from data (Eq.~\ref{eq:rabi_oscillations_mle}). We will denote this by $\hat{p}_\text{rabi}(t)$.

Frequency estimation is typically carried out through the Fourier transform (implemented as Fast Fourier Transform (FFT) \cite{dutt1993fast}). We note this in general assumes the input of a periodic signal which translates into a limited frequency resolution of $F_s/T_p$ where $F_s$ is the sampling frequency and $T_p$ is the period. In order to successfully calculate the frequency, we must set the $\timeSpace$ to be at least have the duration of $T_p$ by the minimum-frequency criterion (defined in Section~\ref{sec:results_performance}). Standard FFTs then allow us to calculate frequency amplitudes at fixed intervals in the spectrum through
\begin{equation}
    F(k) = \sum_{n=0}^{N_t - 1} e^{-i 2\pi \frac{\nu}{N_t}kn} \hat{p}_\text{rabi}(t)
    \label{eq:standard_fft}
\end{equation}
where we have denoted the Fourier coefficients by $F(k)$ and $i=\sqrt{-1}$ is the imaginary unit. However, the frequencies of the Rabi oscillations are not known apriori and hence the $\timeSpace$ may not be adequately set. We would like in practice for the estimation procedure to succeed regardless of the choice of $\timeSpace$. We now detail how the standard FFT may be modified in such a case.
\paragraph{Modification for $\timeSpace$ not satisfying minimum-frequency criteria:}
We could sample frequency amplitudes at the intermediate half steps of $\nu / 2N_t$ by multiplying $\hat{p}_\text{rabi}(n)$ by rotating exponential frequency and then computing a second FFT:
\begin{equation}
    G(k) = F\left(k + \frac{1}{2}\right) = \sum_{n=0}^{N_t - 1} \left[e^{-i2\pi \frac{\nu}{2N_t}kn}\right] e^{-j2\pi \frac{\nu}{N_t}kn}
    \label{eq:shifted_fft}
\end{equation}
where $G(0)$ is the Fourier coefficient at $\nu/2N_t$, $G(1)$ is the Fourier coefficient at $3\nu/2N_t$, etc. However, this only gives us an improvement in resolution from $\nu/N_t$ to $\nu/2N_t$. One can more generally solve the following \textit{normal} equations based on regression of the model Rabi oscillations:
\begin{align}
    \begin{bmatrix}
        \sum_t \cos^2(\omega t) & \sum_t \cos(\omega t) \sin(\omega t) & \sum_t \cos(\omega t) \\
        \sum_t \sum_t \cos(\omega t) \sin(\omega t) & \sin^2(\omega t) & \sum_t \sin(\omega t) \\
        \sum_t \cos(\omega t) & \sum_t \sin(\omega t) \sin(\omega t) & \sum_t 1
    \end{bmatrix}
    \begin{bmatrix}
    A \\ B \\ C
    \end{bmatrix}
    =
    \begin{bmatrix}
        \text{Re}(\mathcal{F}(\omega)) \\ - \text{Im}(\mathcal{F}(\omega)) \\ \mathcal{F}(0)
    \end{bmatrix}
    \label{eq:normal_equations_fft}
\end{align}
where $\mathcal{F}(\omega) = \sum_t p_\text{rabi}(t) e^{-j \omega t}$ is the discrete Fourier transform.

\paragraph{Extension to other multi-parameter Hamiltonian models} In general multi-parameter Hamiltonians, the intial estimation procedure must be modified to ensure that all the frequency components in a Rabi oscillation corresponding to $(M,U)$ can be faithfully extracted. The model to be used is then assume that the Fourier series has up to $K$ modes. Further, normal equations for the same can be setup. This has been employed in Bayesian spectral analysis \cite{bretthorst2013bayesian} and non-stationary time-series estimation  \cite{lange2021fourier,shea2021extraction}.

\subsection{Maximum Likelihood Estimation}
The initial estimation procedure described above produces the estimate of $\hat{\bm{\theta}}^{(i,0)}$ which is used as an initial guess to the MLE. The expectation is that the initial estimation produces an estimate that lies in the same convex basin as the global minimum of the MLE. This allows for a more localized search to be carried out and a stochastic gradient descent procedure should allow us to jump out of any smaller local minima here if present. 

We solve the MLE \ref{eq:general_MLE} through a combination of SGD applied on different parameterizations and the quasi-Newton method for further refinement. Addition of the latter step helps us in saving computationally expensive hyperparameter tuning that is required. 

\begin{enumerate}
    \item MLE solve considering the $\bm{\Lambda}$ parameterization and learning rate of $\eta_{\bm{\Lambda}}^i$ using the input of $\hat{\bm{\theta}}^{(i,0)}$. This returns the output of $\hat{\bm{\theta}}^{(i,1)}$.
    \item MLE solve considering the $\mathbf{J}$ parameterization and learning rate of $\eta_{\mathbf{J}}^i$ using the input of $\hat{\bm{\theta}}^{(i,1)}$. This returns the output of $\hat{\bm{\theta}}^{(i,2)}$.
    \item MLE solve considering the $\mathbf{J}$ parameterization using LFBGS-B using the input of $\hat{\bm{\theta}}^{(i,2)}$. This returns the output of $\hat{\bm{\theta}}^{(i,3)}$.
\end{enumerate}

We set the learning rate $\eta^{i}$ for ADAM \cite{kingma2014adam} in the $i$th round of active learning according to the number of queries already made. We consider the learning rate to be $\eta^{i} \propto \frac{1}{\sqrt{|X^{(i)}|}}$ i.e., the learning rate is reduced inversely to the square root of the number of training examples. This ensures a more localized search as we progress in the learning. We consider $\eta^{0}=10^{-3}$. We found that carrying out step 2 after step 1 gave us more accurate estimates of $\hat{\theta}$ than just carrying out step 1. Moreover, after the first few rounds of HAL, we can skip steps 1 and 2. We can carry out step 3 directly using an initial condition of $\hat{\theta}^{(i,0)}$ from initial estimation or $\hat{\theta}^{(i-1)}$ from the previous round. 

\subsection{Energy Landscapes of Negative Log-Likelihood Loss for Cross-Resonance Hamiltonian}
\begin{figure}[H]
	\centering
    \xincludegraphics[scale=0.25, label=\textbf{(a)}]{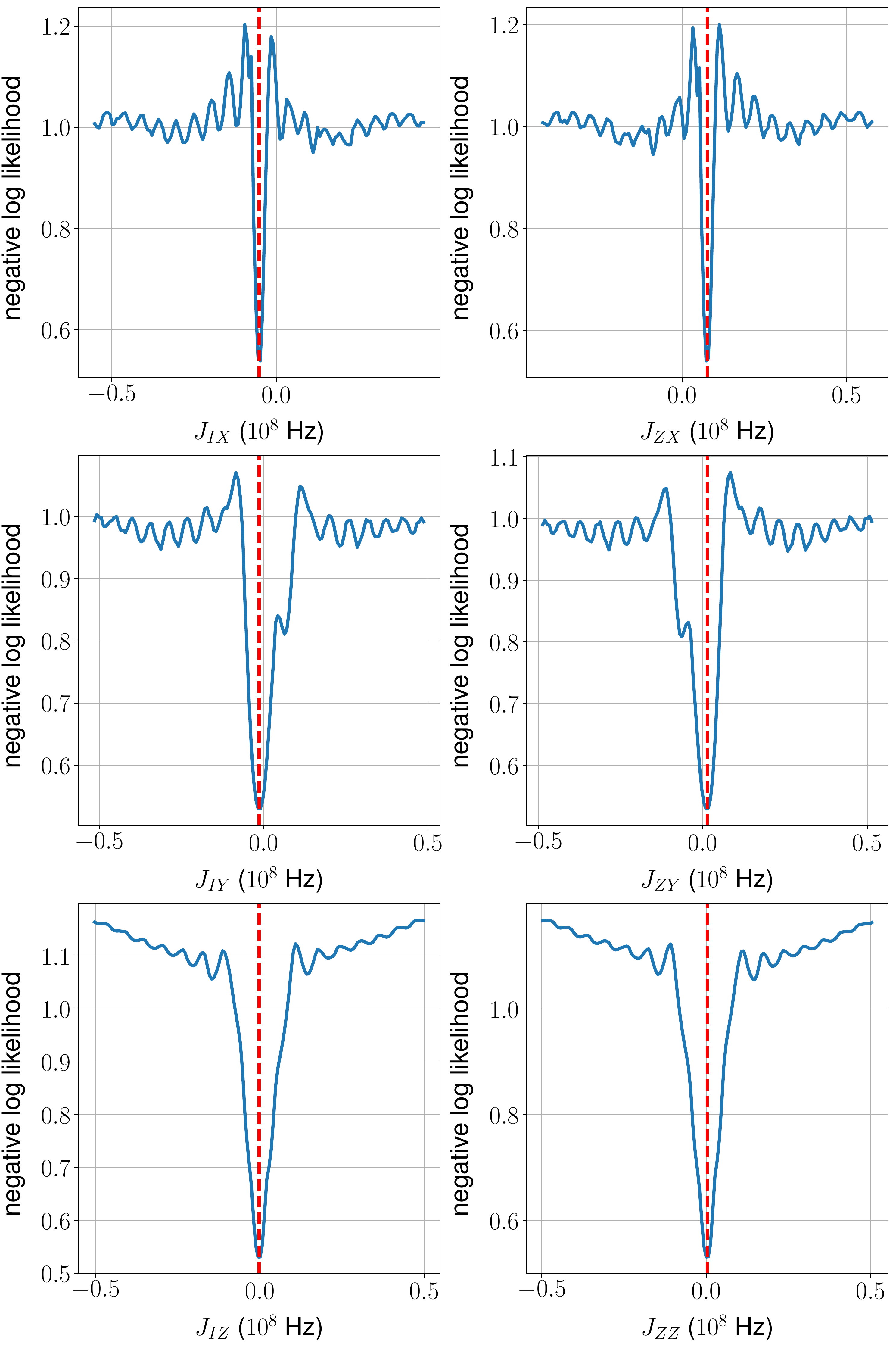}
    \hspace{4em}
    \xincludegraphics[scale=0.25, label=\textbf{(b)}]{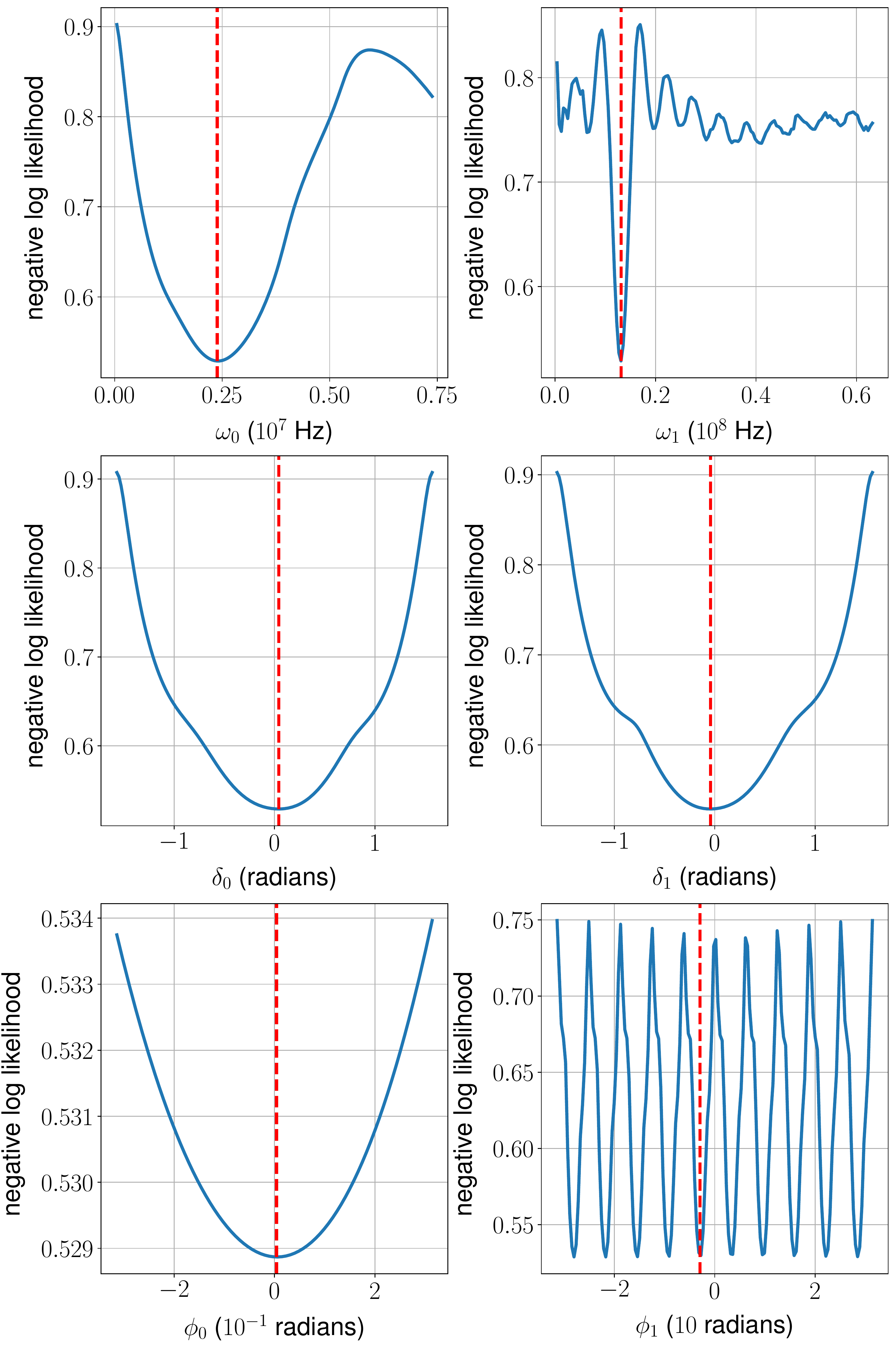}
    \caption{Slices of energy landscapes of the log-likelihood loss function along the different parameter components considering experimental data collected from IBM Quantum device D \textit{ibmq\_boeblingen} under drive configuration 4 (a) using parameterization $\mathbf{J}$, and (b) using parameterization $\bm{\Lambda}$. In each slice of $\theta_i$ (x-axis), we fix the values of the other components to that obtained through estimation and evaluate the negative log-likelihood loss (y-axis) by changing the value of $\theta_i$. We indicate the Hamiltonian parameter estimate $\hat{\theta}_i$ as obtained through our estimation procedure by a dashed red line.}
	\label{fig:results_energy_landscapes}
\end{figure}

We ascertain the efficacy of the estimation procedure by visualizing the energy landscapes of the negative log-likelihood loss function (Eq.~\ref{eq:mle_hl_noise}) and inspecting the location of the $\hat{\bm{\theta}}$ in the landscape. In Figure~\ref{fig:results_energy_landscapes}, we plot the energy landscape obtained from an experimental dataset, for the two different parameterizations $\mathbf{J}$ and $\bm{\Lambda}$. The energy landscapes indicate the nonlinear and non-convex nature of the MLE of Eq.~\ref{eq:mle_hl_noise}. These energy landscapes also indicate why solving the MLE in the parameterization of $\bm{\Lambda}$ using ADAM is carried out before solving the MLE in the parameterization of $\mathbf{J}$. The slices along specific components of $\bm{\Lambda}$ display more convex like nature than those along components of $\mathbf{J}$. It should also be noted that there is a global minimum present in the energy landscapes which we are able to identify using our estimation procedure. 

\subsection{Incorporating uncertainty from shot noise}
The inferred Rabi oscillations $p_{\mathrm{rabi}}(t)$ used for estimation are sensitive to the number of shots made for a given query $x=(M,U,t)$. This variability in the Rabi oscillations leads to a variability in the estimates of $\hat{\bm{\theta}}^{(i,0)}$ produced. This variability is particularly high during the initial rounds of HAL when there are only a few shots of each query present in the set of training examples. In order to accurately account for this variability and hence include the uncertainty in our estimates of the Hamiltonian parameters, we consider the following procedure. Let us consider the initial estimation of Algorithm~\ref{algo:MLE_Estimation_Procedure} as the procedure applied on a particular realization of the Rabi oscillations. We construct $n_\mathrm{rep}$ realizations of the Rabi oscillations considering the observed Rabi oscillation data and sampling according to the binomial distribution associated with the number of shots for each query. For each of these realizations, we obtain frequency estimates of $\omega_{0,1}$ through the initial estimation as described above. We then fit the parametric distributions of log-normal distributions to frequency estimates from each each realization. We then continue with the initial estimation procedure and MLE considering each of these realizations.

Other ways of incorporating uncertainty in the Hamiltonian parameter estimates during the estimation procedure would be to adopt a Bayesian learning framework or stochastic process regression (e.g., Gaussian process regression). This is left for future work.

\section{Learned Hamiltonian Parameters and Learning Error on IBM Quantum Devices} \label{app_sec:details_results_HL}
In this section of the Appendix, we give a summary of the estimated cross-resonance Hamiltonian parameters on different IBM Quantum devices not already discussed in Section~\ref{sec:Results} and lend further support to the performance of different learners on the $20$-qubit IBM Quantum device D \textit{ibmq\_boeblingen} under drive configuration 2 (Table~\ref{sec:results_params_ham_noise_ibmq}). 

\subsection{Summary of Model Parameters on IBM Quantum Devices}
In Section~\ref{sec:results_params_ham_noise_ibmq}, we described the estimated Hamiltonian parameters and the noise model parameters for the IBM Quantum device D \textit{ibmq\_boeblingen}. Here, we give a similar summary for the other IBM Quantum devices (A, B, and, C). 

Considering the entire experimental datasets collected for each of these IBM Quantum devices (Section~\ref{sec:quantum_devices}) as training data, we compute the Hamiltonian parameters using our estimation procedure (Section~\ref{sec:estimation_procedure}), and that of the different noise sources (Section~\ref{sec:noise_sources}). We summarize these parameters for the different devices in Table~\ref{tab:device_ham_params_noise_summary}. These Hamiltonian parameters serve as approximations of the true parameters $\mathbf{J}^\star$. Here, the readout noise parameters $(r_0, r_1)$ are given to indicate the amount of readout noise possible in these devices and serve as a proxy for the conditional distributions of the readout given the measurement outcome used in the final MLE (Eq.~\ref{eq:mle_hl_noise}). The time offset $\Delta t_{\text{eff},i}$ introduced due to imperfect control is also specified and where the subscript $i$ indicates dependence on the preparation operator $U_0 = \sigma_I \sigma_I$ and $U_1 = \sigma_X \sigma_I$.

\begin{table}[H]
\centering
\small
\begin{tabular}{|l|l|ll|ll|}
\hline
\multirow{2}{*}{\textbf{Device}} & \multirow{2}{*}{\begin{tabular}[c]{@{}c@{}}\textbf{Drive} \\ \textbf{Config.}\end{tabular}} & 
\multicolumn{2}{c|}{\textbf{Hamiltonian Parameters [$\mathbf{\times10^6s^{-1}}$]}} & 
\multicolumn{2}{c|}{\textbf{Noise Sources}} \\ \cline{3-6} 
\multicolumn{1}{|c|}{} & \multicolumn{1}{c|}{} & \multicolumn{1}{l|}{$\mathbf{J}=(J_{IX},J_{IY},J_{IZ},J_{ZX},J_{ZY},J_{ZZ})$} & $(\omega_0,\omega_1)$ &
\multicolumn{1}{l|}{\textbf{Readout $(r_0, r_1)$}} & 
\textbf{Time Offset $(\Delta t_{\text{eff},0},\Delta t_{\text{eff},1})$ [ns]} \\ \hline
\multirow{5}{*}{A} & 0 & 
\multicolumn{1}{l|}{$(58.47, 3.68, -5.00, 10.76, 2.29, -0.52)$} & $(69.71, 47.94)$  & \multicolumn{1}{l|}{$(0.160,0.215)$} & $(15, 101)$\\ \cline{2-6}
                   & 1 & 
\multicolumn{1}{l|}{$(38.93, 2.34, 0.26, 11.15, -3.30, -0.30)$} & $(50.09, 28.36)$ & \multicolumn{1}{l|}{$(0.150, 0.210)$} & $(97, 187)$\\ \cline{2-6}
                   & 2 & 
\multicolumn{1}{l|}{$(19.35, 0.12, 0.69, 10.80, -0.66, 0.24)$} & $(30.17, 8.60)$ & \multicolumn{1}{l|}{$(0.220,0.150)$} & $(178, 699)$\\ \cline{2-6}
                   & 3 & 
\multicolumn{1}{l|}{$(-0.21, -1.68, 0.20, 10.47, 1.50, -0.86)$} & $(10.28, 11.20)$ & \multicolumn{1}{l|}{$(0.145, 0.150)$} & $(579, 532)$\\ \cline{2-6}
                   & 4 & 
\multicolumn{1}{l|}{$(-20.11, -1.35, 0.73, 10.55, 0.94, -1.13)$} & $(9.58, 30.80)$ & \multicolumn{1}{l|}{$(0.190 , 0.185)$} & $(594, 174)$\\ \hline

\multirow{5}{*}{B} & 0 & 
\multicolumn{1}{l|}{$(30.03, 3.62, 0.49, 1.75, -0.16, -0.31)$} & $(31.97, 28.54)$ & \multicolumn{1}{l|}{$(0.110, 0.140)$} & $(149, 175)$\\ \cline{2-6}
                   & 1 & 
\multicolumn{1}{l|}{$(15.34, 1.85, 0.19, 1.81, -0.75, -0.59)$} & $(17.19, 13.80)$ & \multicolumn{1}{l|}{$(0.090, 0.070)$} & $(318, 407)$\\ \cline{2-6}
                   & 2 & 
\multicolumn{1}{l|}{$(0.89, 0.72, 0.24, 1.82, -0.54, -0.34)$} & $(2.72, 1.67)$ & \multicolumn{1}{l|}{$(0.120, 0.160)$} & $(2235, 3712)$\\ \cline{2-6}
                   & 3 & 
\multicolumn{1}{l|}{$(-13.71, -2.31, -0.54, 1.78, 0.09, 0.09)$} & $(12.15, 15.69)$ & \multicolumn{1}{l|}{$(0.130, 0.160)$} & $(472, 353)$\\ \cline{2-6}
                   & 4 & 
\multicolumn{1}{l|}{$(-28.45, -2.19, -1.20, 1.56, 2.15, 0.27)$} & $(26.91, 30.35)$ & \multicolumn{1}{l|}{$(0.110, 0.100)$} & $(187, 161)$\\ \hline

\multirow{5}{*}{C/$\text{CR}_{01}$} & 0 & 
\multicolumn{1}{l|}{$(-8.52, -2.15, -0.26, 10.93, 0.85, 0.32)$} & $(2.74, 19.69)$ & \multicolumn{1}{l|}{$(0.200, 0.160)$} & $(2280, 286)$\\ \cline{2-6}
                   & 1 & 
\multicolumn{1}{l|}{$(-3.88, -2.26, -0.35, 10.88, 1.46, 0.43)$} & $(7.04, 15.24)$ & \multicolumn{1}{l|}{$(0.120, 0.160)$} & $(869, 376)$\\ \cline{2-6}
                   & 2 & 
\multicolumn{1}{l|}{$(0.58, -1.81, -0.45, 10.81, 0.83, 1.24)$} & $(11.46, 10.70)$ & \multicolumn{1}{l|}{$(0.080, 0.070)$} & $(518, 563)$\\ \cline{2-6}
                   & 3 & 
\multicolumn{1}{l|}{$(4.86, -1.66, 0.08, 10.85, 0.44, -0.14)$} & $(15.75, 6.35)$ & \multicolumn{1}{l|}{$(0.070, 0.110)$} & $(369, 966)$\\ \cline{2-6}
                   & 4 & 
\multicolumn{1}{l|}{$(9.53, -0.17, 0.29, 10.76, -0.17, -0.32)$} & $(20.29, 1.36)$ & \multicolumn{1}{l|}{$(0.070, 0.120)$} & $(276, 4835)$\\ \hline

\multirow{5}{*}{C/$\text{CR}_{02}$} & 0 & 
\multicolumn{1}{l|}{$(9.42, -0.71, 0.27, 12.21, -0.71, -0.25)$} & $(21.67, 2.84)$ & \multicolumn{1}{l|}{$(0.070, 0.060)$} & $(255, 2123)$\\ \cline{2-6}
                   & 1 & 
\multicolumn{1}{l|}{$(6.17, -0.46, 0.09, 11.96, -0.59, -0.26)$} & $(18.16, 5.81)$ & \multicolumn{1}{l|}{$(0.070, 0.110)$} & $(311, 1038)$\\ \cline{2-6}
                   & 2 & 
\multicolumn{1}{l|}{$(2.59, 0.05, -0.26, 11.90, -1.66, -0.17)$} & $(14.58, 9.46)$ & \multicolumn{1}{l|}{$(0.050, 0.090)$} & $(397, 629)$\\ \cline{2-6}
                   & 3 & 
\multicolumn{1}{l|}{$(-1.03, -0.10, -0.16, 11.99, -0.63, -0.18)$} & $(10.99, 13.03)$ & \multicolumn{1}{l|}{$(0.090, 0.060)$} & $(541, 445)$\\ \cline{2-6}
                   & 4 & 
\multicolumn{1}{l|}{$(-4.53, 0.10, -0.31, 12.04, -0.18, 0.39)$} & $(7.50, 16.59)$ & \multicolumn{1}{l|}{$(0.100, 0.110)$} & $(811, 344)$\\ \hline
\end{tabular}
\caption{\label{tab:device_ham_params_noise_summary} Summary of estimated CR Hamiltonian parameters for the IBM Quantum devices A, B, and C. We give the Hamiltonian parameters in the parameterization $\mathbf{J}$ and the physically relevant frequency components in $\bm{\Lambda}$. The readout noise is defined by the parameters of $r_0$ and $r_1$ which are the conditional probabilities of bit flip given the measurement outcomes are $y=0$ and $y=1$ respectively (see Section~\ref{sec:readout_noise}). We show the results for CR Hamiltonians between two different pairs of qubits on Device C, specified as (control qubit, target qubit): $(0,1)$ ($\text{CR}_{01}$) and $(0,2)$ ($\text{CR}_{02}$).}
\end{table}

\begin{figure}[H]
    \centering
	\xincludegraphics[scale=0.35, label=\textbf{(a)}]{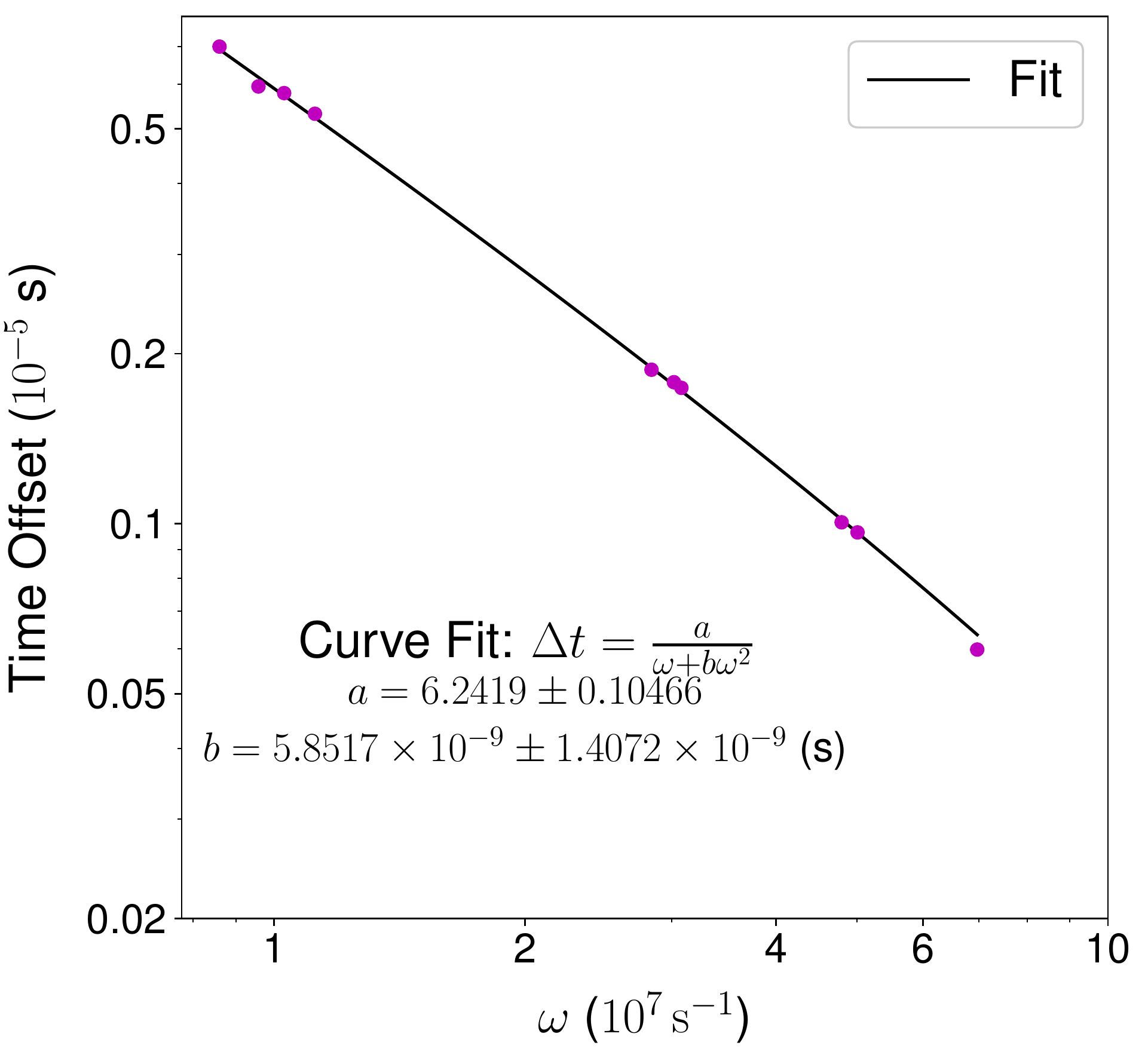}
    \hspace{4em}
    \xincludegraphics[scale=0.35, label=\textbf{(b)}]{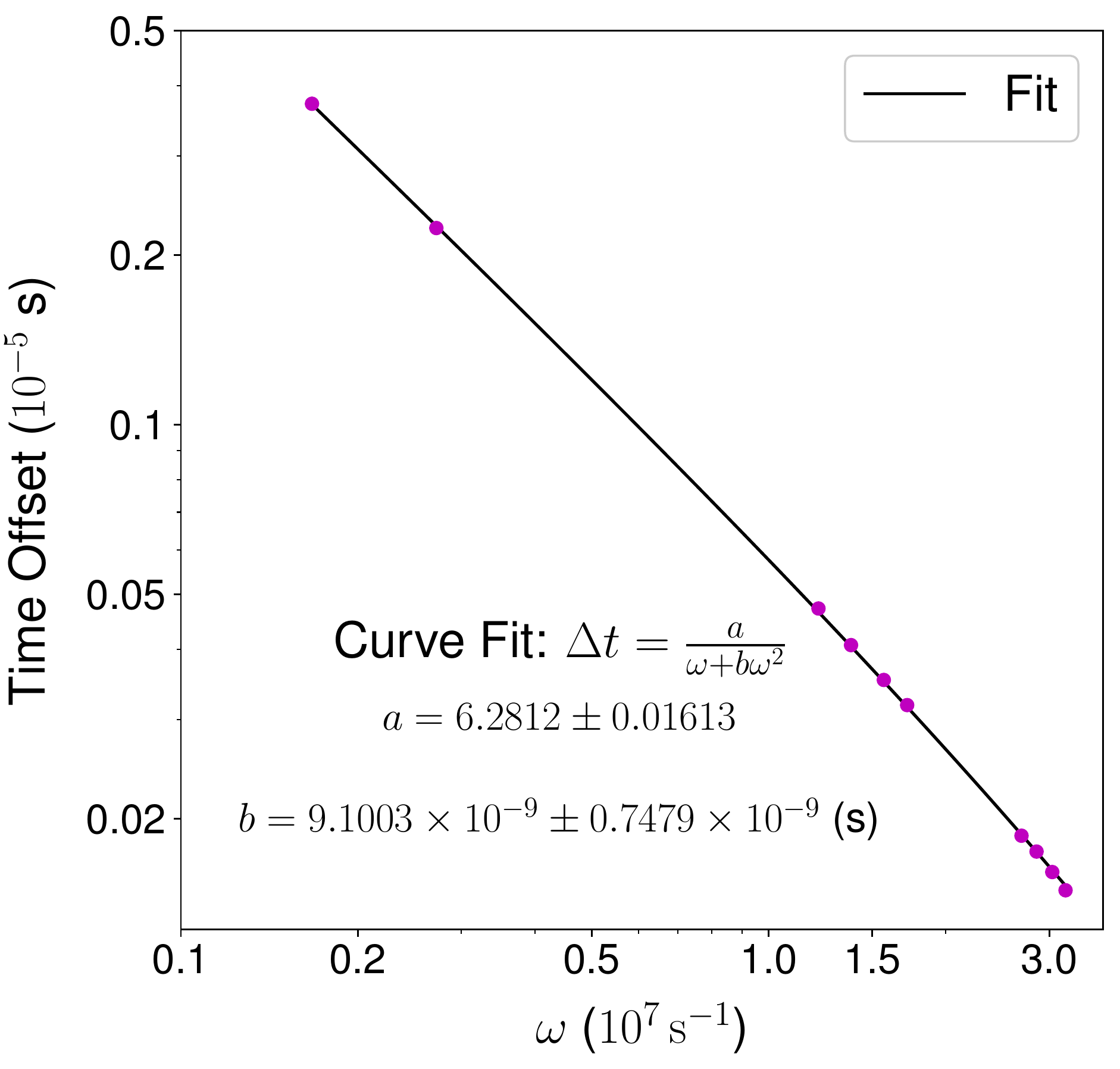}
    
    \xincludegraphics[scale=0.35, label=\textbf{(c)}]{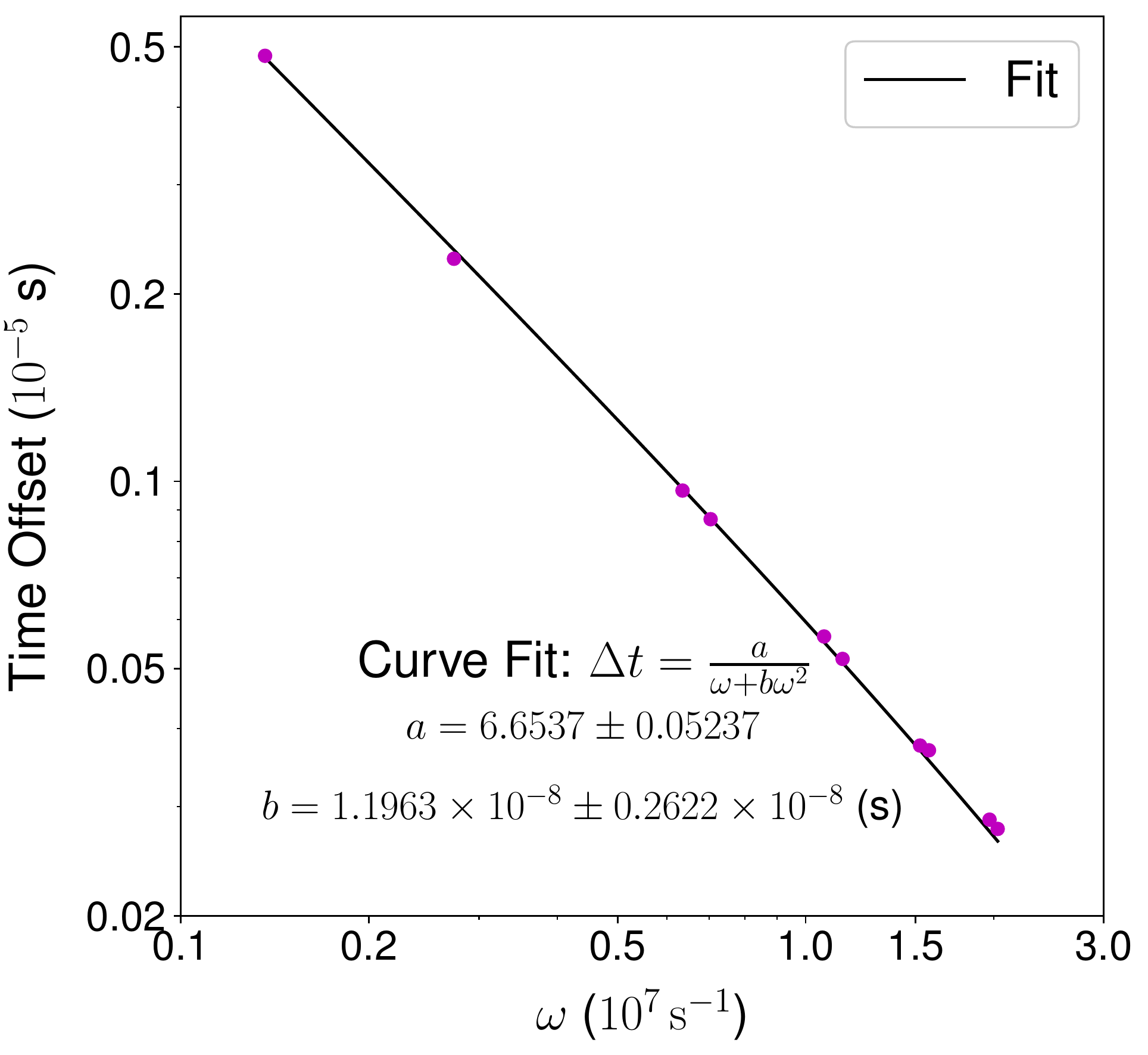}
    \hspace{4em}
    \xincludegraphics[scale=0.35, label=\textbf{(d)}]{Figures/imperfect_pulses/control_noise_model_devC_CR0_1.pdf}
    \caption{Dependence of the time offset $\Delta t$ on parameters $\omega$ for IBM Quantum devices (a) A, (b) B, (c) C $\text{CR}_{01}$, and (d) C $\text{CR}_{02}$. The plotted data points correspond to driving the device under different conditions and hence different cross-resonance Hamiltonians. The imperfect pulse shaping model extracted from these experimental data points is shown by a fit and this is later used in the MLE.}
	\label{fig:data_driven_model_imperfect_pulses}
\end{figure}

As mentioned earlier in Section~\ref{sec:imperfect_pulse_shaping}, we fit the estimated values of $\Delta t_{\text{eff}}$ to the Hamiltonian parameters to obtain a model for the time-offset. This is shown in Figure~\ref{fig:data_driven_model_imperfect_pulses} and is used in the MLE (Eq.~\ref{eq:mle_hl_noise}). 

\subsection{Expected trends of learning error}
In Section~\ref{sec:results_performance}, we assessed the performance of the HAL-FI and HAL-FIR algorithms in different learning scenarios on IBM Quantum device D \textit{ibmq\_boeblingen} under drive configuration 2, where the query distribution was learned in real-time. Here, we lend support that the trends observed in Figure~\ref{fig:CR_Error_Scalings_Simulator_Expt} and Figure~\ref{fig:reduced_HL_sim_expt_data} are expected. 

\begin{figure}[H]
	\centering
    \includegraphics[scale=0.3]{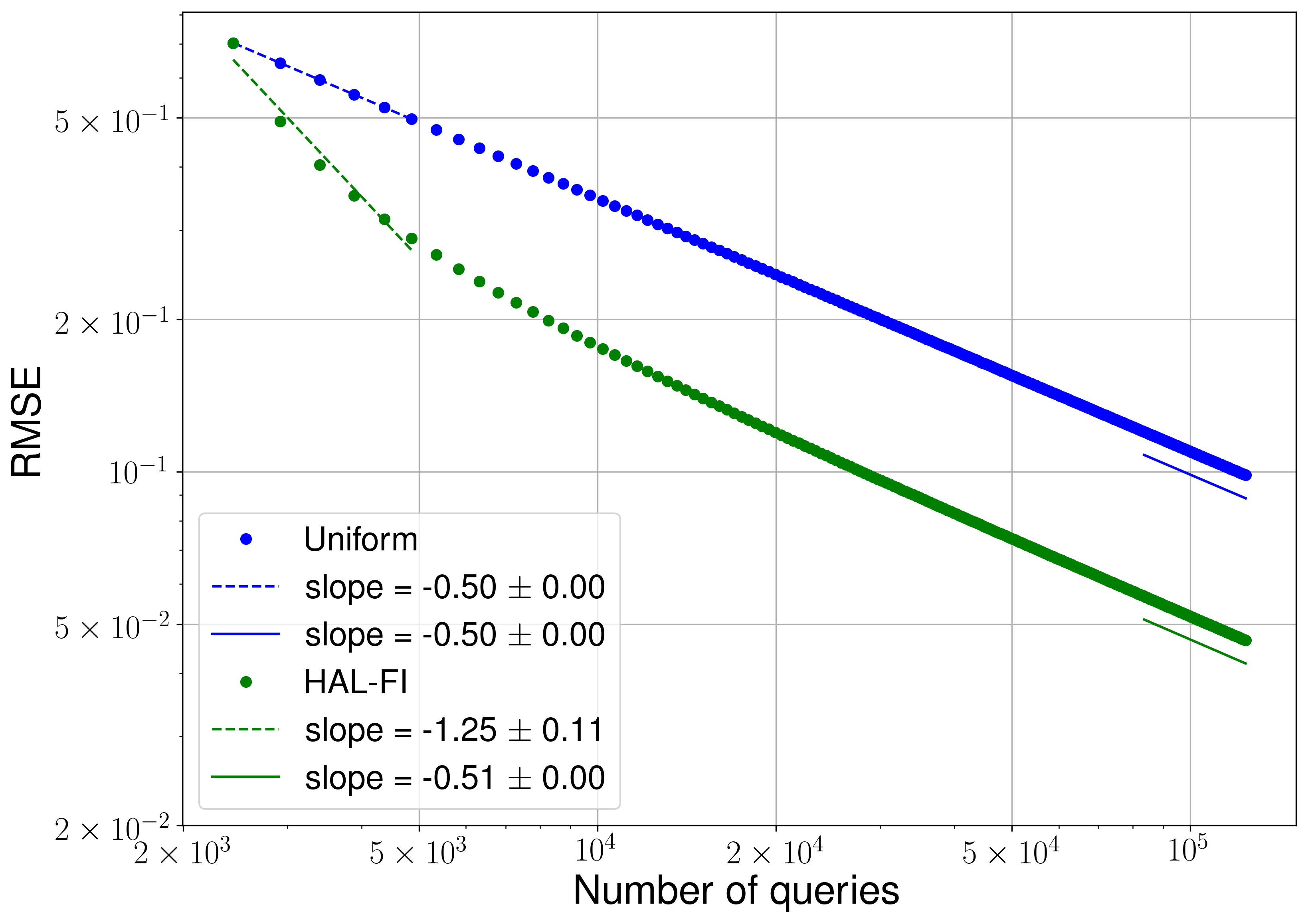}
    \caption{Scaling of RMSE with number of queries for Hamiltonian learning considering access to the asymptotic optimal query distribution $q(\bm{\theta}^\star)$ of HAL-FI and analysis of the Cramer-Rao Bounds. We consider the Hamiltonian parameters of $\bm{\theta}^\star$ as determined from the experimental dataset of IBM Quantum device D \textit{ibmq\_boeblingen} under drive configuration 2. We plot the trend of RMSE with number of queries for HAL-FI against the passive learner which uses the uniform distribution over $\querySpace$.}
	\label{fig:CR_Scalings_HL}
\end{figure}

To determine the behavior of the learners in an idealized setting, we consider the case where we have access to the optimal query distribution during learning. In Figure~\ref{fig:CR_Scalings_HL}, we show the trend of RMSE for HAL-FI with a fixed query space assuming access to the query distribution $q(\bm{\theta}^\star)$ during training and that the Cramer-Rao bound is saturated. We follow the same protocol from Section~\ref{sec:protocol_comparing_HL_methods} as we did for our earlier experiments. For the baseline strategy and a passive learner, the query distribution corresponds to an uniform distribution over the query space. As expected, using a passive learner does not change the scaling in the finite query nor the asymptotic query regimes. A scaling of $\epsilon \sim 1/\sqrt{N}$ or $N \sim \epsilon^{-2}$ is observed which is in line with the standard quantum limit (SQL). For HAL-FI, we observe an initial scaling in RMSE with number of queries which is higher than SQL but this reduces to SQL in the asymptotic query regime. Thus, our results from Section~\ref{sec:results_performance} is in agreement with what we observe here. Asymptotically, we expect a constant savings in the number of queries or resources required when employing an active learner with a fixed query space compared to a passive learner.

\begin{figure}[H]
	\centering
    \includegraphics[scale=0.35]{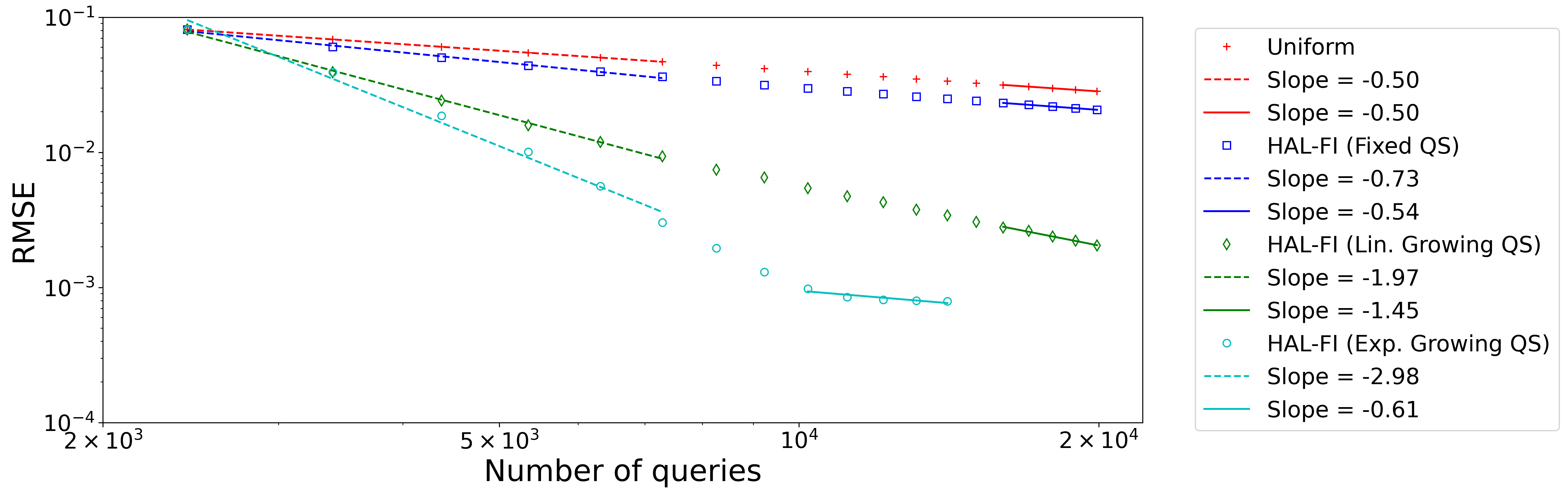}
    \caption{Hamiltonian learning with acess to prior information of subset of parameters during recalibration: Scaling of RMSE with Number of Queries. We assume access to the asymptotic optimal query distribution $q(\bm{\theta}^\star)$ of HAL-FI and analysis of the Cramer-Rao Bounds. We consider the Hamiltonian parameters of $\bm{\theta}^\star$ as determined from the experimental dataset of IBM Quantum device D \textit{ibmq\_boeblingen} under drive configuration 3.}
    \label{subfig:CR_RMSE_Scalings_reduced_HL}
\end{figure}

Likewise for Hamiltonian learning with prior information, we can compute the RMSE with number of queries for HAL-FI with an adaptively growing query space and access to $q^{(i)}(\bm{\theta}^\star)$ for each $i$th batch during learning. In Figure~\ref{subfig:CR_RMSE_Scalings_reduced_HL}, we show the trend of RMSE with number of queries for HAL-FI with a linearly growing query space and an exponentially growing query space. We observe that the passive learner has a scaling of the SQL in the asymptotic query regime. HAL-FI in the linearly growing query space achieves super-Heisenberg scaling. We note that this supports the trend of RMSE achieved during the experiments in Section~\ref{sec:results_analysis_HS}.

As noted in Section~\ref{sec:results_analysis_HS}, HAL-FI with an exponentially growing query space also achieves Heisenberg limited scaling until the evolution times being included in the query space reach the magnitude of the decoherence time $T_1$ and $T_2$. In this case, HAL-FI avoids selecting higher evolution times as the information gained from these measurement outcomes will tend to zero. Thus, we expect that the Heisenberg limit to be achieved for the finite query setting.

\subsection{Sparse Query Distributions}  \label{app_sec:sparse_expt_designs}
\begin{figure}[ht!]
	\centering
    \includegraphics[scale=0.35]{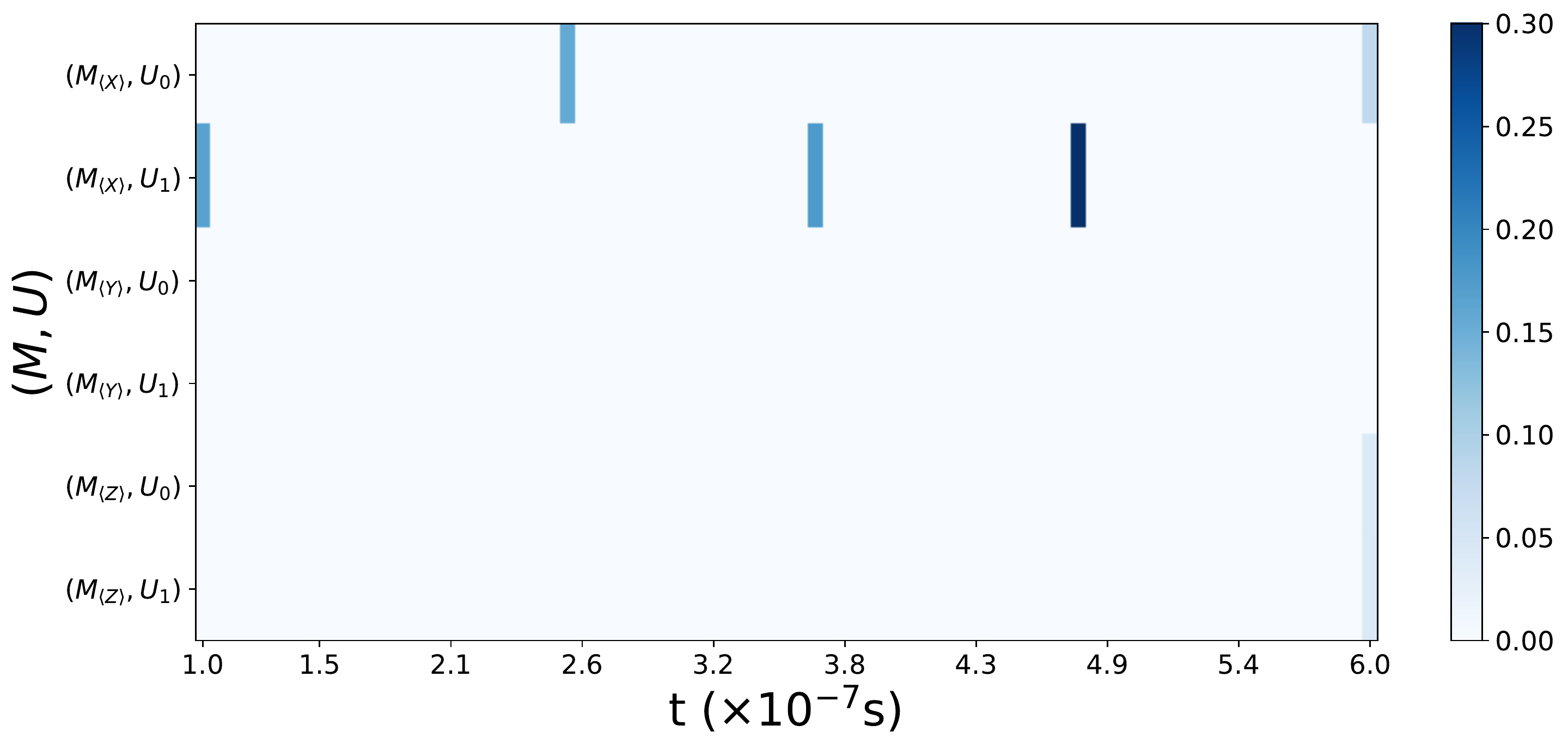}
    \caption{Asymptotic optimal query distribution $q(\bm{\theta}^\star)$ for HAL-FI with a fixed query space (Section~\ref{sec:cr_query_space}) on \textit{ibmq\_boeblingen} under drive configuration 2 (Table~\ref{tab:ibmq_boeb_ham_params_summary}). We consider different noise sources of readout noise, imperfect pulse-shaping, and decoherence. The y-axis indicates the different combinations of measurement operators and preparation operators available for each query in $\querySpace$. The different preparation operators are denoted as $U_0 = \sigma_I\sigma_I$ and $U_1=\sigma_X\sigma_I$. The x-axis corresponds to $\timeSpace$ which is set to $81$ equispaced evolution times in $[10^{-7},6\times 10^{-7}]s$. The query distribution is color-coded according to the colormap on the right.}
    \label{fig:AL_optimal_q_fixed_query_space}
\end{figure}

Another consequence of using HAL-FI is the sparsity of the asymptotic query distribution during learning. This is confirmed by visualizing the optimal query distribution of HAL-FI with the fixed query space (Section~\ref{sec:cr_query_space}) considering IBM Quantum device \textit{ibmq\_boeblingen} under drive configuration 2 in Figure~\ref{fig:AL_optimal_q_fixed_query_space}. It is interesting to note that this was achieved even though sparsity was not incorporated into the learning problem. This can be explained by realizing that the most informative queries are in fact sparse over the query space. It should be noted however that this is typically not the query distribution that HAL-FI has access to during learning as the true parameters $\bm{\theta}^\star$ are not available and the query distribution obtained through optimization (Eq.~\ref{eq:query_optimization_variance_of_params}) is modified by mixing with the uniform distribution (see Section~\ref{sec:HAL-FI_algo_description}).

\section{Heisenberg Limited Scaling in Cross-Resonance Type Hamiltonians} \label{app_sec:HLS_Discovery}
In this section of the Appendix, we discuss if Heisenberg limited scaling (HLS) can be achieved in different quantum systems with simplified cross-resonance Hamiltonians. The simplified cross-resonance Hamiltonians are obtained by removing particular Pauli product terms from the cross-resonance (CR) Hamiltonian (\ref{sec:CR_hamiltonian}) that we have studied so far. We give examples of such Hamiltonians (equipped with the query space described in Section~\ref{sec:CR_Gate_Model_and_Setup}) in Section~\ref{subsec:examples_hls} where HLS is achieved for all the parameters and examples where HLS is not achieved for all the parameters during Hamiltonian learning in Section~\ref{subsec:non-examples_hls}. For examples of HLS, we describe query distributions obtained by HAL-FI and relate them to query distributions obtained through zero crossings of Rabi oscillations or maximum entropy. 

In the following examples, we consider reduction to a single interaction in the two qubit system and a three interaction example which we call the simplified cross-resonance (SCR) Hamiltonian. This reduction from the full CR Hamiltonian is achieved by setting the appropriate Hamiltonian parameters to zero and introducing the matrix $R$ that denotes which parts of the CR Fisher information matrix are involved in the estimation of these respective Hamiltonian parameters. Suppose the reduced set of parameters are collected into the vector $\bm{\theta}_{R}$, then the Cramer-Rao bound is now
\begin{equation}
    \sum_i \Var((\bm{\theta}_R)_i) \geq \frac{1}{N} \Tr(\fisherInfo_q^{-1}(\boldsymbol{\theta}_R)) = \frac{1}{N} \Tr(R^{-T} \fisherInfo_q^{-1}(\boldsymbol{\theta}) R^{-1})
\end{equation}
where $\fisherInfo_q(\bm{\theta}_R)$ is the reduced Fisher information matrix corresponding to the query distribution $q$. In the last step, we noted the relation of the reduced Fisher information matrix with the full Fisher information matrix as $\fisherInfo_q(\bm{\theta}_R) = R \fisherInfo_q(\bm{\theta}) R^T$.

\subsection{Examples} \label{subsec:examples_hls}
\subsubsection{Single Interaction Two-Qubit System}
The Hamiltonian of interest in this case is 
\begin{equation}
    H = J_{ZX} \sigma_Z \otimes \sigma_X
    \label{eq:reduced_Hamiltonian_Jzx}
\end{equation}
This may be obtained by considering the parameter set of $\mathbf{J} = (0,0,0,J_{ZX},0,0)^T$ and $R=[0, 0, 0, 1, 0, 0]$. The values of the parameters in the alternate parameterization of $\bm{\Lambda} = (\omega_0, \delta_0, \phi_0, \omega_1, \delta_1, \phi_1)^T = (J_{ZX}, 0, 0, J_{ZX}, 0, \pi)$ where we have assumed $J_{ZX}>0$. The Rabi oscillations for different queries in this case are as follows:
\begin{align}
    (M_{\braketLR{X}},U_j,t): \, & p_{\mathrm{rabi}}(x) = 0 \\
    (M_{\braketLR{Y}},U_j,t): \, & p_{\mathrm{rabi}}(x) = \sin(2 \omega_j t) \\
    (M_{\braketLR{Z}},U_j,t): \, & p_{\mathrm{rabi}}(x) = \cos(2\omega_j t)
    \label{eq:rabi_oscillations_reduced_Jzx}
\end{align}
where the index $j \in \{0,1\}$ corresponds to different preparation operators $U_0=\sigma_I \sigma_I$ and $U_1=\sigma_X \sigma_I$. We note that the measurement operator of $M_{\braketLR{X}}$ is not informative about the frequency $\omega_0=J_{ZX}$ for this system and can also be noted from considering the corresponding Fisher information. In order to learn the parameter of interest $\omega_0$, it is enough to consider one of the queries in $\{M_{\braketLR{Y}}, M_{\braketLR{Z}}\} \times \{\sigma_I \sigma_I, \sigma_X \sigma_I\}$ and a suitable time range $\timeSpace$.

Let us select the query of $(M_{\braketLR{Z}}, \sigma_I \sigma_I, t)$ and suppose our query distribution over the time range is based on the zeros the Rabi oscillations. Given a time range $\mathcal{T}$, we consider values of $t_k = \frac{\pi}{4 \omega_0} + \frac{k \pi}{2 \omega_0}$ where $k \in \mathbb{N}$. Queries with these system evolution times have the maximum entropy for the considered $(M,U)$.

Note that the Fisher information of a query in this case is given by $\fisherInfo_x(\omega_0) = 4 t^2$. Through the Cramer-Rao bound, we then have
\begin{equation}
    \epsilon^2 = \Var(\omega_0) \geq \sum_{k=1}^N \frac{1}{4 t_k^2} \approx 
    \begin{cases}
        \frac{1}{2} \frac{\omega^2}{N \pi^2} &, \text{ Fixed space of zero crossings} \\
        \frac{3}{4} \frac{\omega^2}{N^3 \pi^2} &, \text{ Linearly spaced zero crossings} \\
    \end{cases}
\end{equation}
where we have set the learning error to be achieved as $\epsilon$. We can thus expect to achieve a scaling of $N \sim \epsilon^{-3/2}$ when using linearly spaced zero crossings. For exponentially spaced zero crossings of the Rabi oscillations, the variance approaches zero at an increasing rate. 

\subsubsection{Two Interaction Systems}
Consider the following Hamiltonian 
\begin{equation}
    H = J_{IX} \sigma_I \otimes \sigma_X + J_{ZX} \sigma_Z \otimes \sigma_X
\end{equation}
The reduced set of parameters is then $\bm{\Lambda}_R = (\omega_0, \omega_1) = (|J_{IX} + J_{ZX}|, |J_{IX} - J_{ZX}|)$. As in the earlier single interaction example, we can choose the queries that contain the zero crossings of the Rabi oscillations. The Rabi oscillations are given by
\begin{align}
    (M_{\braketLR{X}},U_j,t): \, & p_{\mathrm{rabi}}(x) = 0 \\
    (M_{\braketLR{Y}},U_j,t): \, & p_{\mathrm{rabi}}(x) = \sin(2\omega_j t) \\
    (M_{\braketLR{Z}},U_j,t): \, & p_{\mathrm{rabi}}(x) = \cos(2\omega_j t)
    \label{eq:rabi_oscillations_reduced_CR_two_interactions}
\end{align}
where $j\in \{0,1\}$ is used as an index to denote the preparation operators $U_0 = \sigma_I \sigma_I$ and $U_1 = \sigma_X \sigma_I$. A complete set of queries to estimate $(\omega_0, \omega_1)$ with Heisenberg limited scaling would then be 
\begin{equation}
    \mathcal{Q} = \Big\{(M_{\braketLR{Z}}, \sigma_I \sigma_I, t_k) : t_k = \frac{\pi}{4 \omega_0} + \frac{k \pi}{2 \omega_0}, k \in \mathbb{N} \Big\} \bigcup \Big\{(M_{\braketLR{Z}}, \sigma_X \sigma_I, t_k) : t_k =  \frac{\pi}{4 \omega_1} + \frac{k \pi}{2 \omega_1}, k \in \mathbb{N} \Big\}    
\end{equation}

The set of evolution times chosen here also correspond to those with maximum entropy. The Fisher information of a query made through either set of measurement or preparation operators in the query space is given by $\fisherInfo_x(\omega_0,\omega_1) = 4t^2$. Thus, through the argument we made in the previous section, we can also learn the parameters of unitary here with Heisenberg limited scaling.

\subsection{Examples of non-HLS scaling during Hamiltonian learning} \label{subsec:non-examples_hls}
So far, we have given examples of Hamiltonians obtained through simplification of the CR Hamiltonian, that can be learned with HLS scaling. We now give examples of Hamiltonians, which cannot be learned with HLS scaling using the query space described in Section~\ref{sec:cr_query_space}.

\subsubsection{Two Interaction Systems}
Now, let us consider the following alternate Hamiltonian (modified slightly from the previous example discussed)
\begin{equation}
    H = J_{IY} \sigma_I \otimes \sigma_Y + J_{ZX} \sigma_Z \otimes \sigma_X
\end{equation}
where a complete reduced set of parameters is $\bm{\Lambda_R} = (\omega_0, \phi_0) = \left(\sqrt{J_{ZX}^2 + J_{IY}^2}, \tan^{-1}\left( \frac{J_{IY}}{J_{ZX}} \right) \right)$. This set of parameters contains a frequency in addition to a phase. Fisher information matrices in the $\bm{\Lambda}_R$ parameterization considering queries of the form $(M, \sigma_I \sigma_I, t)$ where we select one particular preparation operator is given by
\begin{align}
    M_{\braketLR{X}}: \, & \fisherInfo = \frac{1}{1 - \sin^2(\phi_0) \sin^2(2\omega_0 t) } \begin{bmatrix} 4t^2 \sin^2(\phi_0) \cos^2(2 \omega_0 t) & \frac{1}{2} t \sin(2 \phi_0) \sin(4 \omega_0 t) \\ \frac{1}{2} t \sin(2 \phi_0) \sin(4 \omega_0 t) & \cos^2(\phi_0) \sin^2(2 \omega_0 t) \end{bmatrix} \\
    M_{\braketLR{Y}}: \, & \fisherInfo = \frac{1}{1 - \cos^2(\phi_0) \sin^2(2\omega_0 t) } \begin{bmatrix} 4t^2 \cos^2(\phi_0) \cos^2(2 \omega_0 t) & -\frac{1}{2} t \sin(2 \phi_0) \sin(4 \omega_0 t) \\ -\frac{1}{2} t \sin(2 \phi_0) \sin(4 \omega_0 t) & \sin^2(\phi_0) \sin^2(2 \omega_0 t) \end{bmatrix} \\
    M_{\braketLR{Z}}: \, & \fisherInfo = \begin{bmatrix} 4t^2 & 0 \\ 0 & 0 \end{bmatrix} 
    \label{eq:fisher_info_reduced_Jiy_Jzx}
\end{align}

Fisher information matrices in $\mathbf{J}_R$ is given by
\begin{equation}
    \fisherInfo_J = 
    \begin{bmatrix}
        \frac{J_{IY}^2}{\omega_0^2} \fisherInfo_{11} + \frac{2 J_{IY} J_{ZX}}{\omega_0^3} \fisherInfo_{12} + \frac{J_{ZX}^2}{\omega_0^4} \fisherInfo_{22} & \frac{J_{ZX}}{\omega_0^3} \fisherInfo_{12} - \frac{J_{IY}}{\omega_0^3} \fisherInfo_{12} + \frac{J_{IY} J_{ZX}}{\omega_0^2} (\fisherInfo_{11} - \fisherInfo_{12}) \\
        \cdot & \frac{J_{ZX}^2}{\omega_0^2} \fisherInfo_{11} - \frac{2 J_{IY} J_{ZX}}{\omega_0^3} \fisherInfo_{12} + \frac{J_{IY}^2}{\omega_0^4} \fisherInfo_{22}
    \end{bmatrix}
\end{equation}
where we have related them to elements of $\fisherInfo$ as given above in Eq.~\ref{eq:fisher_info_reduced_Jiy_Jzx}. We observe that in order to obtain HLS in $(J_{IY}, J_{ZX})$, it is necessary to set $\fisherInfo_{22}=0$ and $\fisherInfo_{12} \neq 0$ to ensure $\fisherInfo_J$ is full rank and there is an explicit dependence on the variable $t$ that we can take advantage of, for HLS. However, the required conditions cannot be achieved simultaneously here. This suggests that the current set of queries cannot be used to achieve HLS. 

\subsubsection{Three Interaction Simplified Cross-Resonance Gate}
The Hamiltonian of interest in this case is 
\begin{equation}
    H = J_{IX} \sigma_I \otimes \sigma_X + J_{IY} \sigma_I \otimes \sigma_Y + J_{ZX} \sigma_Z \otimes \sigma_X
    \label{eq:reduced_simple_CR_Hamiltonian}
\end{equation}
This may be obtained by considering the parameter set of $\mathbf{J} = (J_{IX},J_{IY},0,J_{ZX},0,0)^T$. Defining $R$ is a bit more tricky in this case compared to the previous example. Let us first look at the alternate parameterization of $\bm{\Lambda}=\left(\sqrt{(J_{IX} + J_{ZX})^2 + J_{IY}^2}, 0, \tan^{-1} \left(\frac{J_{IY}}{J_{IX} + J_{ZX}}\right), \sqrt{(J_{IX} - J_{ZX})^2 + J_{IY}^2}, 0, \tan^{-1} \left(\frac{J_{IY}}{J_{IX} - J_{ZX}}\right)\right)^T$. Note that the reduced parameterization of $\Lambda_R$ is an over-parameterization with four non-zero components compared to $\mathbf{J}_R$ which only has three non-zero components. The Rabi oscillations in this case are given by
\begin{align}
    (M_{\braketLR{X}},U_j,t): \, & p_{\mathrm{rabi}}(x) = \sin(\phi_j) \sin(2 \omega_j t) \\
    (M_{\braketLR{Y}},U_j,t): \, & p_{\mathrm{rabi}}(x) = \cos(\phi_j) \sin(2 \omega_j t) \\
    (M_{\braketLR{Z}},U_j,t): \, & p_{\mathrm{rabi}}(x) = \cos(2\omega_j t)
    \label{eq:rabi_oscillations_reduced_SCR}
\end{align}
The Fisher information matrices can be obtained by looking at the CR Fisher information matrices (Appendix~\ref{app_sec:details_cr_hamiltonians}) and simplifying them. 
\begin{align}
    M_{\braketLR{X}}: \, & \fisherInfo_j = \frac{1}{1 - \sin^2(\phi_j) \sin^2(2\omega_j t) } \begin{bmatrix} 4t^2 \sin^2(\phi_j) \cos^2(2 \omega_j t) & \frac{1}{2} t \sin(2 \phi_j) \sin(4 \omega_j t) \\ \frac{1}{2} t \sin(2 \phi_j) \sin(4 \omega_j t) & \cos^2(\phi_j) \sin^2(2 \omega_j t) \end{bmatrix} \\
    M_{\braketLR{Y}}: \, & \fisherInfo_j = \frac{1}{1 - \cos^2(\phi_j) \sin^2(2\omega_j t) } \begin{bmatrix} 4t^2 \cos^2(\phi_j) \cos^2(2 \omega_j t) & -\frac{1}{2} t \sin(2 \phi_j) \sin(4 \omega_j t) \\ -\frac{1}{2} t \sin(2 \phi_j) \sin(4 \omega_j t) & \sin^2(\phi_j) \sin^2(2 \omega_j t) \end{bmatrix} \\
    M_{\braketLR{Z}}: \, & \fisherInfo_j = \begin{bmatrix} 4t^2 & 0 \\ 0 & 0 \end{bmatrix} 
    \label{eq:fisher_info_reduced_Jzx}
\end{align}
If we were to consider the zero crossings of the Rabi oscillations as in the previous example, the queries and their corresponding Fisher information matrices are of the following form
\begin{align}
    M_{\braketLR{X}}: \, & t_k(M_{\braketLR{X}}) = \frac{\pi}{2 \omega_j} + \frac{k \pi}{2 \omega_j}, \, \fisherInfo_j = \begin{bmatrix} 4t_k^2 \sin^2(\phi_j) & 0 \\ 0 & 0 \end{bmatrix}\\
    M_{\braketLR{Y}}: \, & t_k(M_{\braketLR{Y}}) = \frac{\pi}{2 \omega_j} + \frac{k \pi}{2 \omega_j}, \, \fisherInfo_j = \begin{bmatrix} 4t_k^2 \cos^2(\phi_j) & 0 \\ 0 & 0 \end{bmatrix}\\
    M_{\braketLR{Z}}: \, & t_k(M_{\braketLR{Z}}) = \frac{\pi}{4 \omega_j} + \frac{k \pi}{2 \omega_j}, \, \fisherInfo_j = \begin{bmatrix} 4t_k^2 & 0 \\ 0 & 0 \end{bmatrix}
    \label{eq:rabi_oscillations_zeros_reduced_SCR}
\end{align}
where $k \in \mathbb{N}$. It should be noted that the evolution times $t_k(M)$ being selected are a function of the measurement operator involved in the query which is made explicit through the argument $M$. As $\bm{\Lambda}_R$ is an over-parameterization, let us look at the query Fisher information matrix $\fisherInfo_q(\mathbf{J}_R)$ for the above set of queries.
\begin{align}
    \fisherInfo_q(\mathbf{J}_R) &= \sum_k \sum_{M \in \{M_{\braketLR{X}}, M_{\braketLR{Z}} \} } 4 t_k^2(M) \begin{bmatrix} \left(\frac{\partial \omega_0}{\partial J_{IX}}\right)^2 + \left(\frac{\partial \omega_1}{\partial J_{IX}}\right)^2 & \left( \frac{\partial \omega_0}{\partial J_{IX}} \frac{\partial \omega_0}{\partial J_{IY}} + \frac{\partial \omega_1}{\partial J_{IX}} \frac{\partial \omega_1}{\partial J_{IY}} \right) & \left( \frac{\partial \omega_0}{\partial J_{IX}} \frac{\partial \omega_0}{\partial J_{ZX}} + \frac{\partial \omega_1}{\partial J_{IX}} \frac{\partial \omega_1}{\partial J_{ZX}} \right) \\ & \left(\frac{\partial \omega_0}{\partial J_{IY}}\right)^2 + \left(\frac{\partial \omega_1}{\partial J_{IY}}\right)^2 & \left( \frac{\partial \omega_0}{\partial J_{IY}} \frac{\partial \omega_0}{\partial J_{ZX}} + \frac{\partial \omega_1}{\partial J_{IY}} \frac{\partial \omega_1}{\partial J_{ZX}} \right) \\ & & \left(\frac{\partial \omega_0}{\partial J_{ZX}}\right)^2 + \left(\frac{\partial \omega_1}{\partial J_{ZX}}\right)^2\end{bmatrix} 
\end{align}
where we have only given the upper-triangular part of the symmetric matrix. It can be shown that for these queries, $\fisherInfo_q(\mathbf{J}_R)$ is rank deficient and thus non-invertible. This was foreshadowed by the fact that $\fisherInfo_q(\bm{\Lambda})$ was informative in $\omega_0$ and $\omega_1$ but not one of the phases $\phi_{0,1}$. Hence, it is more appropriate to consider $R=\begin{bmatrix}1 & 0  & 0 & 0 & 0 & 0 \\ 0 & 0  & 0 & 1 & 0 & 0 \end{bmatrix}$ for these set of queries. It can be verified through an analysis similar to the single interaction system that we can achieve a scaling of $N \sim O(\epsilon^{-3/2})$ and hence make improvements over than SQL. 

If we wish to learn $J_{IY}$ as well, it is necessary to introduce other queries such that the Fisher information matrix is non-zero for the corresponding parameter of interest. Let us start by changing our learning task to the simpler challenge of learning the parameters $(\omega_0, \delta_0)$. In this case, it is enough to consider only queries of the form $(M_{\braketLR{X}},\sigma_I \otimes \sigma_I, t)$ where the time range $t \in \mathcal{T}$ needs to be determined. We immediately observe that 
\begin{equation}
    \fisherInfo_q(\bm{\Lambda}_R)^{-1}  \propto \sum_k \frac{1}{1 - \sin^2(\phi_0) \sin^2(2\omega_0 t_k)} \begin{bmatrix} \cos^2(\phi_j) \sin^2(2 \omega_j t) & -\frac{1}{2} t \sin(2 \phi_j) \sin(4 \omega_j t) \\ -\frac{1}{2} t \sin(2 \phi_j) \sin(4 \omega_j t) & 4t_k^2 \sin^2(\phi_j) \cos^2(2 \omega_j t) \end{bmatrix}
\end{equation}
and the variance of parameter $\phi_0$ 
\begin{equation}
    \Var(\phi_0) \geq \frac{1}{N} \sum_{k=1}^N \frac{1 - \sin^2(\phi_0) \sin^2(2\omega_0 t_k)}{\cos^2(\phi_0) \sin^2(2 \omega_0 t_k)}
\end{equation}
where the term inside the sum on the right hand side is fixed for any periodic or equi-spaced set of evolution times $t_k$ and thus HLS cannot be achieved using such a set of queries. One key to ensure achieving Heisenberg limited scaling is to introduce an explicit dependence on the variable of system evolution time $t$ into the corresponding Fisher information. We note that this is not followed by the different set of measurement operators considered here. 


\clearpage

\bibliographystyle{IEEEtran}
\bibliography{references}

\end{document}